\documentclass[trackchanges,twocolumn]{aastex701}
\usepackage{placeins}
\usepackage{textcomp}
\usepackage{newunicodechar}
\newunicodechar{−}{\ensuremath{-}}
\usepackage{amsmath}
\usepackage{adjustbox}

\begin{document}

\title{HETDEX [\ion{O}{2}] galaxies at $z \le 0.48$: Volume-limited samples and their power spectra}

\author[0009-0009-2135-7061]{Jeongin Moon}
\affiliation{Max-Planck-Institut f\"{u}r Astrophysik, Karl-Schwarzschild-Str. 1, 85741 Garching, Germany}
\affiliation{Ludwig-Maximilians-Universität München, Schellingstr. 4, 80799 München, Germany}
\email{jmoon@mpa-garching.mpg.de}

\author[0000-0002-0136-2404]{Eiichiro Komatsu}
\affiliation{Max-Planck-Institut f\"{u}r Astrophysik, Karl-Schwarzschild-Str. 1, 85741 Garching, Germany}
\affiliation{Ludwig-Maximilians-Universität München, Schellingstr. 4, 80799 München, Germany}
\affiliation{Kavli Institute for the Physics and Mathematics of the Universe (WPI), The University of Tokyo Institutes for Advanced Study (UTIAS), The University of Tokyo, Chiba 277-8583, Japan}
\email{komatsu@mpa-garching.mpg.de}

\author[0000-0002-1328-0211]{Robin Ciardullo}
\affiliation{Department of Astronomy \& Astrophysics, The Pennsylvania State University, University Park, PA 16802, USA}
\affiliation{Institute for Gravitation and the Cosmos, The Pennsylvania State University, University Park, PA 16802, USA}
\email{rbc3@psu.edu}

\author[orcid=0000-0002-0212-4563,sname='North America']{Olivia Curtis}
\affiliation{Department of Astronomy \& Astrophysics, The Pennsylvania State University, University Park, PA 16802, USA}
\affiliation{Institute for Gravitation and the Cosmos, The Pennsylvania State University, University Park, PA 16802, USA}
\affiliation{Penn State Extraterrestrial Intelligence Center, 525 Davey Laboratory, 251 Pollock Road, Penn State, University Park, PA 16802, USA}
\email{ocurtis@psu.edu}

\author[0000-0002-8925-9769]{Dustin Davis}
\affiliation{Department of Astronomy, The University of Texas at Austin, 2515 Speedway Boulevard, Austin, TX 78712, USA}
\email{dustin@astro.as.utexas.edu}

\author[0000-0003-2575-0652]{Daniel J. Farrow}
\affiliation{E. A. Milne Centre for Astrophysics, University of Hull, Cottingham Road, Hull, HU6 7RX, UK}
\affiliation{Centre of Excellence for Data Science, Artificial Intelligence \& Modelling (DAIM), University of Hull, Cottingham Road, Hull, HU6 7RX, UK}
\email{D.J.Farrow@hull.ac.uk}

\author[0000-0002-8433-8185]{Karl Gebhardt}
\affiliation{Department of Astronomy, The University of Texas at Austin, 2515 Speedway Boulevard, Austin, TX 78712, USA}
\email{gebhardt@astro.as.utexas.edu}

\author[0000-0001-6842-2371]{Caryl Gronwall}
\affiliation{Department of Astronomy \& Astrophysics, The Pennsylvania State University, University Park, PA 16802, USA}
\affiliation{Institute for Gravitation and the Cosmos, The Pennsylvania State University, University Park, PA 16802, USA}
\email{cag18@psu.edu}

\author[0000-0001-9054-1414]{Laura Herold}
\affiliation{William H. Miller III Department of Physics and Astronomy, Johns Hopkins University, 3400 North Charles Street, Baltimore, MD 21218, USA}
\email{lherold@jhu.edu}

\author[0000-0001-6717-7685]{Gary J. Hill}
\affiliation{McDonald Observatory, The University of Texas at Austin, 2515 Speedway Boulevard, Stop C1402, Austin, TX 78712, USA}
\affiliation{Department of Astronomy, The University of Texas at Austin, 2515 Speedway Boulevard, Austin, TX 78712, USA}
\email{hillgj@eid.utexas.edu}

\author[0000-0002-8434-979X]{Donghui Jeong}
\affiliation{Department of Astronomy \& Astrophysics, The Pennsylvania State University, University Park, PA 16802, USA}
\affiliation{Institute for Gravitation and the Cosmos, The Pennsylvania State University, University Park, PA 16802, USA}
\affiliation{School of Physics, Korea Institute for Advanced Study, 85 Heogiro, Dongdaemun-gu, Seoul, 02455, Republic of Korea}
\email{djeong@psu.edu}

\author[0000-0001-5561-2010]{Chenxu Liu}
\affiliation{South-Western Institute for Astronomy Research, Key Laboratory of Survey Science of Yunnan Province, Yunnan University, Kunming, Yunnan 650500, People's Republic of China}
\email{lorenaustc@gmail.com}

\author[0000-0002-6907-8370]{Maja {Lujan Niemeyer}}
\affiliation{Max-Planck-Institut f\"{u}r Astrophysik, Karl-Schwarzschild-Str. 1, 85741 Garching, Germany}
\affiliation{Ludwig-Maximilians-Universität München, Schellingstr. 4, 80799 München, Germany}
\email{maja@MPA-Garching.MPG.DE}

\author[0000-0002-2307-0146]{Erin {Mentuch Cooper}}
\affiliation{Department of Astronomy, The University of Texas at Austin, 2515 Speedway Boulevard, Austin, TX 78712, USA}
\email{erin.hetdex@gmail.com}

\author[0000-0003-3823-8279]{Shiro Mukae}
\affiliation{Department of Physics, School of Advanced Science and Engineering, Faculty of Science and Engineering, Waseda University, 3-4-1 Okubo, Shinjuku, Tokyo 169-8555, Japan}
\affiliation{MIRAI Technology Institute, Shiseido Co., Ltd., 1-2-11, Takashima, Nishi-ku, Yokohama, Kanagawa, 222-0011, Japan}
\email{shiromukae88@gmail.com}

\author[0000-0002-6186-5476]{Shun Saito}
\affiliation{Institute for Multi-messenger Astrophysics and Cosmology, Department of Physics, Missouri University of Science and Technology, 1315 N. Pine St., Rolla MO 65409, USA}
\affiliation{Kavli Institute for the Physics and Mathematics of the Universe (WPI), The University of Tokyo Institutes for Advanced Study (UTIAS), The University of Tokyo, Chiba 277-8583, Japan}
\email{saitos@mst.edu}

\author[0000-0003-1198-831X]{Ariel G. S\'anchez}
\affiliation{Max-Planck-Institut f\"ur extraterrestrische Physik, Postfach 1312, Giessenbachstr., 85748 Garching, Germany}
\email{arielsan@mpe.mpg.de}

\author[0000-0001-7240-7449]{Donald P. Schneider}
\affiliation{Department of Astronomy \& Astrophysics, The Pennsylvania State University, University Park, PA 16802, USA}
\affiliation{Institute for Gravitation and the Cosmos, The Pennsylvania State University, University Park, PA 16802, USA}
\email{dps7@psu.edu}


\begin{abstract}
The catalog from the Hobby-Eberly Telescope Dark Energy Experiment (HETDEX) Public Data Release 1 (PDR1) contains half a million emission-line-selected [\ion{O}{2}] galaxies spread across $540~\mathrm{deg}^2$ at $z \le 0.48$ from HETDEX's unprecedented untargeted spectroscopic survey. In this paper, we construct volume-limited samples from PDR1 in three luminosity bins across the two main fields: ``Spring'' and ``Fall''. 
The numbers of galaxies in the bins range from 11,354 to 64,794
and number densities, $\bar{n}\simeq (2-5)\times10^{-3}~h^3~\mathrm{Mpc}^{-3}$, are higher than those of typical cosmological spectroscopic surveys of emission-line galaxies by a factor of five to ten.
The monopole and quadrupole power spectra derived from these samples are in excellent agreement with the mock power spectra from the Uchuu simulation based on a flat $\Lambda$CDM model and the cosmological parameters from the \textit{Planck} cosmic microwave background data, at all wavenumbers used for the measurement ($0.01<k<0.7~h~\mathrm{Mpc}^{-1}$). We find that the power spectrum amplitudes are consistent with a characteristic dark matter halo mass of $\log(M_0~[h^{-1}M_\sun])\simeq 11.9$--$12.3$, with the halo mass showing a weak dependence on [\ion{O}{2}] luminosity, $M_0\propto L^a$, increasing with a slope of $a = 0.37\pm0.10$. The best-fit mock suggests that approximately 13 percent of the [\ion{O}{2}] galaxies in our sample reside in subhalos.
The new, high-density tracers of the underlying matter distribution presented in this paper provide precise measurements of clustering in a low-redshift regime sensitive to the late-time growth of structures. These samples will form the basis for forthcoming analyses of the redshift-space distortion effect, galaxy--halo connection, and cross-correlations with external low-redshift probes.

\end{abstract}

\keywords{\uat{Cosmology}{343} --- \uat{Observational cosmology}{1146} --- \uat{Large-scale structure of the universe}{902} --- \uat{Galaxies}{573}}

\section{Introduction} 
\label{sec:intro}
The Hobby-Eberly Telescope Dark Energy Experiment \citep[HETDEX;][]{Gebhardt_2021} has delivered a catalog of more than two million \textit{emission-line-selected} galaxies in a cosmological volume through its unprecedented \textit{untargeted} integral field unit spectroscopic survey using the Visible Integral-field Replicable Unit Spectrograph~\citep[VIRUS;][]{Hill_2021}, on the 10m Hobby-Eberly Telescope~\citep[HET;][]{Ramsey_1998,Hill_2021}.

The catalog from Public Data Release 1~\citep[PDR1;][]{pdr1} contains half a million [\ion{O}{2}] emitters at $z \leq 0.48$ and 1.6 million Lyman-$\alpha$ emitter (LAE) candidates at $1.88 \leq z \leq 3.52$. In this paper, we focus on the former catalog.

The low-redshift universe has become an increasingly important regime for cosmological studies. \citet{Nguyen_2023} found evidence for a suppressed growth rate of large-scale structure during the dark-energy-dominated era compared to the prediction of general relativity in a flat $\Lambda$ cold dark matter ($\Lambda$CDM) cosmology. There is also a mild discrepancy in the $S_8$ parameter, which characterizes the present-day amplitude of linear matter density fluctuations~\citep{Jain1997}: some studies report a $\simeq 2\sigma$ tension between the $S_8$ values inferred from the cosmic microwave background (CMB) and weak gravitational lensing surveys~\citep{HSCY3_shear,DESY6_shear}, while others find them to be in good agreement~\citep{KIDSLegacy_shear,ACT2025}. Furthermore, analyses combining Dark Energy Spectroscopic Instrument (DESI) Baryonic Acoustic Oscillation measurements with Type Ia supernovae data have suggested evidence for dynamical dark energy~\citep{DESIDR2_2025}.

To understand better about these tensions and findings, independent probes of matter clustering in the low-redshift universe are essential. Several cosmological spectroscopic surveys have targeted galaxies at $z \gtrsim 0.5$~\citep{Blake_2011,Takada_2014,Alam_2017,Alam_2021,Wang_2022,DESIDR2_2025,euclid_2025}, including the LAE component of the HETDEX survey, leaving the low-redshift regime comparatively less explored. At lower redshifts, the Galaxy And Mass Assembly survey~\citep[GAMA;][]{GAMADR1,GAMADR2}, the DESI Bright Galaxy Survey~\citep{DESIBGS_2023}, and the DESI Peculiar Velocity Survey~\citep{DESIPV_2025} probe the growth of structures at $z \lesssim 0.5$, $0.6$, and $0.1$, respectively, using photometrically selected galaxies. In this context, the HETDEX [\ion{O}{2}] sample offers a complementary and independent probe of structure growth using untargeted emission-line-selected galaxies at $z \leq 0.48$. Combining observations from surveys with different target selection criteria and systematics is crucial for obtaining robust cosmological constraints.

[\ion{O}{2}] galaxies are relatively unbiased tracers of the large-scale structure since their emission lines provide accurate spectroscopic redshifts, and their clustering traces the underlying matter distribution with a relatively modest bias compared to more massive galaxy populations such as luminous red galaxies~\citep{Tadaki2012,Kajisawa2013,Jimenez2020}.
The HETDEX [\ion{O}{2}] samples complement those from other spectroscopic surveys targeting [\ion{O}{2}] galaxies at $z\gtrsim 0.5$~\citep{Takada_2014,Alam_2021,DESIDR2_2025}. However, note that the surveys listed above are continuum-selected, meaning that galaxies must exceed a certain apparent magnitude threshold in order to be targeted for spectroscopy. In contrast, HETDEX [\ion{O}{2}] galaxies are emission-line-selected, without requiring continuum pre-selection, yielding a galaxy population fundamentally different in nature. We refer the reader to~\citet{Ciardullo2013} for a related discussion of emission-line-selected galaxy samples in the context of the HETDEX Pilot Survey.

Since HETDEX use integral field unit (IFU) spectroscopy, it is different from other cosmological spectroscopic surveys: HETDEX has measured all the galaxy redshifts above the survey sensitivity with some flags due to bad data.
While HETDEX is primarily designed to detect LAEs in the distant universe, the experiment is also capable to produce an unbiased dataset of [\ion{O}{2}] emitters in the low-redshift universe. This allows us to construct \textit{volume-limited} samples from the HETDEX [\ion{O}{2}] emitting galaxy catalog. 

In a flux-limited sample, only galaxies brighter than a given flux threshold are detected: while both faint and bright galaxies are observed at low redshifts, only the most luminous galaxies are observable at high redshifts, introducing a redshift-dependent selection bias. The resulting catalog is also sensitive to flux limits, which depend on complex factors such as observing conditions and sky subtraction residuals. This complicates the radial selection function and consequently the interpretation of the clustering results. 
In contrast, a volume-limited sample is constructed by selecting galaxies within a specified luminosity range such that all galaxies are detected across the given redshift range. 
This approach produces a spatially homogeneous selection, which gives a flat radial selection function and an unbiased representation of the galaxy population with given luminosity cuts, making the interpretation of the clustering results more straightforward and reliable.

As the HETDEX survey has been optimized to obtain sufficient number densities of LAEs in the distant universe, the number densities of our volume-limited samples of HETDEX [\ion{O}{2}] galaxies, $\bar{n}\simeq (2-5)\times10^{-3}~h^3~\mathrm{Mpc}^{-3}$ depending on luminosity bins, are higher than those of typical cosmological spectroscopic surveys of emission-line galaxies (ELGs), a few to several times $10^{-4}~h^3~\mathrm{Mpc}^{-3}$, by a factor of five to ten~\citep{Blake_2011,Takada_2014,eBOSS_2021,Wang_2022,DESIDR2_2025,euclid_2025}. 
Here, the Hubble constant is given by the standard notation, $H_0=100~h~\mathrm{km~s^{-1}~Mpc^{-1}}$. This makes our sample useful not only for cosmology but also for astrophysical studies of galaxy formation and evolution and cross-correlation studies with other tracers in the low-redshift universe.

In this work, we present the first robust clustering measurements of emission-line-selected [\ion{O}{2}] galaxies in the low-redshift universe using the HETDEX PDR1 catalog. By constructing highly complete volume-limited samples across distinct luminosity bins, we measure the monopole and quadrupole power spectra down to non-linear scales. To interpret these measurements and investigate the [\ion{O}{2}] galaxy--halo connection, we model the clustering signal using mock catalogs based on the Uchuu N-body simulation~\citep{Uchuu_dr1, Uchuu_nu2GCC, Uchuu_UM, Uchuu_RMG,Uchuu_SDSS} combined with a halo occupation distribution~\citep[HOD; e.g.,][]{Zheng2005,Geach2012} framework. The high-density tracers and clustering properties established here validate the HETDEX [\ion{O}{2}] dataset as a precise probe of late-time structure growth, providing the critical foundation for forthcoming analyses of the redshift-space distortions, galaxy--halo connection, and cross-correlations with independent low-redshift probes.

The rest of this paper is organized as follows.
Section~\ref{sec:hetdex_data} describes the HETDEX data, including the public catalog of emission-line-selected [\ion{O}{2}] galaxies. Section~\ref{sec:v-lim_samples} explains how we construct volume-limited samples in three luminosity bins. Section~\ref{sec:pk} presents the power spectrum multipoles of the volume-limited [\ion{O}{2}] galaxies. Section~\ref{sec:interpretation} discusses the interpretations of the monopole power spectra and we conclude in Section~\ref{sec:conclusions}. 

Throughout this paper, we use $\Omega_\mathrm{m}=0.311$ when converting redshift $z$ to comoving distance in units of $h^{-1}~\mathrm{Mpc}$, assuming a flat $\Lambda$CDM model~\citep{Planck2020cosmo}.

\section{HETDEX Data}
\label{sec:hetdex_data}

HETDEX~\citep{Gebhardt_2021} has completed its seven-year observing program in 2024. The PDR1 catalog~\citep{pdr1} contains LAEs at $1.88 \le z \le 3.52$ and [\ion{O}{2}] emitting galaxies at $z \le 0.48$, as well as stars, low-redshift continuum selected galaxies, and active galactic nuclei, over $540~\mathrm{deg}^2$ of the sky. HETDEX has observed its survey area with VIRUS~\citep{Hill_2021}, a set of 78 IFUs, covering the wavelength range of $3470\,\mbox{\AA}\leq\lambda\leq5540\,\mbox{\AA}$ with a spectral resolving power of $750\lesssim R\lesssim 950$. 

Each IFU spans $51''\times51''$ on the sky and contains 448 fibers of $1''.5$ in diameter, enabling the simultaneous acquisition of approximately 35,000 spectra. With over 600 million spectra collected, HETDEX is the largest spectroscopic sky survey ever conducted.

The HETDEX sample of emission-line galaxies is unique in its combination of a large area, depth, and untargeted emission-line search without any requirement on a continuum magnitude compared to the most spectroscopic cosmology surveys. The observed spectra are searched for emission lines (and the continuum for low-redshift galaxies) without any photometric preselection.
Using several successful classification approaches, HETDEX provides robust spectroscopic redshifts and classifications for the entire catalog~\citep{Dustin_2023, Mentuch_Cooper_2023}. Compared to spectroscopic redshifts from external catalogs, 94.1$\%$ of the sources are within $|\Delta z|<0.02$~\citep{pdr1}.

The IFUs are distributed across a $18'$-diameter field of view on a $100''$-grid. The gaps between adjacent IFUs are roughly equal to the size of each IFU, resulting in a fill factor of $\simeq 1/4.6$ per observation. The sky is therefore sparsely sampled within the footprint of the survey rather than being uniformly covered with fibers. However, this sparse sampling has been shown to provide a constraining power on cosmological models similar to that of complete imaging of the same area with a lower-number-density tracer. As discussed by \citet{Chiang2013}, complete sky coverage within this area is not required to achieve the HETDEX's scientific goals; a fill factor of $1/4.6\simeq 0.22$, which optimizes the packing of the IFUs within the area of the HET focal plane, is sufficient.

\subsection{HETDEX PDR1}
\label{subsec:pdr1}

We use the HETDEX Public Source Catalog 2 (HPSC2) from the PDR1~\citep{pdr1}. HPSC2 consists of all HETDEX survey data from 2017-Jan-01 until 2024-Jul-31, which is an extension of the HETDEX Public Source Catalog 1~\citep[HPSC1;][]{Mentuch_Cooper_2023} collected from 2017-Jan-01 to 2020-June-26, and contains four additional years of data and several updates in masking. Alongside the catalog, 431\,K datacubes are included in PDR1 and can be accessed through the HETDEX data portal (\url{https://hetdex.org/data-results/}). HPSC2 is also publicly available via Zenodo (\href{https://doi.org/10.5281/zenodo.19581262}{DOI: 10.5281/zenodo.19581262}).
These data consist of two primary observational fields: an equatorial “Fall” field, whose footprint is centered at $(\mathrm{RA_{J2000}},\mathrm{DEC_{J2000}})=(22.50^\circ,+0.00^\circ)$ and covers $150~\mathrm{deg}^2$, and a high declination “Spring” field, which centers at $(195.00^\circ,+51.00^\circ)$ and covers $390~\mathrm{deg}^2$. Although HPSC2 also contains other fields, we only use the Spring and Fall fields in this paper due to their large sky coverages.

\subsection{HETDEX \texorpdfstring{[\ion{O}{2}]}{[O II]} Galaxies}
\label{subsec:hetdex_oii}

\begin{figure*}
    \centering 
    \includegraphics[width=0.6\textwidth]{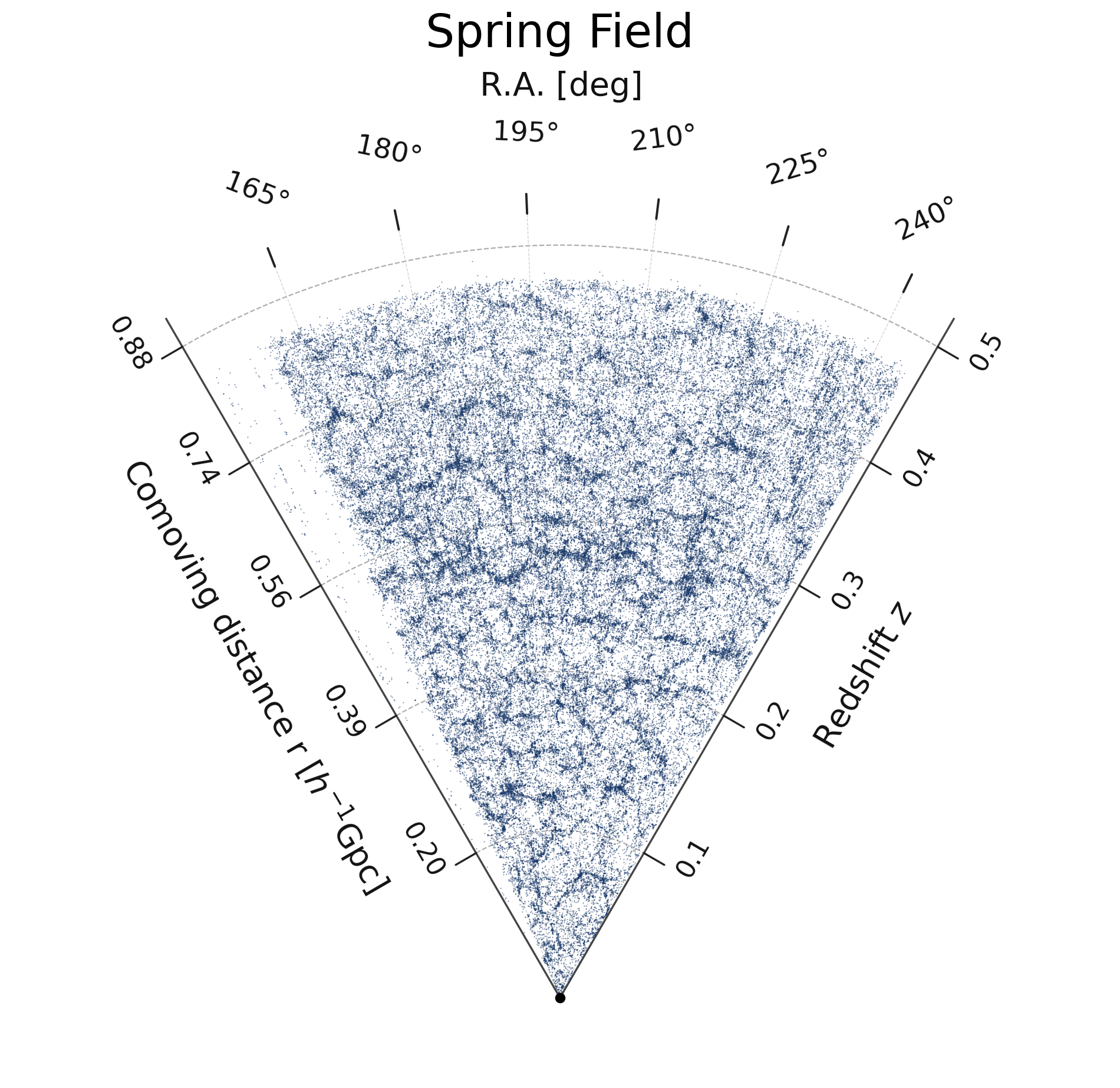}
    \includegraphics[width=0.35\textwidth]{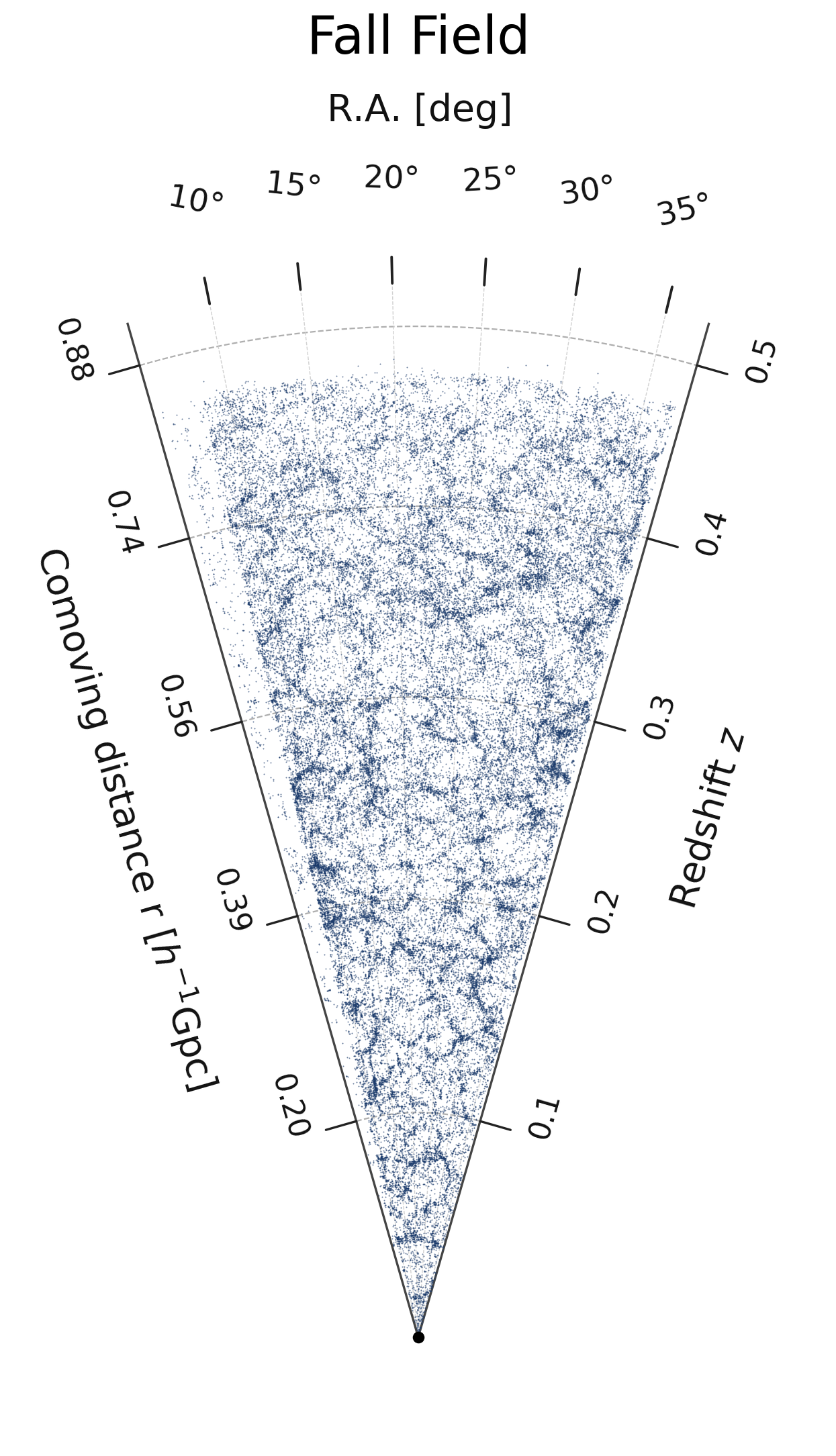}
    \caption{Projected distributions of the HETDEX [\ion{O}{2}] galaxies from HPSC2 in the plane of right ascension and comoving distance in units of $h^{-1}~\mathrm{Gpc}$ with corresponding redshifts, collapsed along the declination axis with a width of $2^\circ$. The large-scale structure is clearly visible across all observational fields. There are 286,992 and 140,122 galaxies with $\mbox{\texttt{z\_{hetdex}}}>0$ in comoving volumes of $4.7\times10^7$ and $2.6\times10^7~h^{-3}~\mathrm{Mpc}^3$ in the Spring and Fall fields, respectively.}
    \label{fig:plot_oii_full_distribution}
\end{figure*}

Although the primary science driver of HETDEX is the detection of high-redshift LAEs, the sensitivity of the VIRUS spectrograph is sufficient to ensure high completeness in the detection of [\ion{O}{2}] galaxies at $z \le 0.48$. Since HETDEX's spectral resolution is too low to resolve the [\ion{O}{2}] doublet, we use a mean rest-frame air wavelength of $\lambda = 3727.8$\,\AA~\citep{pdr1} in our analysis.

In this paper, we use the catalog of [\ion{O}{2}] galaxies from the PDR1 v1.3\footnote{\texttt{hetdex\_sc2\_v1.3.fits}}.
For each source observation, the catalog provides the object's J2000 equatorial coordinates and redshift. 
For sources with either Ly$\alpha$ or [\ion{O}{2}] line emission, the catalog also provides the optimal measurement for the dust- and aperture-corrected flux and luminosity. 
We apply no selection cuts other than the choice of fields and object classification.

Figure~\ref{fig:plot_oii_full_distribution} shows the projected distribution of all HPSC2 [\ion{O}{2}] galaxies in the Spring and Fall fields in the plane of right ascension and comoving distance with corresponding redshifts, collapsed in declination with a width of $2^\circ$. There are 286,992 and 140,122 galaxies with $\mbox{\texttt{z\_{hetdex}}}>0$ in comoving volumes of $4.7\times10^7$ and $2.6\times10^7~h^{-3}~\mathrm{Mpc}^3$ in the Spring and Fall fields, respectively. The large-scale structure is clearly visible across both observational fields. The entire catalog of the HETDEX [\ion{O}{2}] emitting galaxies has a high number density of about $6\times10^{-3}~h^3~\mathrm{Mpc}^{-3}$. Note that this number density is approximately $4.6$ times lower than the underlying density (i.e., the number density derived from the luminosity function) due to the angular sparse sampling of the HETDEX survey, as described in Section~\ref{sec:hetdex_data}.

\section{Volume-limited Samples}
\label{sec:v-lim_samples}

Several previous studies~\citep{Norberg_2001, Zehavi_2002, Tegmark_2004, Deng_2007, Tempel_2014,Farrow_2015,Favole2017} have utilized volume-limited samples to investigate the luminosity dependence of galaxy bias. These studies compared flux-limited samples with volume-limited samples or galaxy groups extracted from the volume-limited samples. \citet{Tegmark_2004} reported that the power spectra of the Sloan Digital Sky Survey (SDSS) from different volume-limited samples are consistent in shape at large scales but differ in amplitude and the power spectrum of each sample is free of the overall red-tilting effect that affects the power spectrum of flux-limited samples. \citet{Norberg_2001} found that the projected two-point correlation functions of volume-limited samples from the Two-degree-Field Galaxy Redshift Survey (2dFGRS) reveal a slowly increasing clustering amplitude for galaxies brighter than the $\mathrm{b_J}$-band magnitude of $−19.7$, with a more pronounced rise for the brightest galaxies. \citet{Farrow_2015} reported that the projected two-point correlation functions of volume-limited samples from the GAMA survey exhibit stronger clustering of $r$-band luminous, more massive and redder galaxies.
\citet{Favole2017} studied the dependence of galaxy clustering on the [\ion{O}{2}] emission line luminosity using volume-limited samples from the SDSS DR7 Main galaxy sample for five different minimum luminosities, $L_{\text{[\ion{O}{2}]}}^\mathrm{min} = (0.1, 0.3, 1, 3, 10) \times 10^{40}~\mathrm{erg~s^{-1}}$, which correspond to maximum redshifts of $z_\mathrm{max} = (0.05, 0.09, 0.14, 0.17, 0.20)$, respectively.

Our analysis is based on the methodology developed in these previous studies.
To satisfy the volume-limited criterion and to utilize as much data as possible, we divide the galaxy catalog into five luminosity bins. Then, we apply additional redshift cuts to each bin to extract volume-limited samples from the HETDEX [\ion{O}{2}] catalog in HPSC2, as described in Section~\ref{subsec:hetdex_oii_v-lim_cat}.

\subsection{HETDEX \texorpdfstring{[\ion{O}{2}]}{[O II]} Volume-limited Catalog}
\label{subsec:hetdex_oii_v-lim_cat}

\begin{figure}
\centering 
\includegraphics[width=0.9\columnwidth]{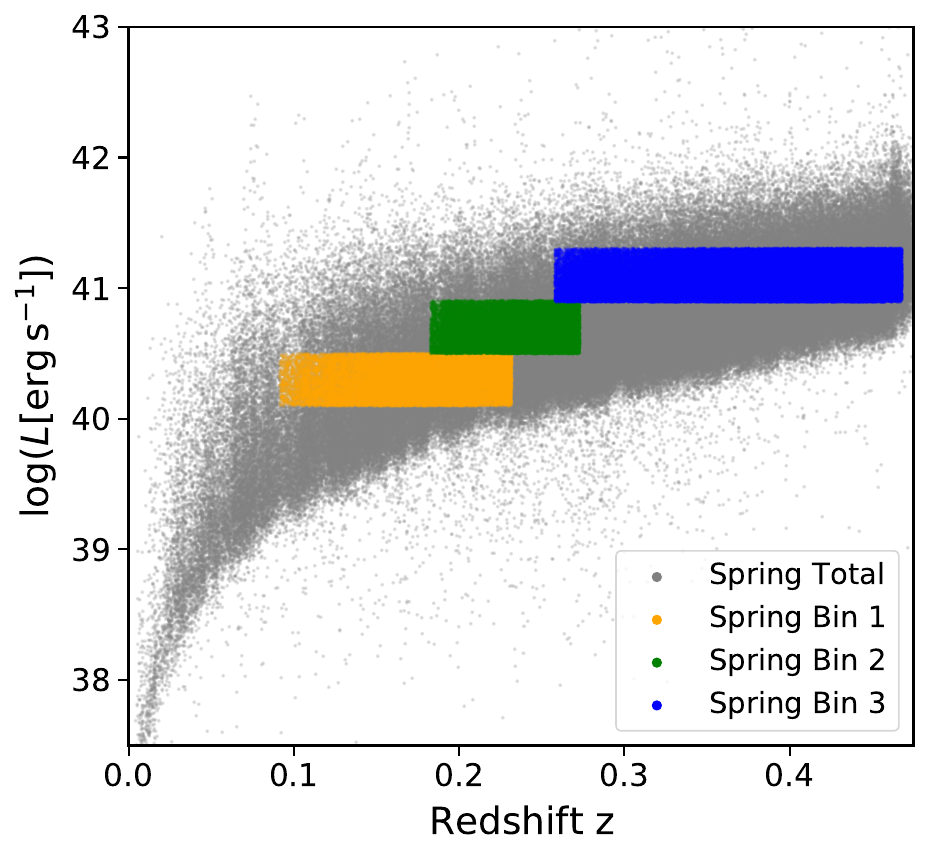}
\includegraphics[width=0.9\columnwidth]{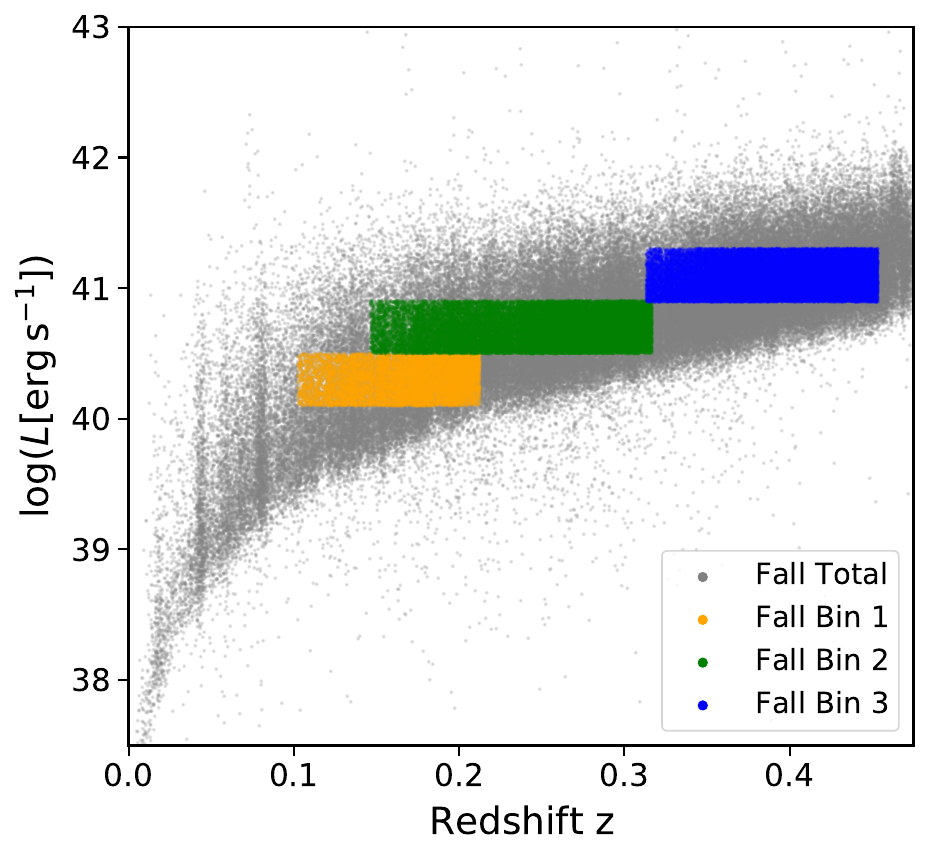}
\caption{Luminosity distribution of the HETDEX [\ion{O}{2}] emission-line as a function of redshift, shown as grey dots, with volume-limited samples shown as colored dots, for the Spring (top) and Fall (bottom) fields.}
\label{fig:plot_lum_z}
\end{figure}

\begin{figure}
\centering 
\includegraphics[width=0.9\columnwidth]{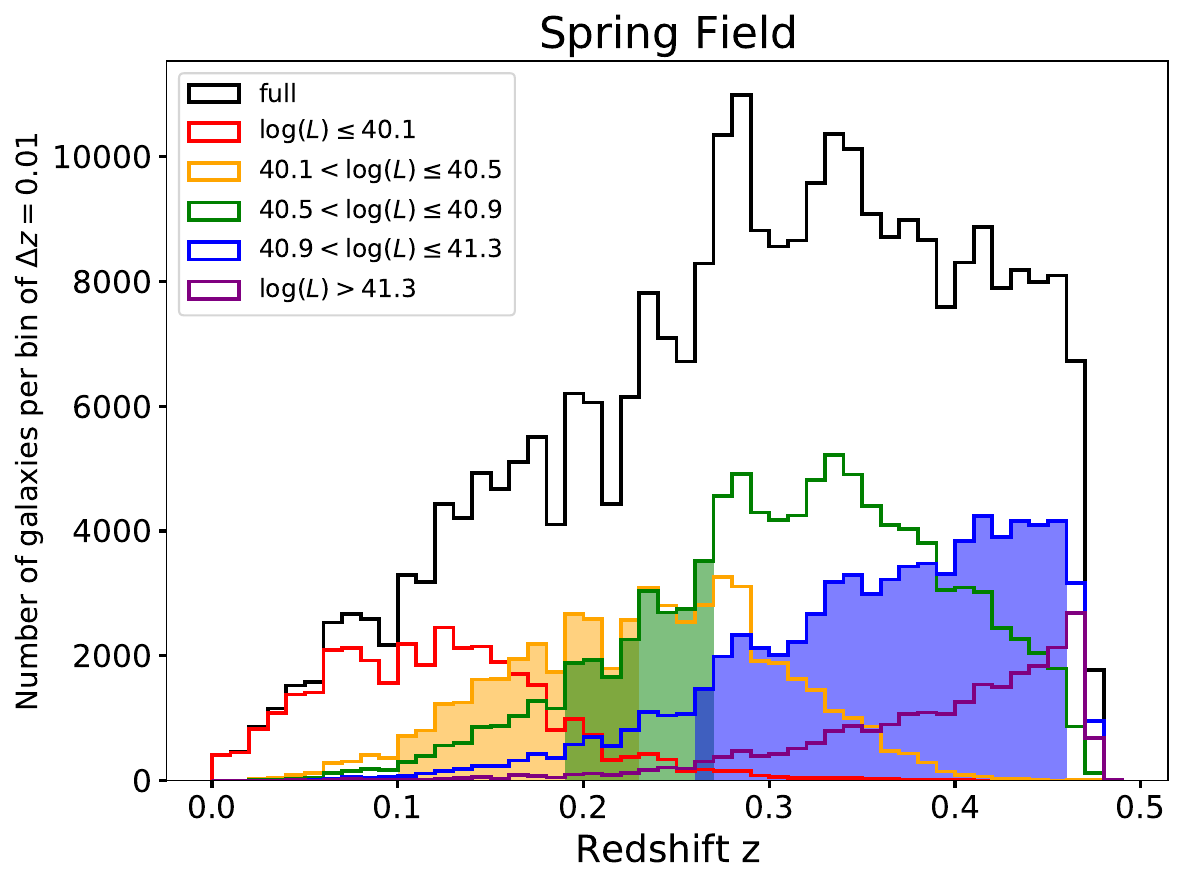}
\includegraphics[width=0.9\columnwidth]{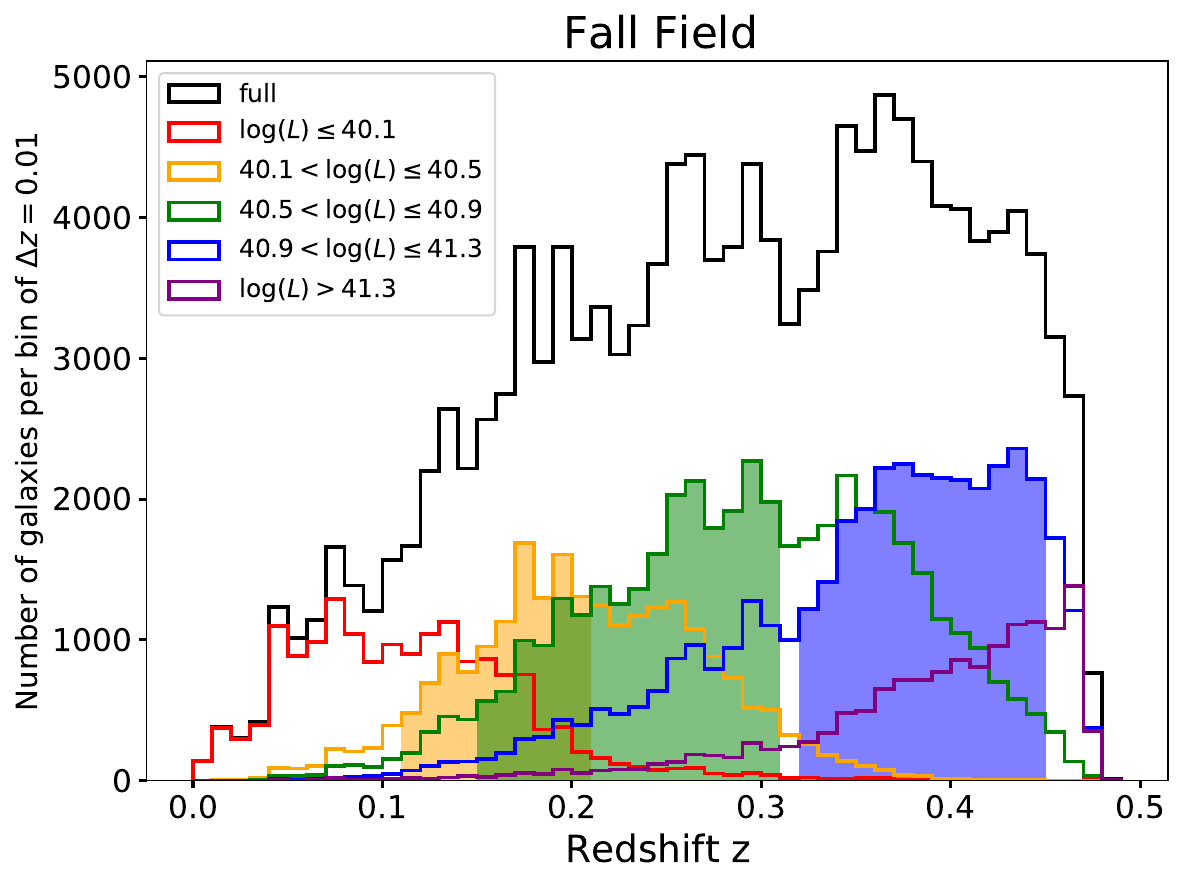}
\caption{The number of [\ion{O}{2}] emitting galaxies per bin of $\Delta z=0.01$ in the Spring (top) and Fall (bottom) fields. The black line represents all [\ion{O}{2}] emitting galaxies in HPSC2, while the colored lines and shaded regions indicate the number and the applied redshift cuts for different luminosity bins used to extract volume-limited samples. Bins 1 (red) and 5 (purple) are excluded from the cosmological analysis and thus the redshift cuts are not shown.}
\label{fig:plot_dNdz}
\end{figure}

\begin{figure}
\centering 
\includegraphics[width=0.9\columnwidth]{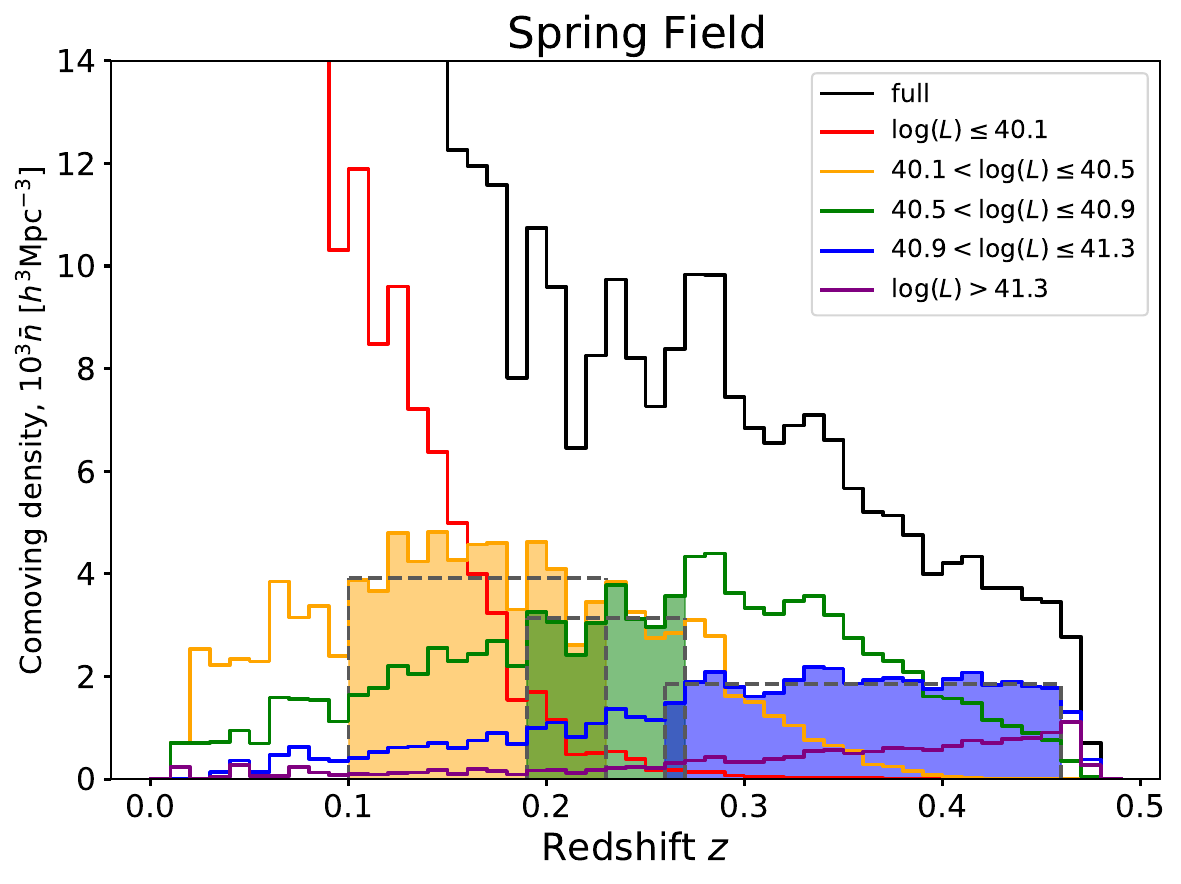}
\includegraphics[width=0.9\columnwidth]{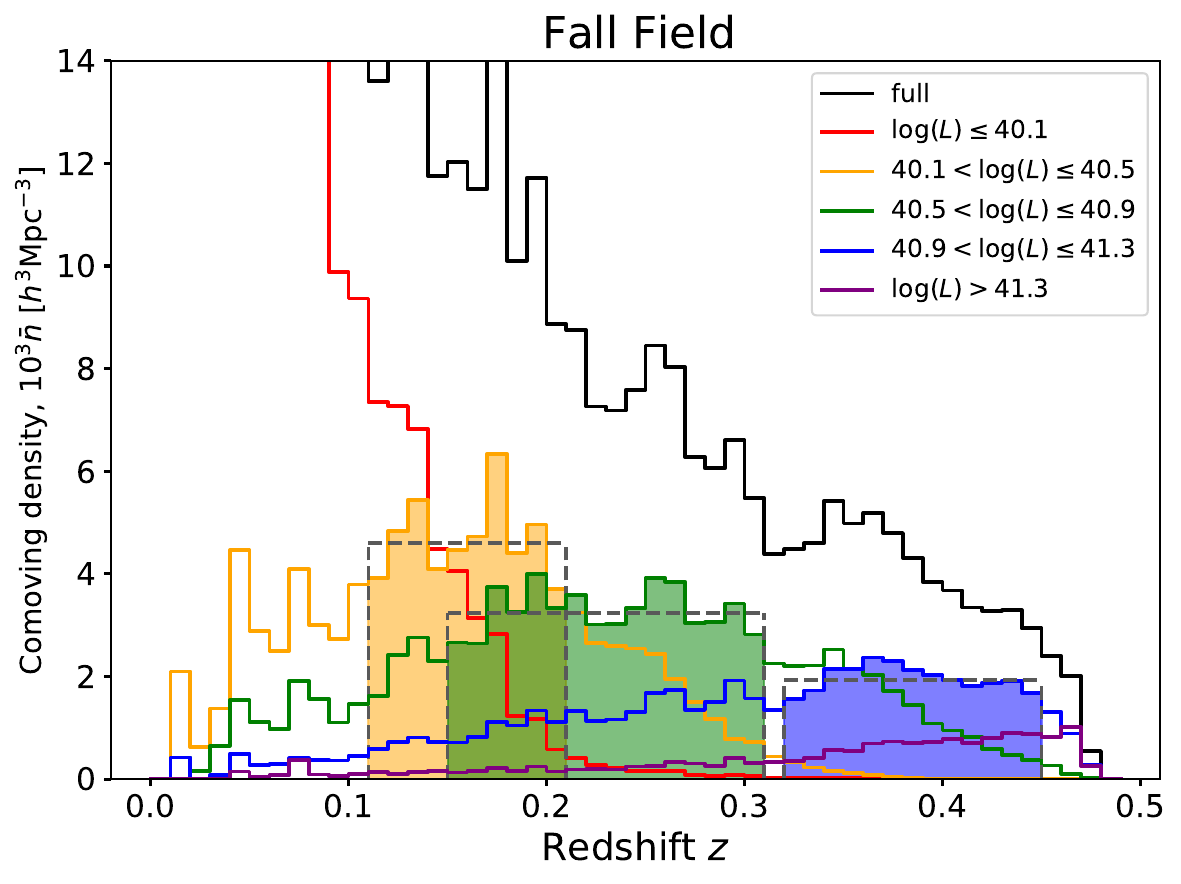}
\caption{Same format as Figure~\ref{fig:plot_dNdz} but for the comoving number density times $10^3$, $10^3\bar{n}$, per bin of $\Delta z=0.01$. The dashed lines indicate the constant mean number densities of the different bins.}
\label{fig:plot_dndz}
\end{figure}

\startlongtable
\begin{deluxetable*}{cccccccc}
\tabletypesize{\small}
\tablecaption{Key properties of the [\ion{O}{2}] volume-limited samples in the 
Spring and Fall fields. \label{tab:table_oii_sub}}
\tablehead{
\colhead{Samples} & \colhead{Field} & \colhead{Luminosity range} & \colhead{$z$ range} & 
\colhead{$z_\mathrm{eff}$} & \colhead{$N_\mathrm{gal}$} & \colhead{$10^3\bar{n}$} & \colhead{Volume} \\
\colhead{} & \colhead{} & \colhead{$\log(L~[\mathrm{erg~s}^{-1}])$} & \colhead{} & 
\colhead{} & \colhead{} & \colhead{[$h^3~\mathrm{Mpc}^{-3}$]} & \colhead{[$h^{-3}~\mathrm{Mpc}^3$]}
}
\startdata
Bin 2 & Spring & 40.1 $<$ $\log(L)$ $\leq$ 40.5 & 0.091 $\leq z \leq$ 0.231 & 0.176 & 23347 & 3.9 & $6.0\times 10^6$\\
Bin 3 & Spring & 40.5 $<$ $\log(L)$ $\leq$ 40.9 & 0.183 $\leq z \leq$ 0.273 & 0.235 & 21573 & 3.1 & $6.9\times 10^6$ \\
Bin 4 & Spring & 40.9 $<$ $\log(L)$ $\leq$ 41.3 & 0.258 $\leq z \leq$ 0.468 & 0.378 & 64794 & 1.9 & $3.5\times 10^7$ \\
\hline
Bin 2 & Fall & 40.1 $<$ $\log(L)$ $\leq$ 40.5 & 0.102 $\leq z \leq$ 0.212 & 0.168 & 11354 & 4.6 & $2.5\times 10^6$ \\
Bin 3 & Fall & 40.5 $<$ $\log(L)$ $\leq$ 40.9 & 0.146 $\leq z \leq$ 0.316 & 0.247 & 24536 & 3.2 & $7.6\times 10^6$ \\
Bin 4 & Fall & 40.9 $<$ $\log(L)$ $\leq$ 41.3 & 0.313 $\leq z \leq$ 0.453 & 0.389 & 27494 & 1.9 & $1.4\times 10^7$ \\
\enddata
\end{deluxetable*}

The high number density of HETDEX [\ion{O}{2}] galaxies (see Figure~\ref{fig:plot_oii_full_distribution}) ensures that each luminosity bin retains a sufficient number density for meaningful cosmological analyses, even after subsampling into luminosity bins and applying redshift cuts according to the literature on volume-limited sample analyses, as discussed in Section~\ref{sec:v-lim_samples}.

To maximize both the luminosity range and the number density of galaxies in each sample, we verify the volume-limited nature of each subsample by comparing the observed redshift distribution to that of a random catalog constructed without any radial selection function. Doing this confirms that galaxies within a given luminosity range are uniformly detected across the corresponding redshift interval. The result is a series of volume-limited galaxy samples, each within a well-defined luminosity bin and redshift range. We note that the redshift ranges of different bins may partially overlap.

We divide the samples into five luminosity bins: $\log(L)\le 40.1$ (Bin 1),  $40.1<\log(L)\le 40.5$ (Bin 2), $40.5<\log(L)\le 40.9$ (Bin 3), $40.9<\log(L)\le 41.3$ (Bin 4), and $\log(L)>41.3$ (Bin 5), where $L$ is in units of $\mathrm{erg~s^{-1}}$.
To assess the robustness of the power spectrum results with respect to the choice of luminosity bins, we tested different luminosity cuts with several bin sizes and found consistent results.
The final luminosity cuts are chosen to maximize the number of galaxies in Bins 2, 3, and 4 within their respective redshift ranges. For Bin 4, the upper redshift boundary is kept low enough to avoid possible LAE contamination due to a misclassification~\citep{Dustin_2023}, which is briefly described in the later paragraphs.

To determine the redshift ranges of the luminosity bins, we perform a two-sample Kolmogorov-Smirnov (KS) test for each pair of redshift boundaries with a given trial bin width, $\Delta z=0.01$, quantifying the statistical difference between the redshift distributions of the data and random catalogs.
For each pair of redshift boundaries, we compute the probability density functions (PDF) of the redshift distributions of both catalogs.
We then compute the cumulative distribution functions (CDF) by taking the cumulative sum of each PDF multiplied by $\Delta z=0.01$. We apply the two-sample KS test by taking the two CDFs, measuring the maximum absolute difference between them, and calculating the $p$-value that represents the probability that the two distributions are drawn from the same underlying distribution. This procedure is repeated for all possible combinations of redshift bin pairs within the range of interest.
For each luminosity bin, we identify the ($z_\mathrm{min}$, $z_\mathrm{max}$) pair that yields a sufficiently high $p$-value ($p>0.05$), ensuring the volume-limited feature, while also providing an adequate galaxy number density and volume coverage for meaningful cosmological analyses.

Figure~\ref{fig:plot_lum_z} shows the luminosity distribution of the HETDEX [\ion{O}{2}] galaxies as a function of redshift. Approximately 1\% of the galaxies in the full HETDEX [\ion{O}{2}] catalog have a large aperture correction to their luminosity of more than one dex, which may be too large due to an aperture that keeps growing because of a nearby bright object. This issue will be addressed in more detail in a future HETDEX data paper. Fortunately, our luminosity and redshift cuts eliminate almost all such galaxies, leaving only five in our volume-limited subsamples.

Figure~\ref{fig:plot_dNdz} presents the number of [\ion{O}{2}] emitting galaxies per bin of $\Delta z=0.01$ in the Spring (top panel) and Fall (bottom panel) fields, while Figure~\ref{fig:plot_dndz} shows the number density distribution, $\bar{n}$. 
The distribution of $\bar{n}$ within a luminosity bin is consistent with a constant number density, as shown by the dashed lines. 
This is consistent with the findings of~\citet{Ciardullo2013}, who reports a weak time evolution in the galaxy number density.

Bins 1 and 5 are excluded from the cosmological analysis. Bin 1 is excluded due to its small volume and Bin 5 is dropped because of its lower number density ($\bar{n}\simeq 0.8\times10^{-3}~h^3~\mathrm{Mpc}^{-3}$), significant overlap with Bin 4's redshift range, and potential contamination at the highest $z$ due to a misclassification. 
Briefly explaining, in this bin, the flux levels are faint enough so that the two populations, LAEs and [\ion{O}{2}] galaxies, have a large overlap in luminosity, and even a small error in the classifications can significantly change the inferred [\ion{O}{2}] galaxy number density.
Since HETDEX is designed to be extremely pure in Lyman-$\alpha$ classification, our classifier, ELiXer, purposefully classifies ambiguous Lyman-$\alpha$ lines as [\ion{O}{2}] lines and the contamination of Lyman-$\alpha$ into the [\ion{O}{2}] sample is larger than the contamination of [\ion{O}{2}] into the LAE sample~\citep{Dustin_2023}.
Thus, their redshift cuts are not displayed in Figures~\ref{fig:plot_dNdz} and \ref{fig:plot_dndz}.
Table~\ref{tab:table_oii_sub} summarize the key properties of the HETDEX [\ion{O}{2}] volume-limited samples extracted from the Spring and Fall fields. The effective redshift $z_\mathrm{eff}$ is computed as the number-weighted mean redshift of the galaxy sample. 

The fact that $z_{\mathrm{min}}$ does not extend to zero for all volume-limited samples, that the exact boundaries differ between the Spring and Fall fields, and that there are overlaps between bins are all consequences of the KS test. At very low redshifts, the small comoving volume leads to enhanced Poisson noise and local cosmic variance, which breaks the statistical uniformity required by the test. Similarly, the field-to-field differences in bin boundaries simply reflect the KS test responding to local cosmic variance rather than differences in instrumental sensitivity. This interpretation is supported by the highly consistent number densities across both fields, as shown in Table~\ref{tab:table_oii_sub}.

\begin{figure*}
    \centering
    \includegraphics[width=0.5\textheight]{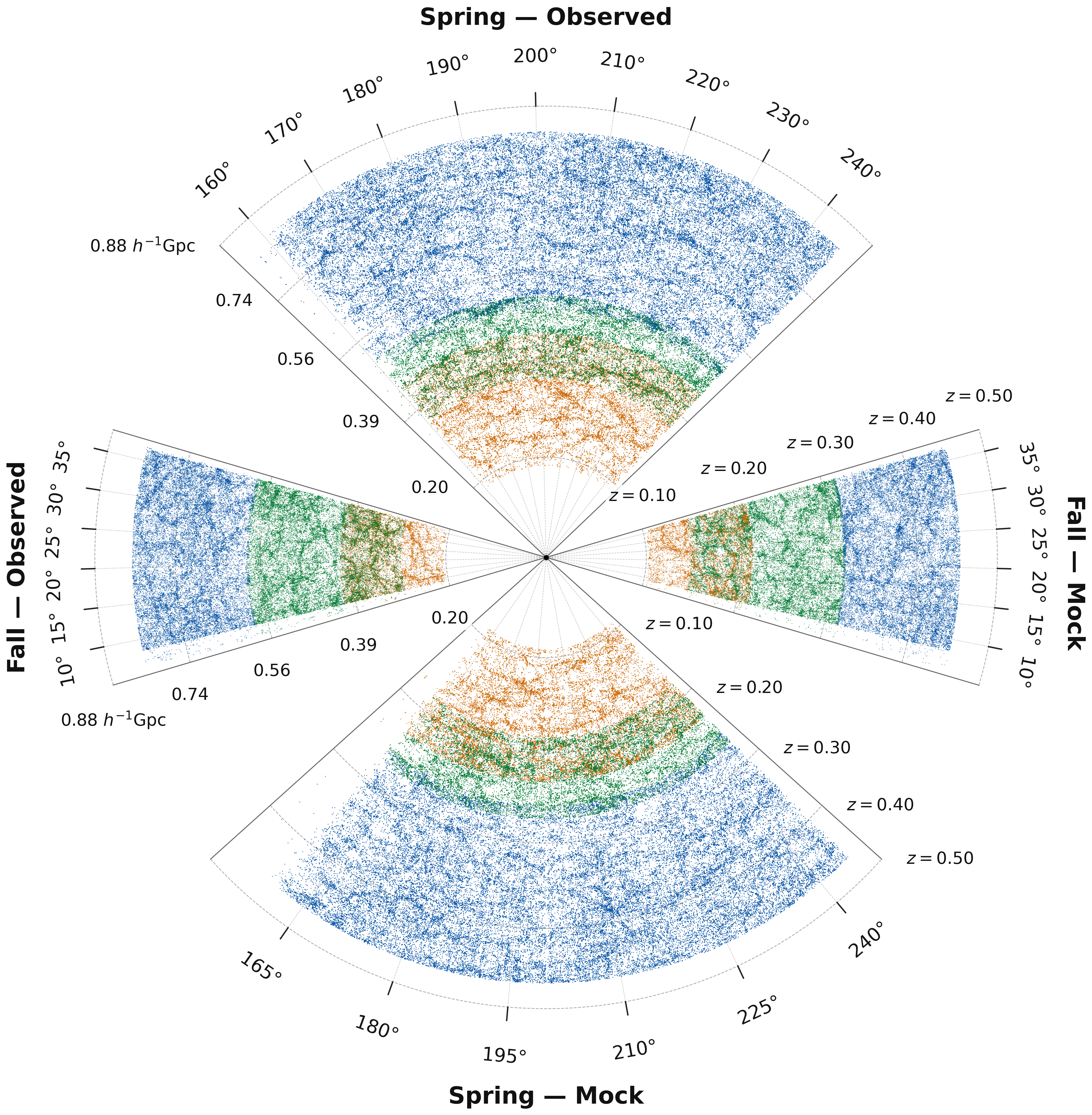}
    \caption{Sky distribution of the observed and mock [\ion{O}{2}] volume-limited samples in the Spring and Fall fields, projected onto the right ascension–comoving distance plane by collapsing along the declination axis with a width of $2^\circ$. Comoving distances in units of $h^{-1}~\mathrm{Gpc}$ are indicated by the radial labels, along with their corresponding redshifts. The orange, green, and blue dots represent Bins 2, 3, and 4, respectively. The large-scale structures are clearly visible across both fields and all luminosity bins, and the observed and mock distributions show statistically similar structures.}
    \label{fig:plot_oii_v-lim_distribution}
\end{figure*}

The upper and left wedges of Figure~\ref{fig:plot_oii_v-lim_distribution} present the spatial distributions of the observed volume-limited samples in the Spring and Fall fields, respectively, with luminosity bins colored as in Figures~\ref{fig:plot_dNdz} and \ref{fig:plot_dndz}. The lower and right wedges show one realization of the corresponding mocks, generated using the best-fit parameters described in Section~\ref{sec:interpretation}.
The statistical distribution of mocks is consistent with the observed data as discussed in Section~\ref{subsec:results}.

One can easily generate a volume-limited sample tailored to their specific purposes using HETDEX [\ion{O}{2}] emitting galaxies. 
For astrophysical purposes that require a stricter volume-limited criterion, one can apply a higher $p$-value threshold at the cost of a lower number density. 
On the other hand, for statistical and cosmological purposes, a higher number density and larger volume coverage are preferable. In this case, one can lower the $p$-value threshold to retain more galaxies and widen the redshift ranges in the catalog. 

Since the luminosity is correlated with the redshift in this volume-limited catalog by construction, analyzing different luminosity bins, which depend on redshifts, does not allow us to distinguish between intrinsically luminosity-dependent and time-dependent effects. Nevertheless, by modeling both effects simultaneously, we can study the luminosity and redshift dependence of quantities related to galaxies, such as the halo mass, linear bias, stellar mass, and the growth of structures. The first two items are discussed in Section~\ref{subsec:results}, while the latter two will be addressed in upcoming papers.

\section{Power Spectrum Multipoles}
\label{sec:pk}
\subsection{Estimator}
\label{subsec:pk_estimator}
Our power spectrum estimation pipeline\footnote{\url{https://github.com/jeongin-moon/MoonPower.git}} is built from a Fast Fourier Transform (FFT)-based approach \citep{Scoccimarro2015,Bianchi2015,Hand_2017}.
We begin by defining the Feldman-Kaiser-Peacock (FKP) overdensity function~\citep{FKP_1994},
\begin{equation}
    F(\mathbf{r}) = \frac{w(\mathbf{r})}{I^{1/2}}\left[n(\mathbf{r}) - \alpha \, n_\mathrm{s}(\mathbf{r})\right],
    \label{eq:FKP_func}
\end{equation}
where $w(\mathbf{r})$ is a weight at position $\mathbf{r}$, $I$ is a normalization factor defined later, 
$n(\mathbf{r})$ and $n_s(\mathbf{r})$ are the number densities of galaxies in the data and random catalogs at position $\mathbf{r}$, respectively, and $\alpha$ is the ratio of the weighted number of galaxies in the data to that in the random catalog. For the $i$-th cubic cell of a three-dimensional grid, this becomes
\begin{equation}
    F_i \equiv \frac{1}{I^{1/2}}
    \left[\frac{\tilde{N}_d^i - \alpha \, \tilde{N}_r^i}{dV}\right],
    \label{eq:FKP_func_grid}
\end{equation}
where $\tilde N^i_\mathrm{d}$ and $\tilde N^i_\mathrm{r}$ are the weighted galaxy counts in the data and the random catalog, respectively, in the $i$-th grid cell, and $dV$ is the volume of a single cubic cell. 

The normalization factor $I$ is given by
\begin{equation}
    I = \int d\mathbf{r}~w^2(\mathbf{r})\bar{n}^2(\mathbf{r}) 
    = \frac{\alpha}{dV}\sum_{i=1}^{N_\mathrm{grid}} \tilde{N}_\mathrm{d}^i \tilde{N}_\mathrm{r}^i,
    \label{eq:pk_norm}
\end{equation}
where $N_\mathrm{grid}$ is the total number of grid cells and the overbar denotes an averaged quantity.

To assign galaxies to the grid, we use the Cloud-in-Cell (CIC) scheme~\citep{Hockney1988}, where each galaxy count is weighted by a CIC kernel and $w(\mathbf{r}) = N_\mathrm{overlap}^{-1}$, where $N_\mathrm{overlap}$ is the number of repeated observational shots overlapping at the same sky position. This weight is applied not only to the data catalogs, but also to the mock and random catalogs, since applying a survey mask to the catalogs involves counting the points several times when multiple shots overlap at the same sky position.
In Appendix~\ref{appendix:N_overlap}, we show the sky distribution of $N_\mathrm{overlap}$ and the effect of $w(\mathbf{r})$ on the power spectrum measurements.

The weighted galaxy count in the $i$-th grid cell is given by
\begin{equation}
    \tilde{N}^i_\mathrm{d,r} = \sum_{j=1}^{N_\mathrm{gal,ran}} w(\mathbf{r}_j)\,\mathcal{W}(\mathbf{r}_j-\mathbf{r}_i),
    \label{eq:N_i_grid}
\end{equation}
where $N_\mathrm{gal}$ and $N_\mathrm{ran}$ are the total number of galaxies in the data and random catalogs, respectively, and $\mathcal{W}(\mathbf{r}_j - \mathbf{r}_i)$ is the CIC assignment kernel. In three dimensions, the kernel takes the form
\begin{equation}
    \mathcal{W}(\mathbf{r}) = W_1(x_1)\,W_1(x_2)\,W_1(x_3),
    \label{eq:CIC_3D_kernel}
\end{equation}
where each one-dimensional kernel is defined as
\begin{equation}
    W_1(x) =
    \begin{cases}
        1 - \dfrac{|x|}{H} & \text{if } |x| < 
        H\\
        0 & \text{otherwise}.
    \end{cases}
    \label{eq:CIC_1D_kernel}
\end{equation}
Here, $H$ is the linear size of a cubic grid cell.
Note that we do not apply FKP weights here, as the galaxy number density is nearly constant across the survey volume, rendering the FKP weighting scheme ineffective.

We then estimate the power spectrum multipoles following~\citet{Yamamoto_2006}, which accounts for varying line-of-sight (LOS) directions,
\begin{equation}
\begin{split}
    \hat{P}_{\ell}^\mathrm{Yama}(k) = (2l+1)\int\frac{d\Omega_k}{4\pi}\biggl[\int d\mathbf{r}_1~F(\mathbf{r}_1)e^{i\mathbf{k}\cdot\mathbf{r}_1}\\
    \times\int d\mathbf{r}_2~F(\mathbf{r}_2)e^{-i\mathbf{k}\cdot\mathbf{r}_2}\mathcal{L}_{\ell}(\hat{\mathbf{k}}\cdot{\hat{\mathbf{r}}_2})-P_{\ell}^\mathrm{noise}(\mathbf{k})\biggr],
\end{split}
\label{eq:Yamamoto_estimator}
\end{equation}
where $d\Omega_k$ is the solid angle element, $\mathcal{L}_\ell$ is the Legendre polynomial of order $\ell$, and $P_{\ell}^\mathrm{noise}$ is the Poisson shot noise term,
\begin{equation}
    P_{\ell}^{\mathrm{noise}}(\mathbf{k}) = \frac{(1+\alpha)}{I}\,C(\mathbf{k})
    \int d\mathbf{r}\; \bar{n}(\mathbf{r})\,w^2(\mathbf{r})\,
    \mathcal{L}_{\ell}(\hat{\mathbf{k}}\cdot\hat{\mathbf{r}}).
    \label{eq:shotnoise}
\end{equation}
Here, $C(\mathbf{k})$ is the analytical expression of the CIC kernel squared in Fourier space, $\sum_{\mathbf{n}} |\mathcal{W}_\mathrm{CIC}(\mathbf{k} + 2k_N \mathbf{n})|^2$, following the derivation, conventions and expressions in~\citet{Jing_2005},
\begin{equation}
      C(\mathbf{k})= \prod_{i} \left[ 1 - \dfrac{2}{3} \sin^2 \left( \dfrac{\pi k_i}{2k_N} \right) \right].
    \label{eq:shotnoise_C}
\end{equation}

Since $P_{\ell}^\mathrm{noise}\ll\hat{P_\ell}$ for multipoles of order $\ell>0$, the shot noise correction is negligible for these higher multipoles. We therefore subtract the shot noise only for the monopole and the shot noise is evaluated on the Fourier grid as
\begin{equation}
    P_0^\mathrm{noise}(\mathbf{k}) = \frac{C(\mathbf{k})}{I}\left[\sum_{j=1}^{N_\mathrm{gal}}w^2(\mathbf{r}_j) 
    + \alpha^2\sum_{j=1}^{N_\mathrm{ran}}w^2(\mathbf{r}_j)\right].
    \label{eq:shotnoise_grid}
\end{equation}
The CIC assignment kernel is subsequently deconvolved following Equation~(2.9) of~\citet{Wang_2024}.

This estimator retains the relevant LOS information by approximating the LOS of each galaxy pair with that of one of the two galaxies. This approximation is reliable on the scales of interest and is well-suited for HETDEX, given its moderate sky coverage. And the estimator is implemented using multiple FFTs~\citep{Hand_2017}.

We use random catalogs of a number density of $\bar{n}_\mathrm{s}(\mathbf{r})=\alpha^{-1}\bar{n}(\mathbf{r})$, with $\alpha^{-1}$ ranging from 54 to 164 depending on the luminosity bins and observational fields. 
The only selection function applied to the random catalogs is the fiber and sky emission line masks,
which remove points that fall outside the observed regions or flagged due to bright sky emission lines or bad data~\citep{pdr1}.

We adopt a grid-cell resolution for power spectrum estimation with a linear binning of $H=2.2~h^{-1}~\mathrm{Mpc}$ for all bins.
The grid-cell size was chosen to be sufficiently larger than the separation between the IFUs ($\simeq 100''$), which corresponds to $0.22$, $0.30$, and $0.45~h^{-1}~\mathrm{Mpc}$ for Bins 2, 3, and 4, respectively. This minimizes the window function effects arising from a sparse sampling~\citep{Chiang2013}. 
In our analysis, we calculate the power spectra up to a maximum wavenumber of $k_\mathrm{max} = 0.7~h~\mathrm{Mpc}^{-1}$, which is approximately half of the Nyquist wavenumber.

\subsection{Mock: Simulation-based Model}
\label{subsec:mock}

Interpreting the clustering of galaxies requires a model for their connection to the underlying dark matter halo population.
A wide variety of approaches have been proposed to describe this connection, ranging from perturbative methods to empirical models and numerical simulations, each with varying levels of physical assumptions and computational complexity. Examples include perturbation theory with bias expansion \citep[e.g.,][]{Desjacques2018, Philcox2022}, HOD \citep[e.g.,][]{Zheng2005}, subhalo abundance matching \citep[e.g.,][]{Kravtsov2004,Behroozi2019}, and hydrodynamical simulations \citep[e.g.,][]{Springel2005, Pakmor2023}.
Traditionally, the galaxy--halo connection has been described using the HOD framework, in which the probability of a halo hosting central and satellite galaxies is parameterized as a function of halo mass~\citep[e.g.,][]{Zheng2005}. However, galaxies selected through their [\ion{O}{2}] emission trace ongoing star formation and are known to exhibit occupation statistics that differ from those of mass-selected samples. Semi-analytic models and direct clustering measurements from Subaru Hyper Suprime-Cam (HSC), eBOSS, and DESI surveys indicate that [\ion{O}{2}] emitters preferentially reside in halos of characteristic mass $M_{\rm h} \simeq 10^{12}\, h^{-1}M_\odot$ and exhibit relatively low satellite fractions~\citep[e.g.,][]{Favole2016,Gonzalez-Perez2018,Alam2020,Avila2020,Okumura2021,Gao2023,Osato2023,Ortega-Martinez2025,Yuan2025,Favole2026,Ortega-Martinez2026}. 
The HETDEX [\ion{O}{2}] sample offers a complementary perspective: unlike higher-redshift emission-line galaxy samples targeted by other surveys, HETDEX selects [\ion{O}{2}] galaxies through a well-defined flux threshold at $z \leq 0.48$, probing a lower-redshift, star-forming galaxy population.

A key advantage of the HETDEX data is that, unlike surveys that rely on color or stellar mass cuts in photometric data as a pre-selection strategy, HETDEX performs an untargeted spectroscopic survey and obtains redshifts for \textit{all} galaxies with emission lines brighter than its sensitivity limit, without any prior target selection. This approach eliminates an additional layer of complexity in modeling the galaxy--halo connection that would otherwise arise from such selection effects. In this paper, we employ HOD modeling to interpret the power spectra of the HETDEX [\ion{O}{2}] volume-limited samples. We find that both the monopole and quadrupole power spectra can be successfully described by a simple lognormal HOD model with only three free parameters.

We follow Equation~(9) of~\citet{Geach2012} for the HOD, but exclude the error function term. \citet{Geach2012} constructed a HOD model specifically tailored for emission-line galaxies, motivated by the agreement between observations of H$\alpha$ emitters (HAEs) at $z = 2.23$ and GALFORM semi-analytic model predictions.
A key feature of their model is that the central galaxy occupation distribution follows an approximately lognormal distribution at low halo masses, with a characteristic host mass $M_0$ and width $\sigma_{\log M}$, rather than the standard step function used in mass-limited HODs. 
This lognormal component reflects the fact that the [\ion{O}{2}] emitting galaxies, like HAEs, are star-forming galaxies whose occupation number depends on the star formation efficiency as a function of halo mass, and therefore is not a simple monotonically increasing function of halo mass. 
We therefore adopt this form as the basis of our HOD model, which is well-suited for modelling emission-line galaxy populations such as [\ion{O}{2}] emitters. 
Since our mock dataset resolves subhalos, we populate galaxies to all halos including subhalos. We demonstrate in Appendix~\ref{appendix:subhalo_exclusion} that excluding galaxies in subhalos in the mock leads to a poorer agreement with the data, especially the quadrupole power spectra, justifying their inclusion.

The expectation value for the number of galaxies in a host dark matter halo of mass $M_\mathrm{h}$ is given by
\begin{equation}
N_\mathrm{g}(M_\mathrm{h}) = F_\mathrm{g}\exp\biggl[-\frac{\log(M_\mathrm{h}/M_0)^2}{2\sigma_{\log M}^2}\biggr],
\label{eq:HOD}
\end{equation}
where $F_\mathrm{g}$ is the mean fraction of [\ion{O}{2}] emitting galaxies hosted by dark matter halos with a characteristic mass of $M_0$, and $\sigma_{\log M}$ is the standard deviation of the Gaussian distribution.

Since the three parameters are degenerate with the host halo mass $M_\mathrm{h}$ and galaxy number density, we fix $\sigma_{\log M}=0.6$ and set $F_\mathrm{g}$ to match the number density of each sample. This leaves $M_0$ as the only free parameter in this analysis. This strategy allows easy interpretation of the results from different volume-limited samples. We chose a value for $\sigma_{\log M}$ that is neither too large nor too small, so that $M_\mathrm{h}$ remains larger than the smallest halo mass resolved by the mock and $\sigma_{\log M}$ is not too small to be physically implausible.

We apply this HOD model to the Uchuu simulation~\citep{Uchuu_dr1, Uchuu_nu2GCC, Uchuu_UM, Uchuu_RMG,Uchuu_SDSS}, which provides a sufficient resolution of dark matter halos and subhalos to reproduce the high number densities of the HETDEX [\ion{O}{2}] volume-limited samples. The cosmological parameters are taken from the 2015 release of the \textit{Planck} CMB data~\citep{planck2015cosmo} for a flat $\Lambda$CDM model: $\Omega_\mathrm{m} = 0.3089$, $\Omega_\Lambda = 0.6911$, $h = 0.6774$, $\sigma_8 = 0.8159$, $\Omega_\mathrm{b} = 0.0486$, and $n_\mathrm{s} = 0.9667$. We use $M_\mathrm{h}=M_\mathrm{200b}~[h^{-1}M_{\sun}]$ which is the halo mass enclosed within the mean overdensity of $200\rho_\mathrm{m}$ in the Uchuu halo catalog. The variable $\rho_\mathrm{m}$ is the mean background mass density of the universe. The minimum mass of the halos resolved in the simulation is $1.3\times 10^{10}~h^{-1}~M_\sun$, sufficient to identify low-mass star-forming galaxies such as [\ion{O}{2}] emitting galaxies. We use halo catalogs at snapshot redshifts of $z=0.19$, 0.25, and 0.36 for Bins 2, 3 and 4, respectively.
We have confirmed that the effects of small differences between the snapshot redshifts and $z_\mathrm{eff}$ of the data (see Table~\ref{tab:table_oii_sub}) are negligible in the power spectrum. The fractional difference  is less than one percent.

After selecting halos and subhalos using the HOD with a given value of $M_0$, they are moved along the LOS to account for peculiar velocities, using the corresponding scale factors and Hubble parameters from the aforementioned cosmological parameters. They are then projected onto sky coordinates (right ascension, declination and redshift).
Since only one Uchuu realization with a box size of $L_\mathrm{box}=2~h^{-1}~\mathrm{Gpc}$ is available per redshift snapshot, it is divided into 50 sub-boxes for statistical analysis.
Since Spring Bin 4 has the largest volume of the bins, 21 sub-boxes of this bin contain slightly overlapping regions, which amount to 9\% of the simulation box.

\subsection{Monopole and Quadrupole Power Spectra of HETDEX Volume-limited  \texorpdfstring{[\ion{O}{2}]}{[O II]} Samples}
\label{subsec:powerspec}
\begin{figure*}
    \centering 
    \includegraphics[width=0.69\columnwidth]{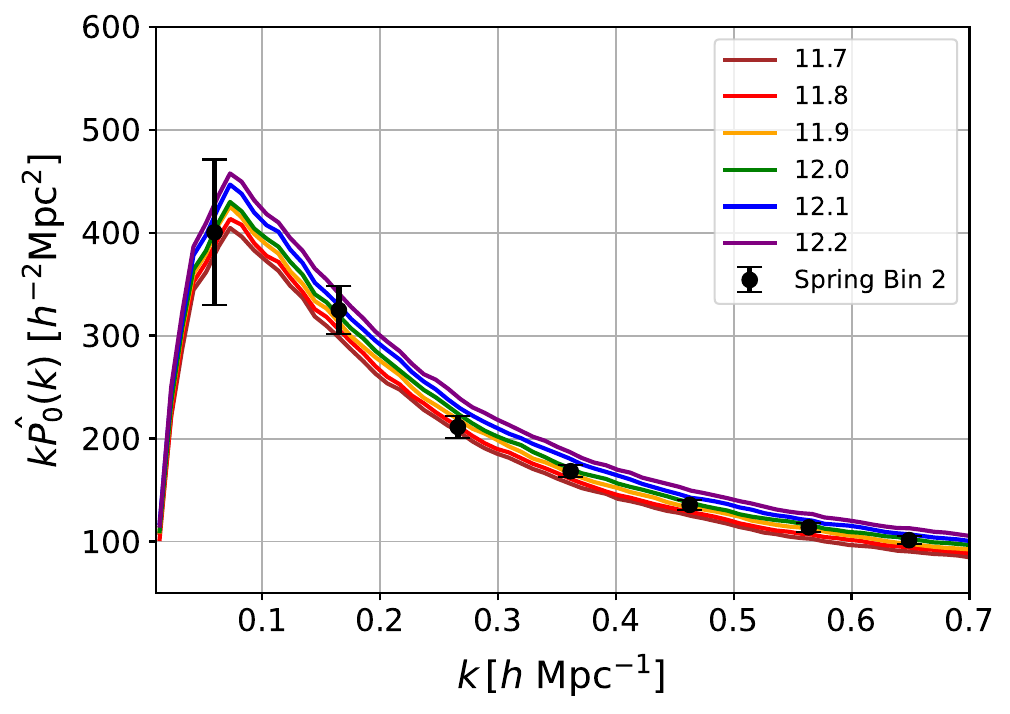}
    \includegraphics[width=0.69\columnwidth]{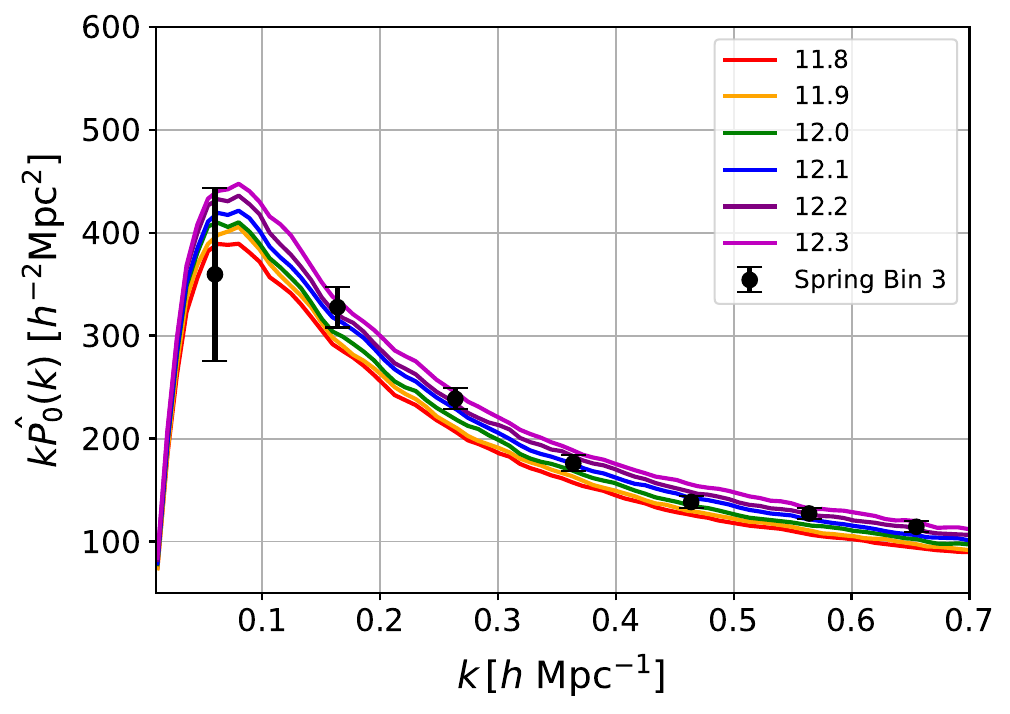}
    \includegraphics[width=0.69\columnwidth]{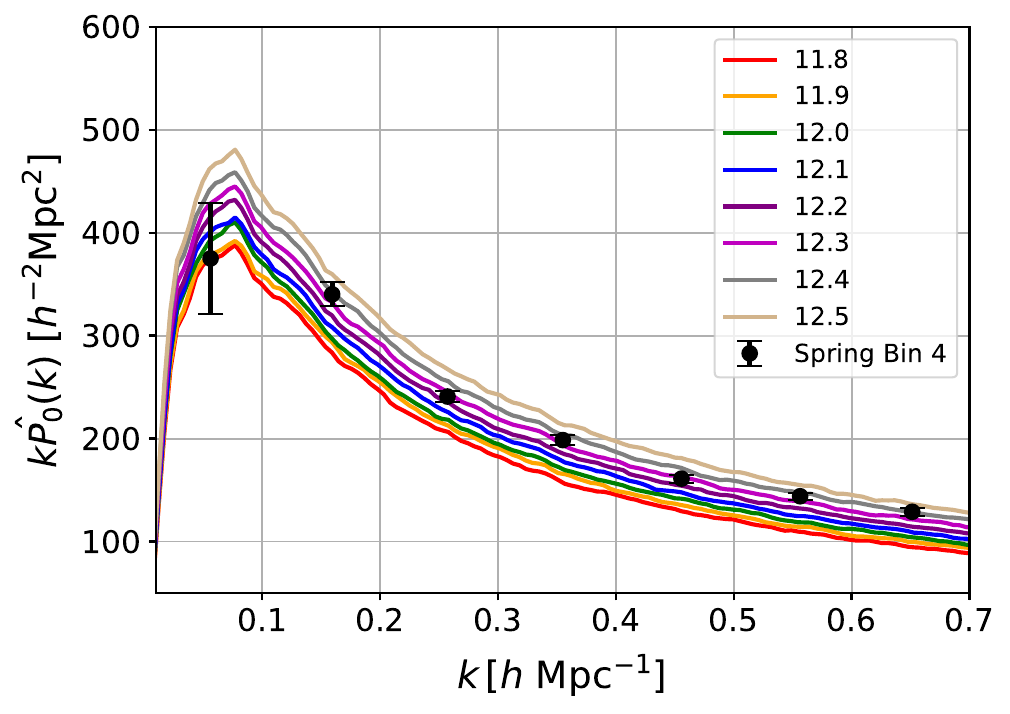}
    \includegraphics[width=0.69\columnwidth]{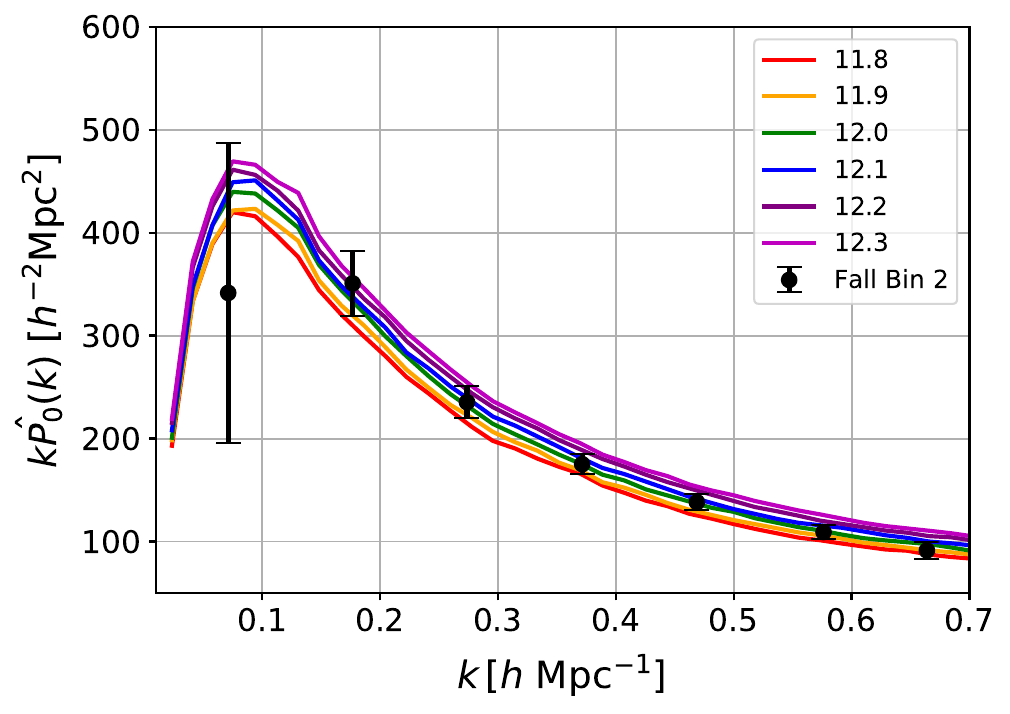}
    \includegraphics[width=0.69\columnwidth]{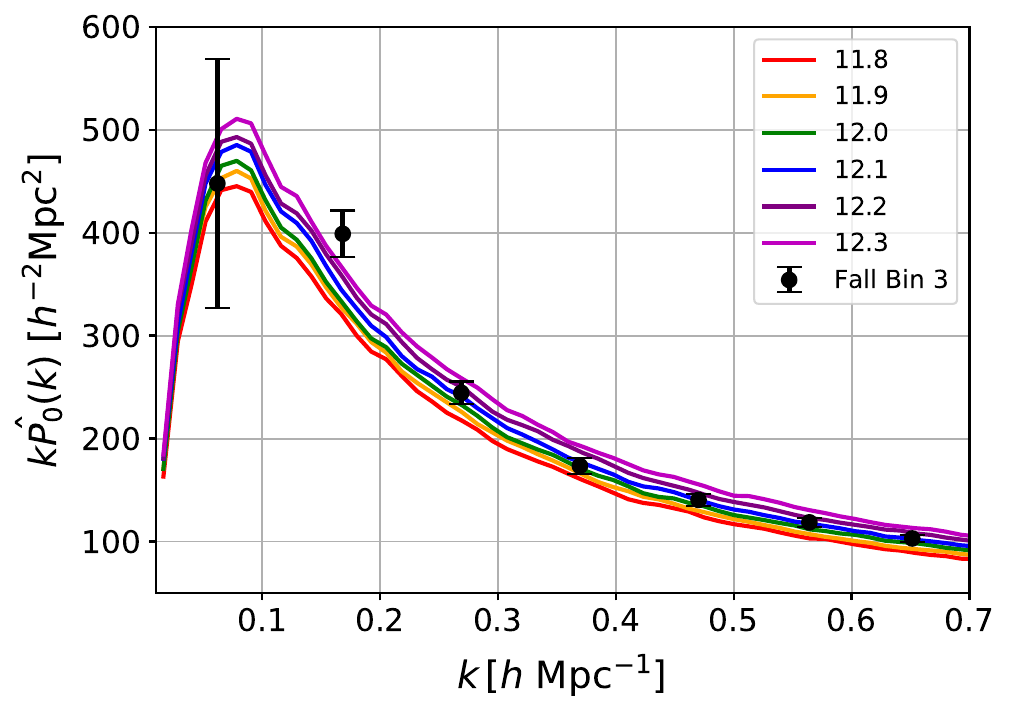}
    \includegraphics[width=0.69\columnwidth]{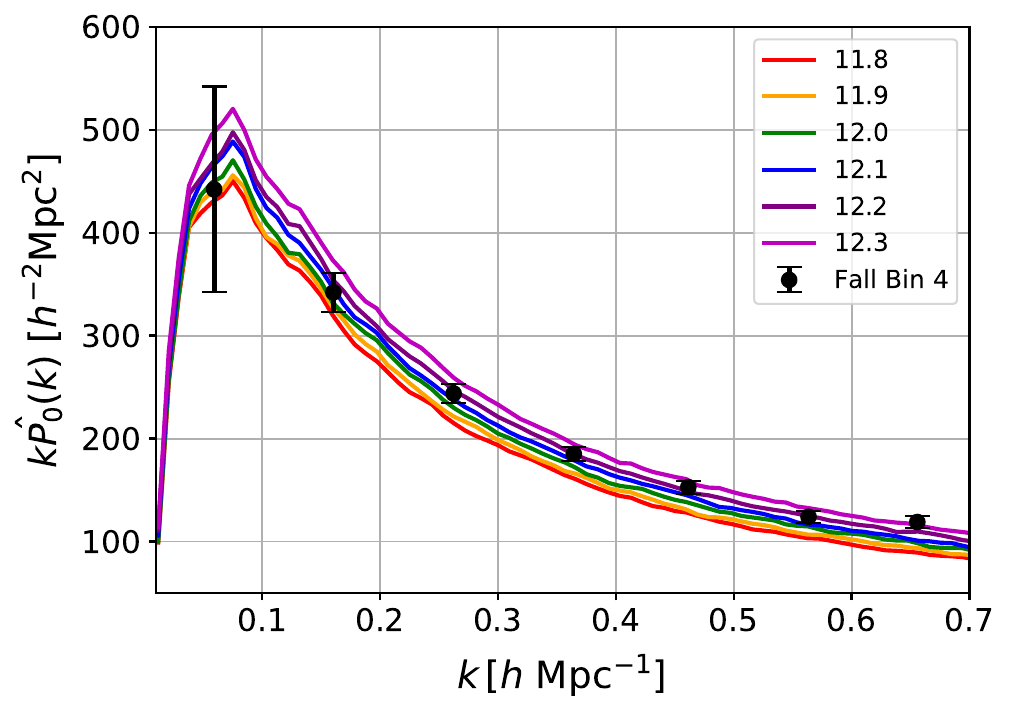}
    \caption{Monopole power spectra of HETDEX [\ion{O}{2}] volume-limited samples in Bins 2 (left panels), 3 (middle panels), and 4 (right panels) in the Spring (top panels) and Fall (bottom panels) fields. The points with the error bars show the HETDEX data binned with a width $\Delta k=0.1~h~\mathrm{Mpc}^{-1}$, with $1\sigma$ uncertainties estimated from 50 mock realizations. The colored solid lines represent the mean power spectra from the mocks for different values of $\log(M_0)$ in finer bins of $\Delta k\simeq0.01~h~\mathrm{Mpc}^{-1}$.
    Overall amplitudes increase with $\log(M_0)$ and this result suggests the existence of an optimal $\log(M_0)$ value that provides an excellent description of the data at all wavenumbers up to $k_\mathrm{max}= 0.7\,h\,\mathrm{Mpc}^{-1}$, which justifies our choice of a lognormal HOD.
    }
    \label{fig:plot_p0_M0_comparison}
\end{figure*}

\begin{figure*}
    \centering 
    \includegraphics[width=0.69\columnwidth]{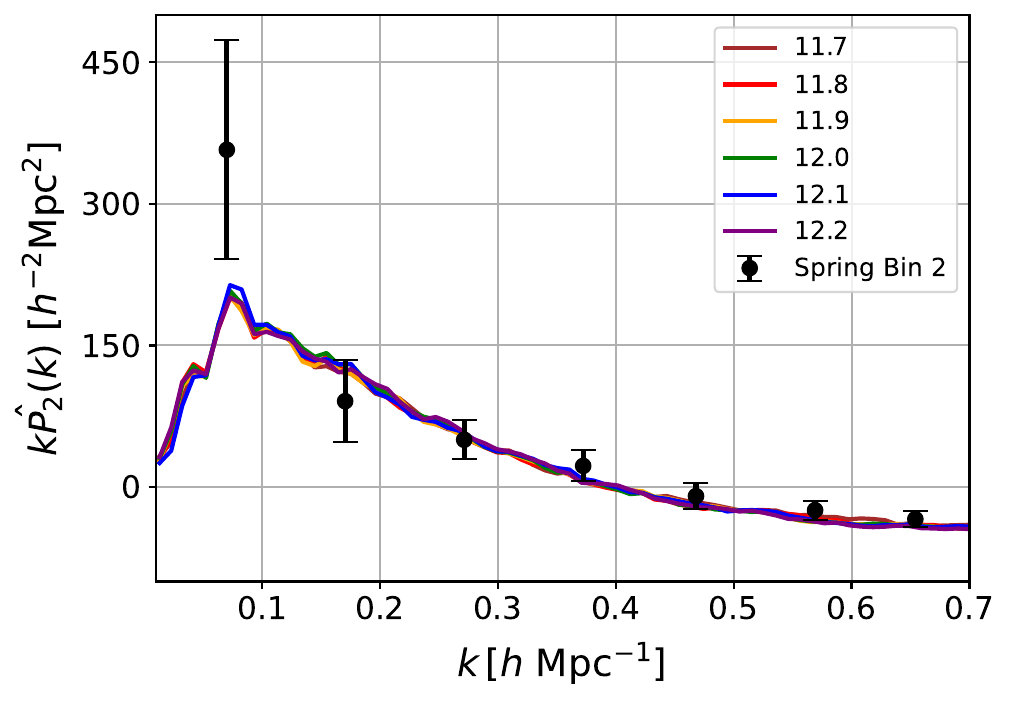}
    \includegraphics[width=0.69\columnwidth]{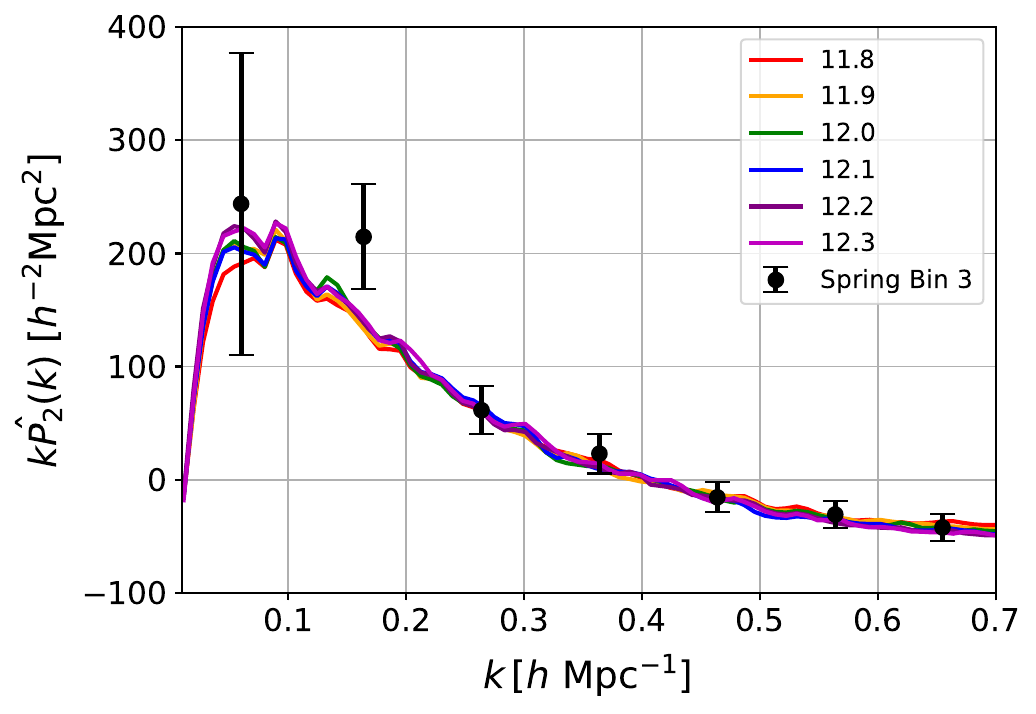}
    \includegraphics[width=0.69\columnwidth]{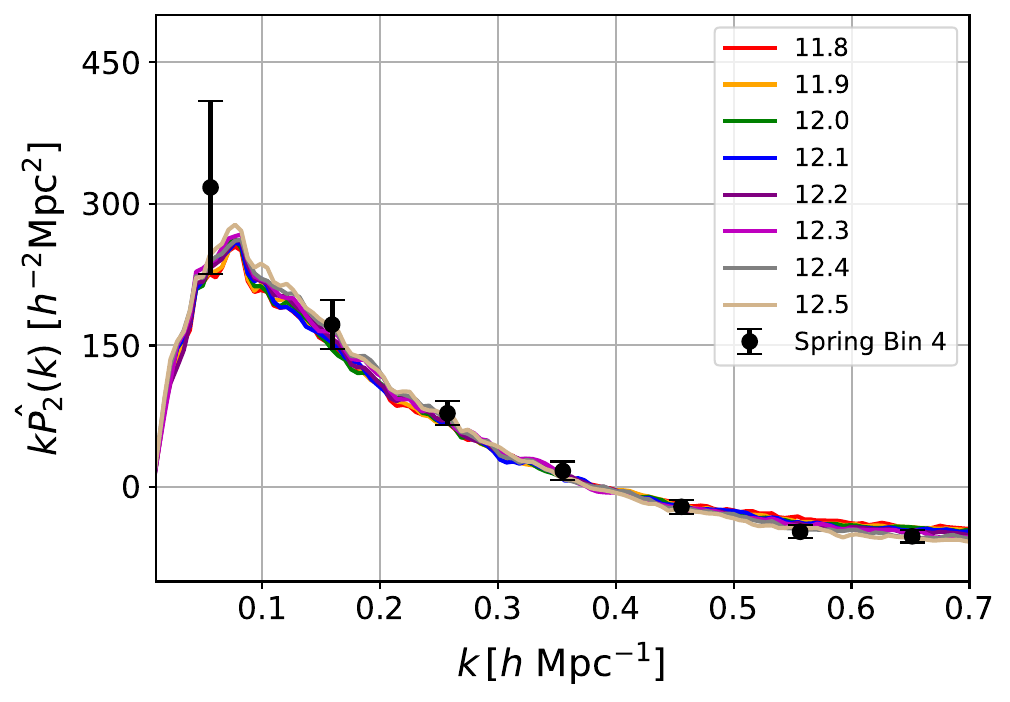}
    \includegraphics[width=0.69\columnwidth]{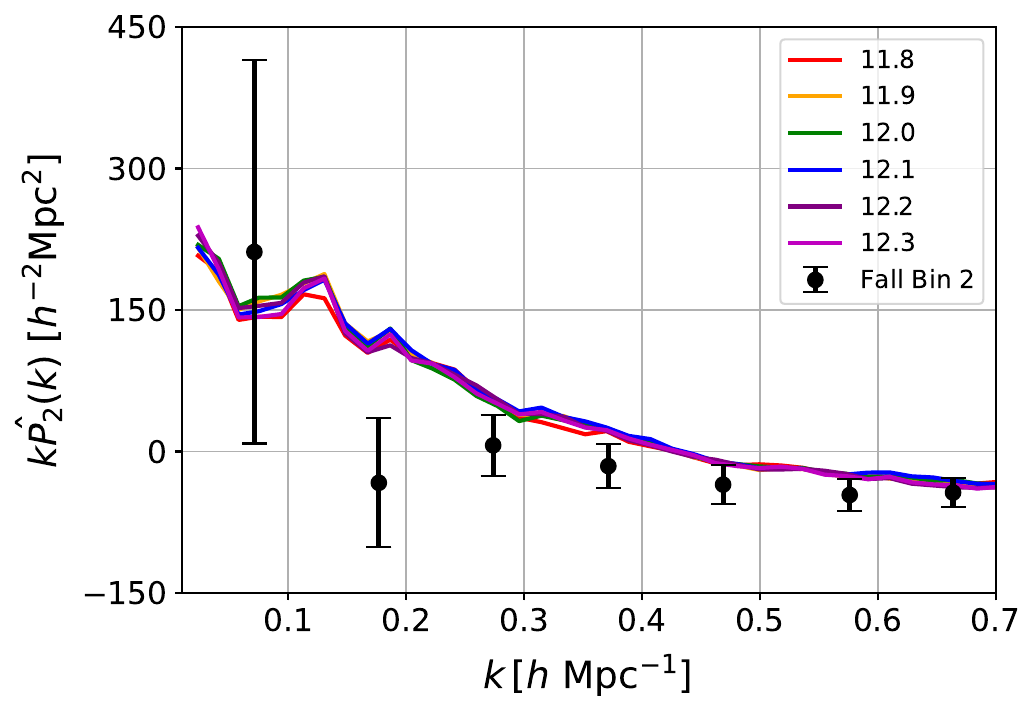}
    \includegraphics[width=0.69\columnwidth]{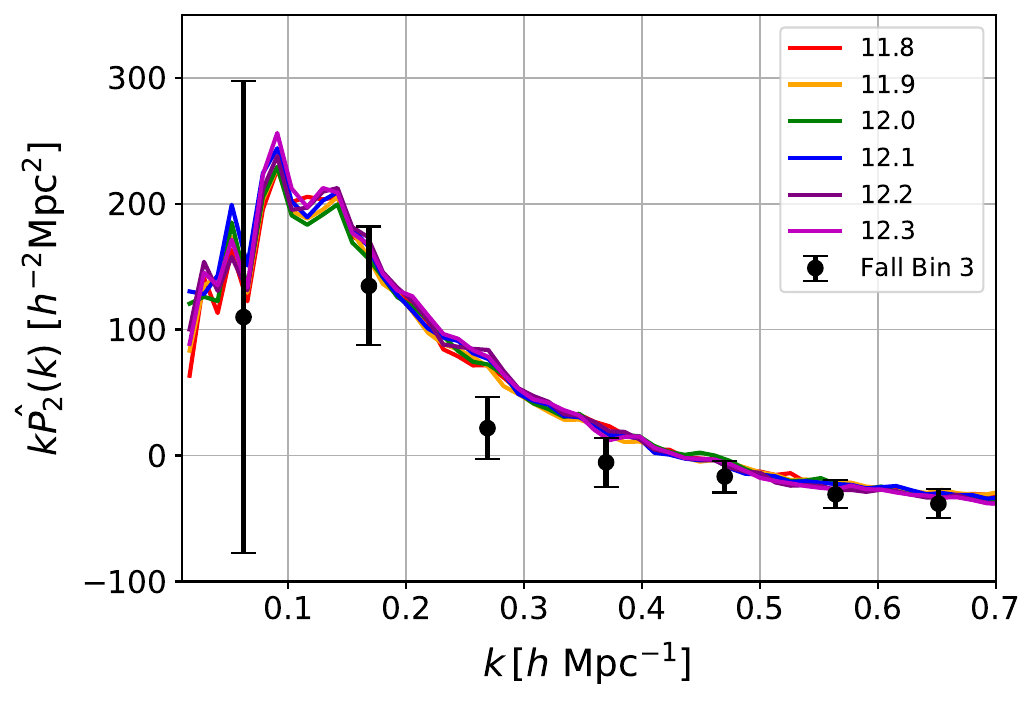}
    \includegraphics[width=0.69\columnwidth]{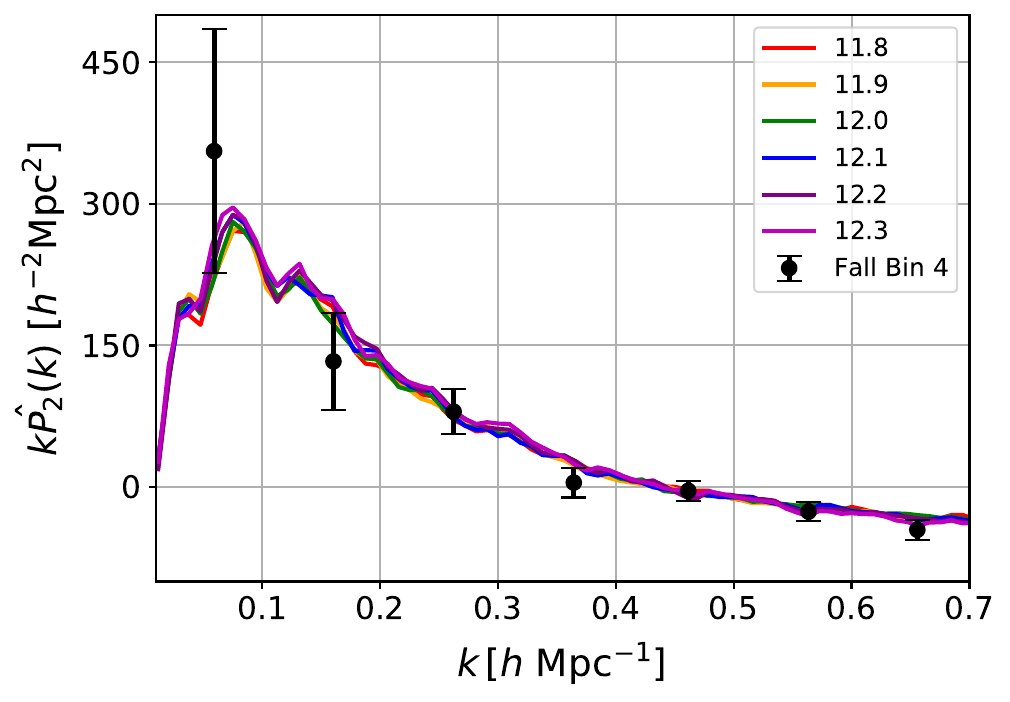}
    \caption{Same format as Figure~\ref{fig:plot_p0_M0_comparison} but for the quadrupole power spectra.
    Overall amplitudes remain similar across different $\log(M_0)$ values and a lognormal HOD provides excellent description of the data at all wavenumbers up to $k_\mathrm{max}= 0.7\,h\,\mathrm{Mpc}^{-1}$.
    }
    \label{fig:plot_p2_M0_comparison}
\end{figure*}

Figures \ref{fig:plot_p0_M0_comparison} and \ref{fig:plot_p2_M0_comparison} show the monopole and quadrupole power spectra of the HETDEX [\ion{O}{2}] volume-limited samples, respectively, and compare them with those of the mock dataset for different values of $\log(M_0)$. 
We find that the overall amplitude of the monopole power spectrum increases with $\log(M_0)$, while that of the quadrupole power spectrum remains similar across different $\log(M_0)$ values. The mock provide an excellent description of the data, which justifies our choice of a lognormal HOD. The data do not require a more complicated HOD in this analysis.

In this work, we restrict our likelihood analysis to the monopole power spectra. Higher-order multipoles will be utilized in future  work to constrain the growth of structures.
We estimate the best-fit $\log(M_0)$ values using the Sellentin-Heavens likelihood \citep{Sellentin2016}, as described in Section~\ref{subsec:likelihood}.

\section{Interpretation of the monopole power spectra}
\label{sec:interpretation}

\subsection{Likelihood analysis}
\label{subsec:likelihood}
\begin{figure*}
    \centering 
    \includegraphics[width=0.69\columnwidth]{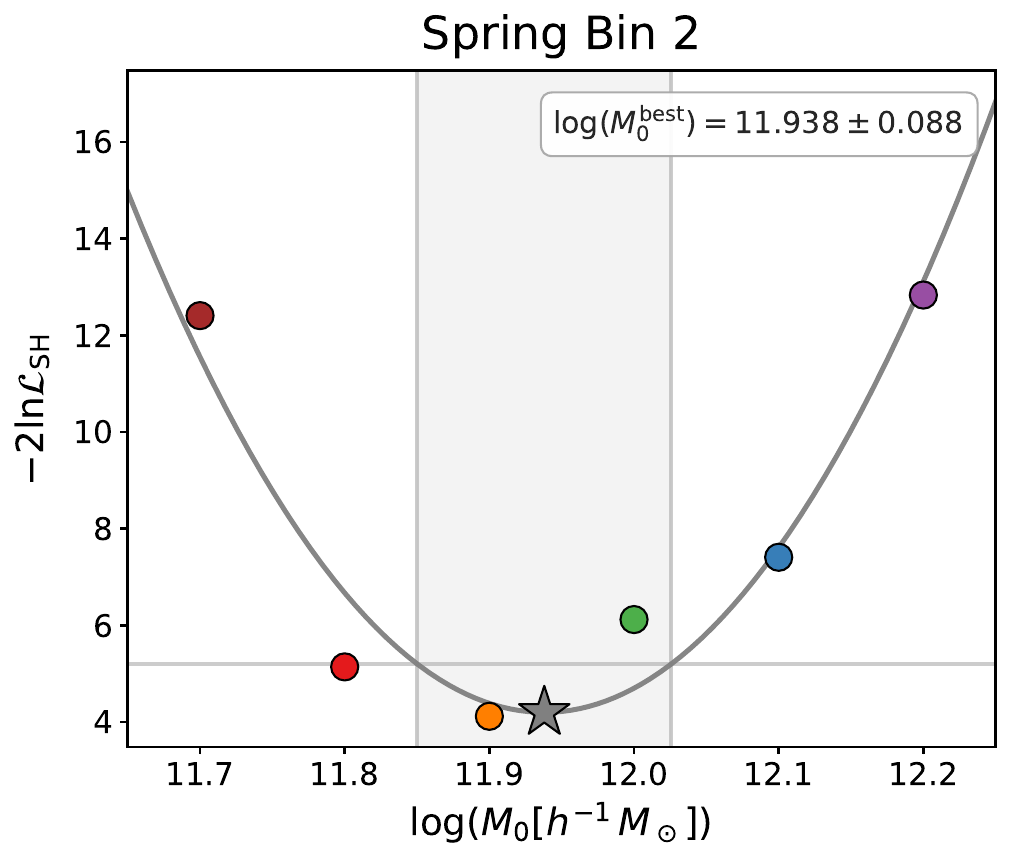}
    \includegraphics[width=0.69\columnwidth]{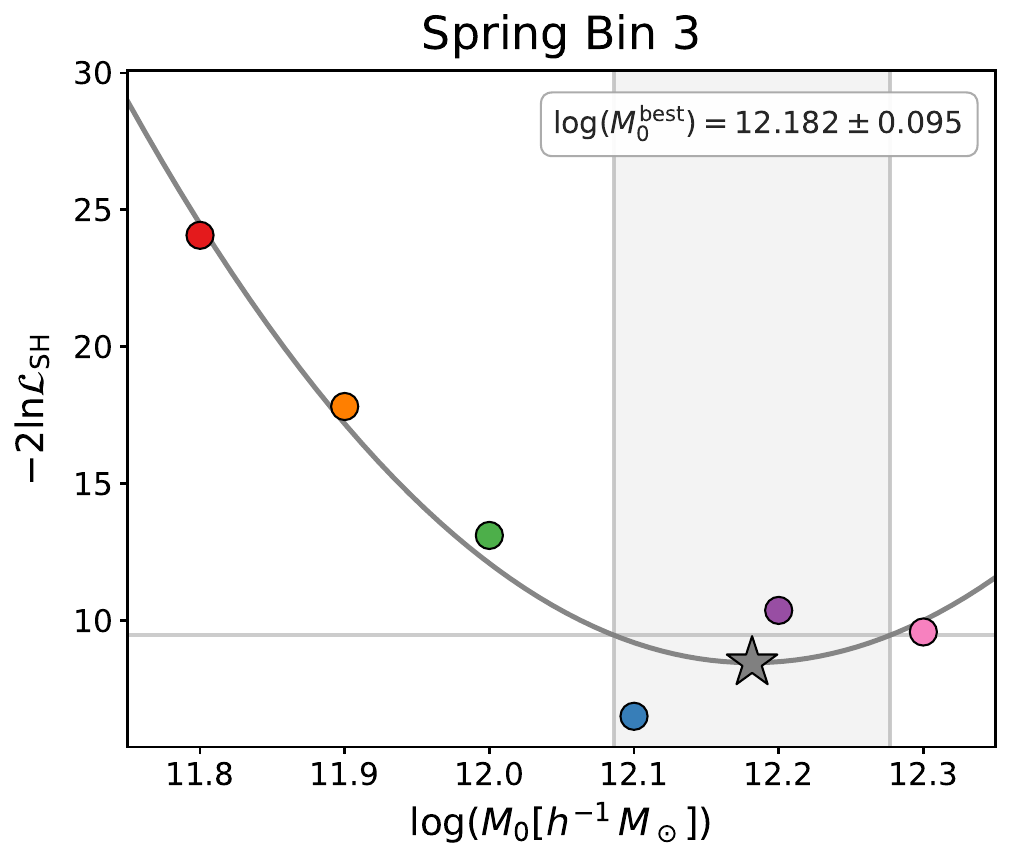}
    \includegraphics[width=0.69\columnwidth]{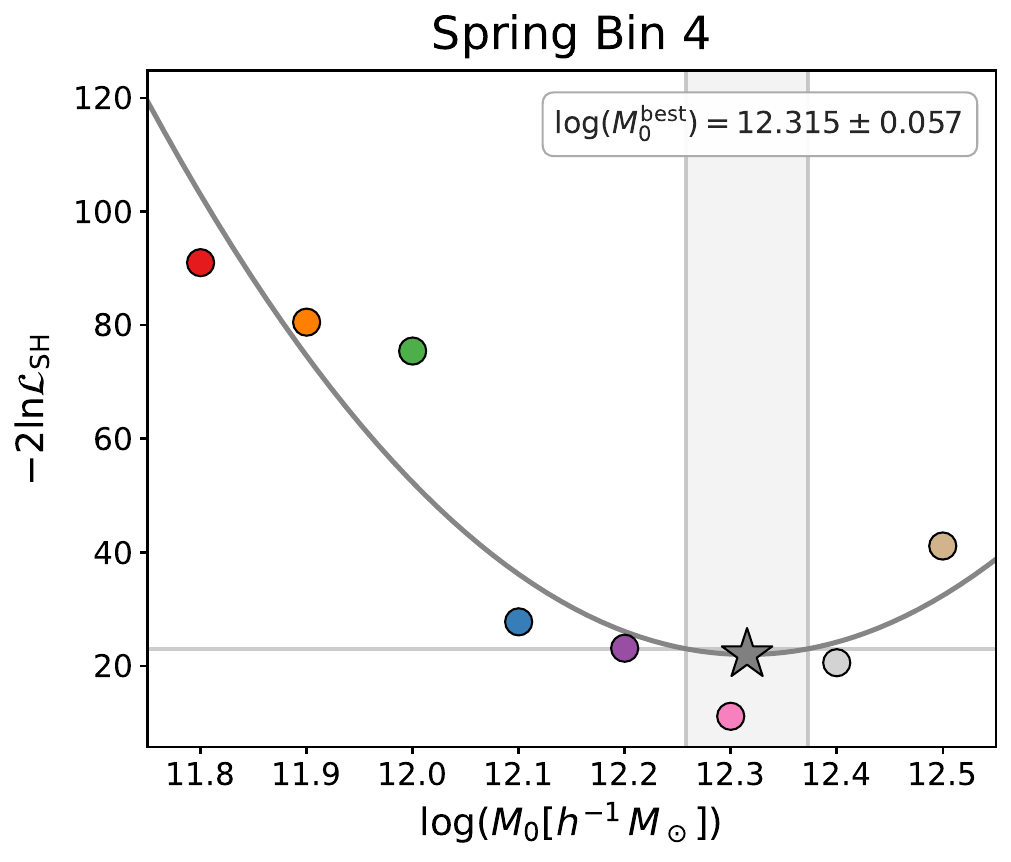}
    \includegraphics[width=0.69\columnwidth]{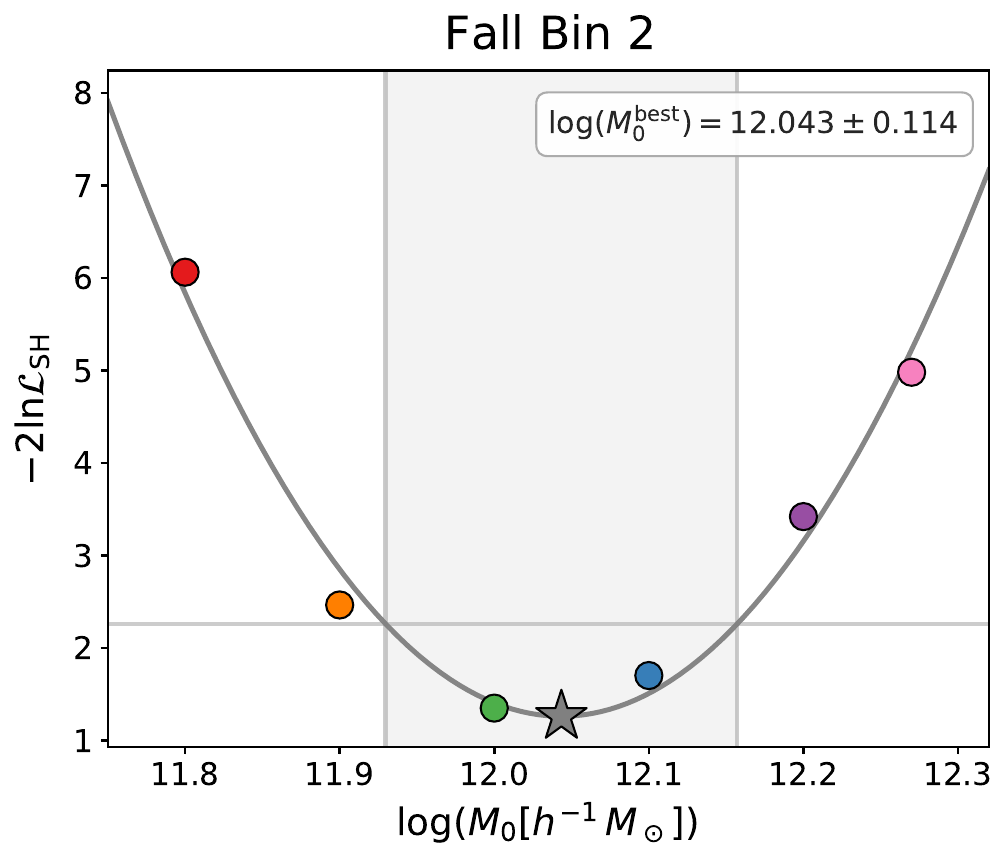}
    \includegraphics[width=0.69\columnwidth]{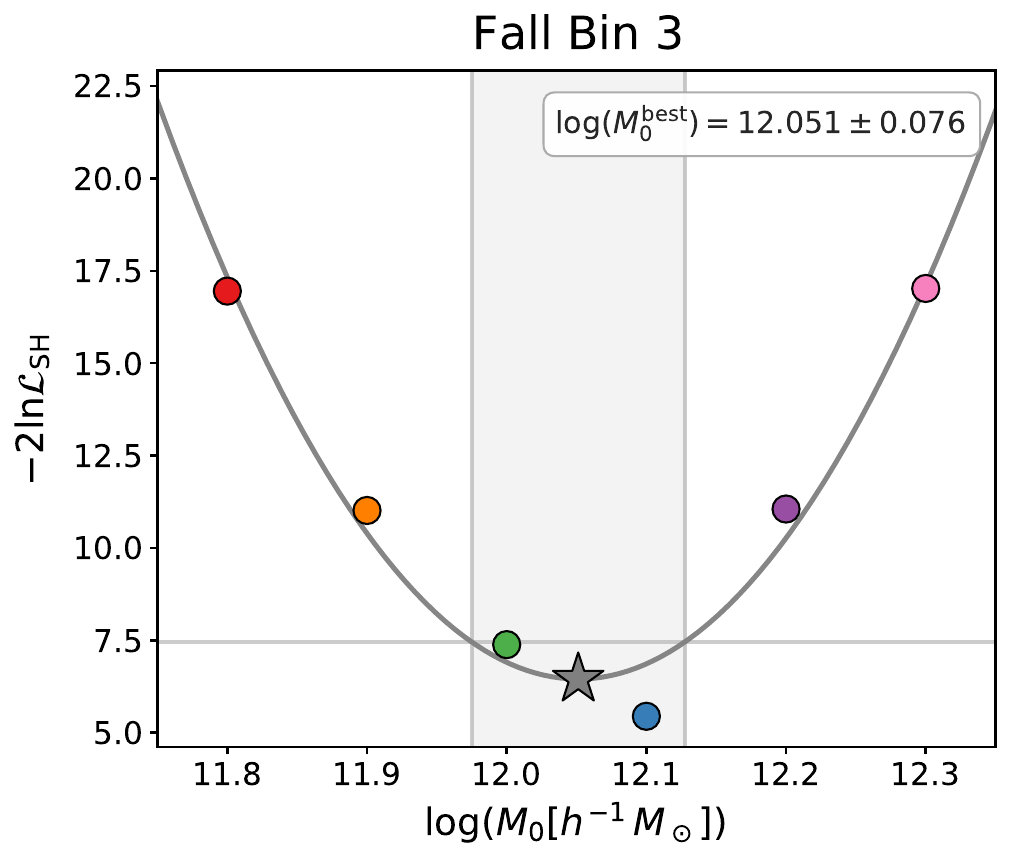}
    \includegraphics[width=0.69\columnwidth]{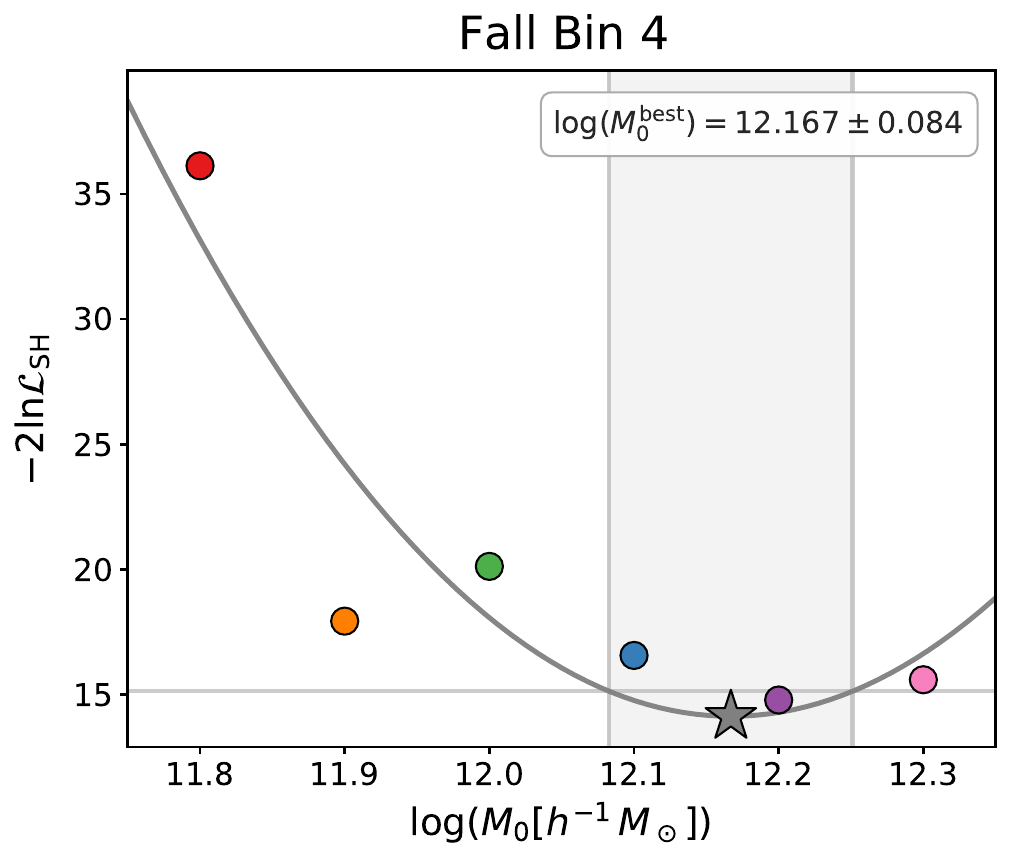}
    \caption{Parabolic fits to $\chi_\mathrm{SH}^2({\cal M})=-2\ln{\cal L}_\mathrm{SH}$ [Equation~\eqref{eq:chisqSH}] estimating the best-fit log characteristic halo mass, $\mathcal{M}=\log(M_0)$, where $M_0$ is in units of $h^{-1}M_\sun$, from Bins 2 (left panels), 3 (middle panels), and 4 (right panels) in the Spring (top panels) and Fall (bottom panels) fields. The filled circles represent $\chi_\mathrm{SH}^2({\cal M})$ evaluated at different $\mathcal{M}$ values, the stars denote the best-fit ${\cal M}$, and the horizontal and vertical lines mark the $1\sigma$ uncertainty, defined by $\Delta \chi^2_\mathrm{SH} = 1$. The best-fit $\mathcal{M}$ values are quoted in the upper-right corner of each panel.}
    \label{fig:plot_chi2_interpolation}
\end{figure*}

\begin{figure*}
    \centering 
    \includegraphics[width=0.69\columnwidth]{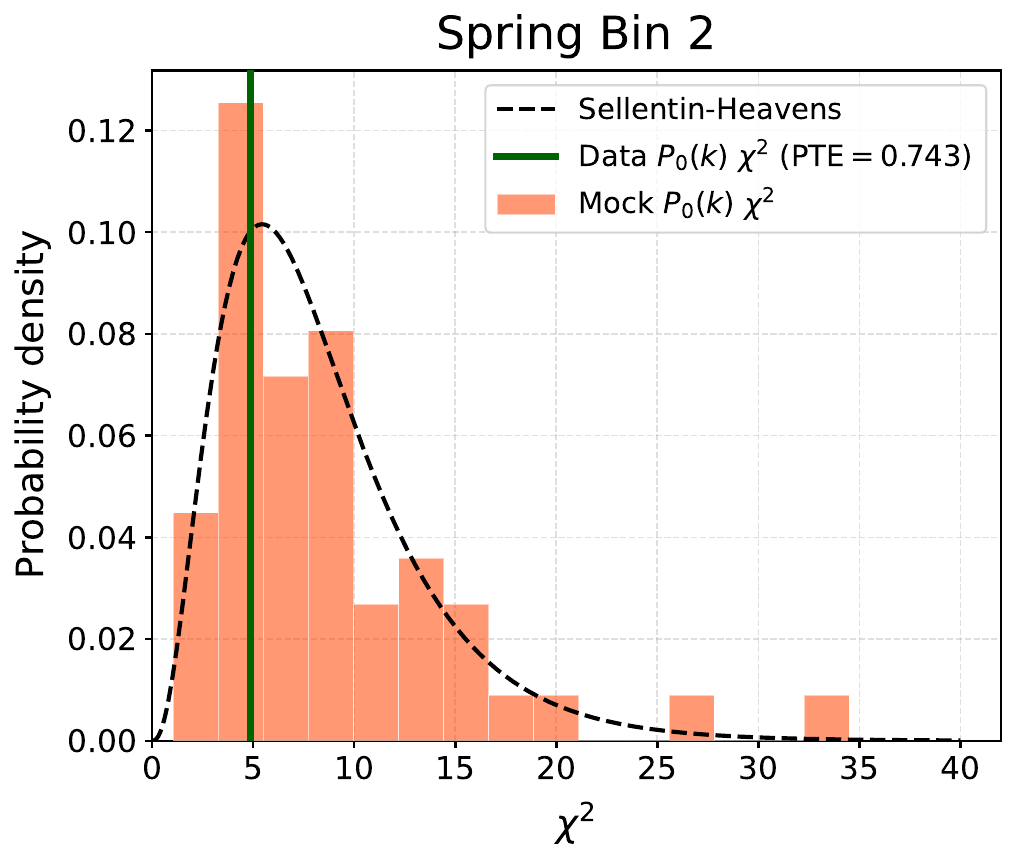}
    \includegraphics[width=0.69\columnwidth]{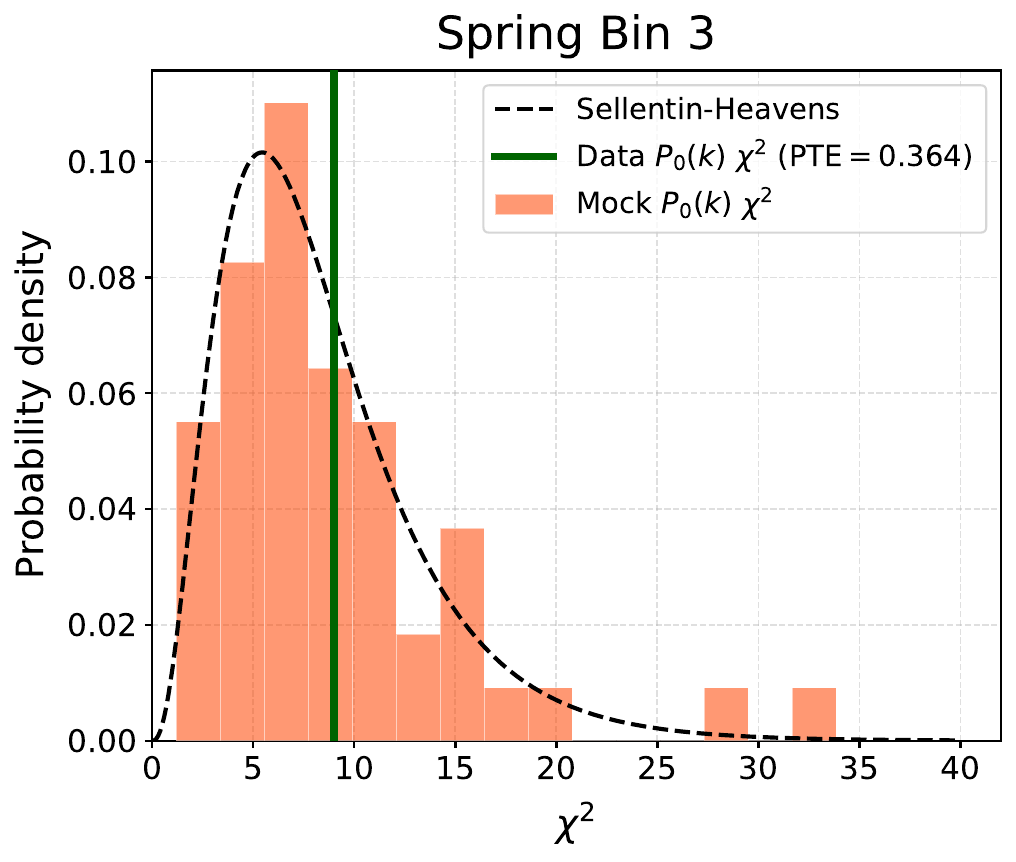}
    \includegraphics[width=0.69\columnwidth]{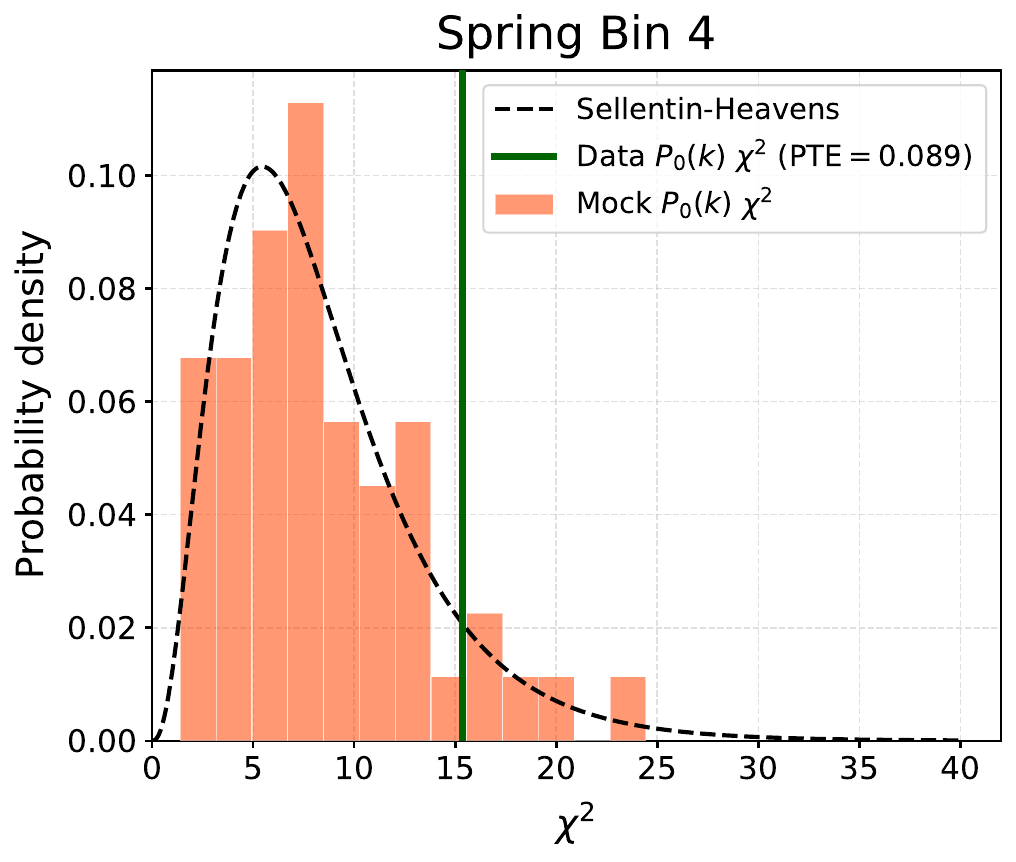}
    \includegraphics[width=0.69\columnwidth]{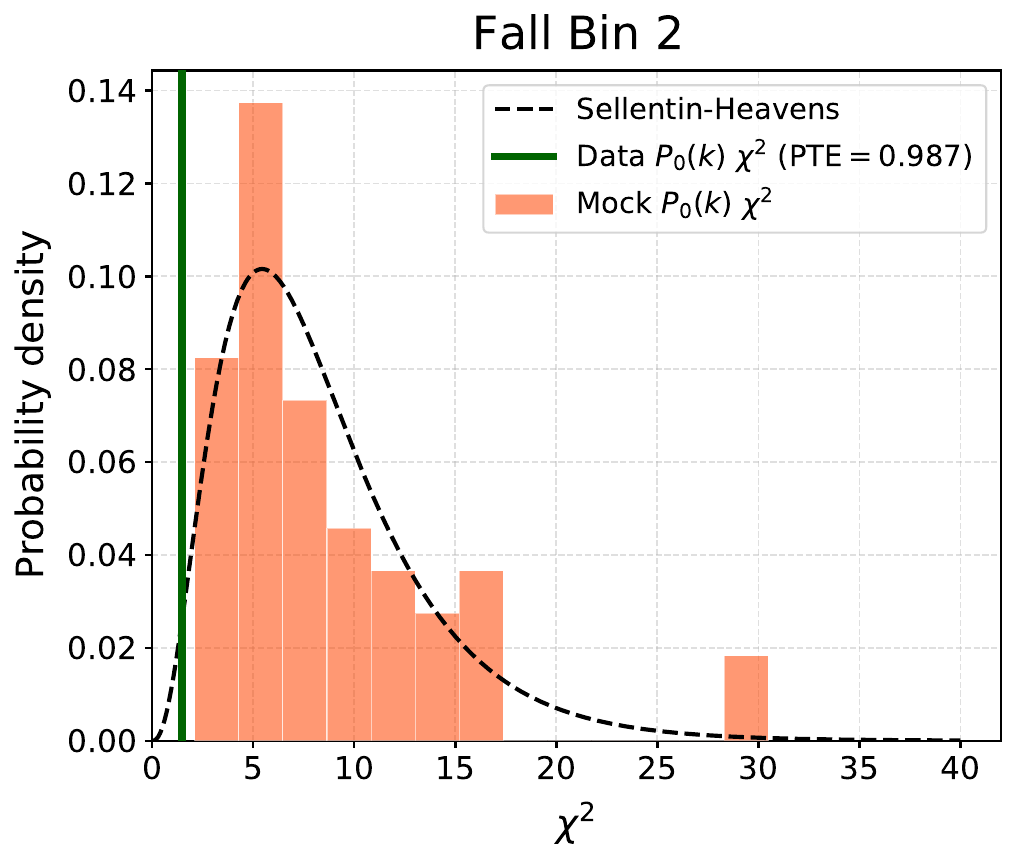}
    \includegraphics[width=0.69\columnwidth]{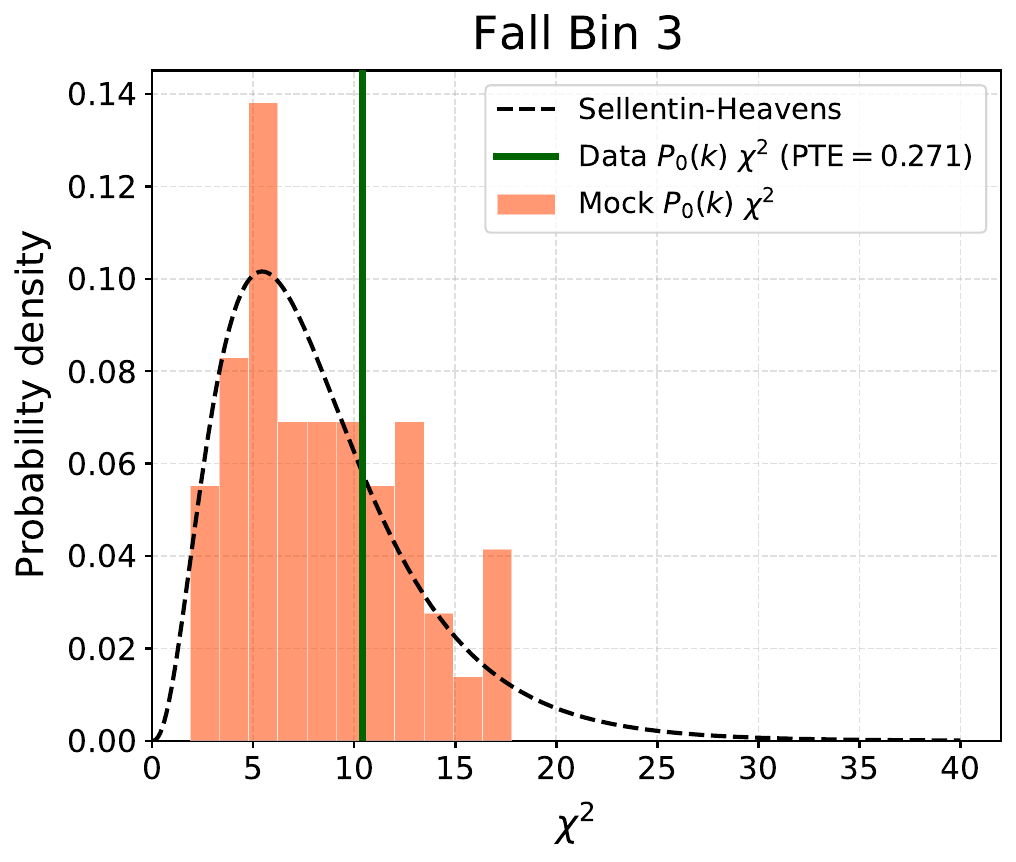}
    \includegraphics[width=0.69\columnwidth]{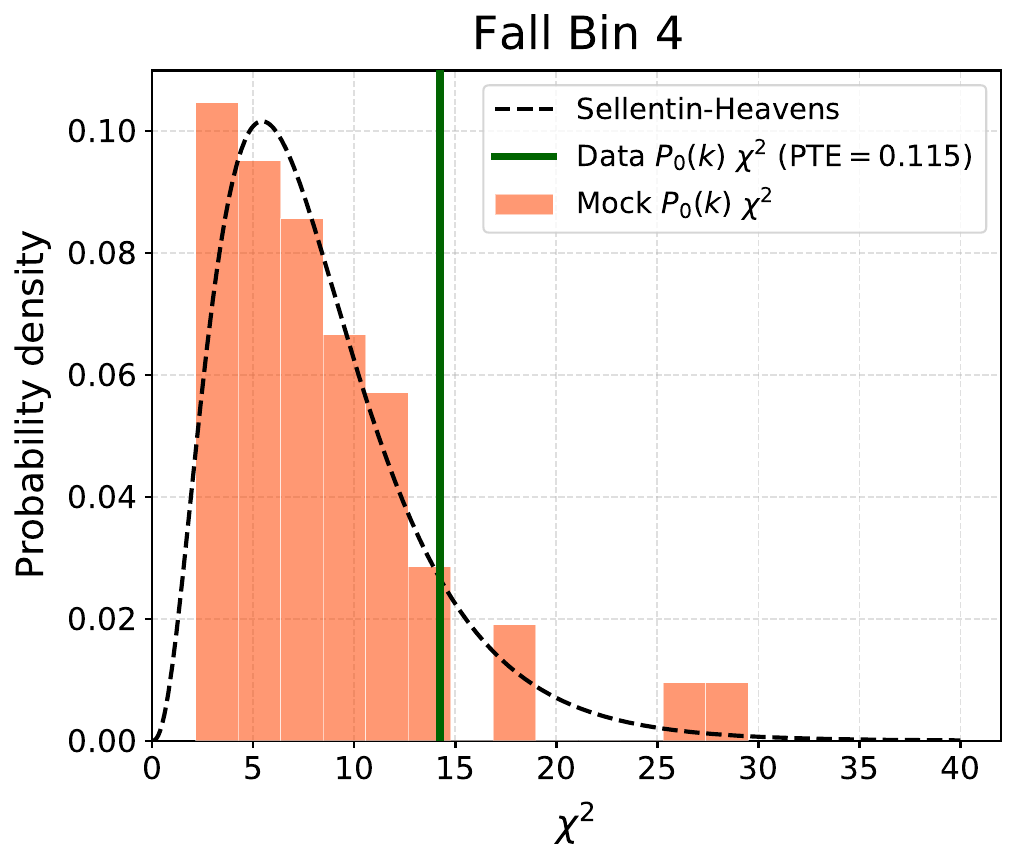}
    \caption{Distribution of $\chi^2$ [Equation~\eqref{eq:chi2_full}] of the monopole power spectra calculated from the mock realizations with the best-fit value of $\mathcal{M}$ for Bins 2 (left panels), 3 (middle panels), and 4 (right panels) in the Spring (top panels) and Fall (bottom panels) fields. 
    The histograms show the distribution of $\chi^2$ from the mock, the vertical lines show the data $\chi^2$ values, and the dashed lines show the S\&H PDFs.
    Empirical distributions of $\chi^2$ from the mock realizations are consistent with the theoretical expectations.
    Excellent PTE values are identified for data and their $\chi^2$ values remain well within the mock distribution.
    }
    \label{fig:plot_P0_chi2_distribution_best-fit_Mcut}
\end{figure*}

\begin{figure*}
    \centering 
    \includegraphics[width=0.69\columnwidth]{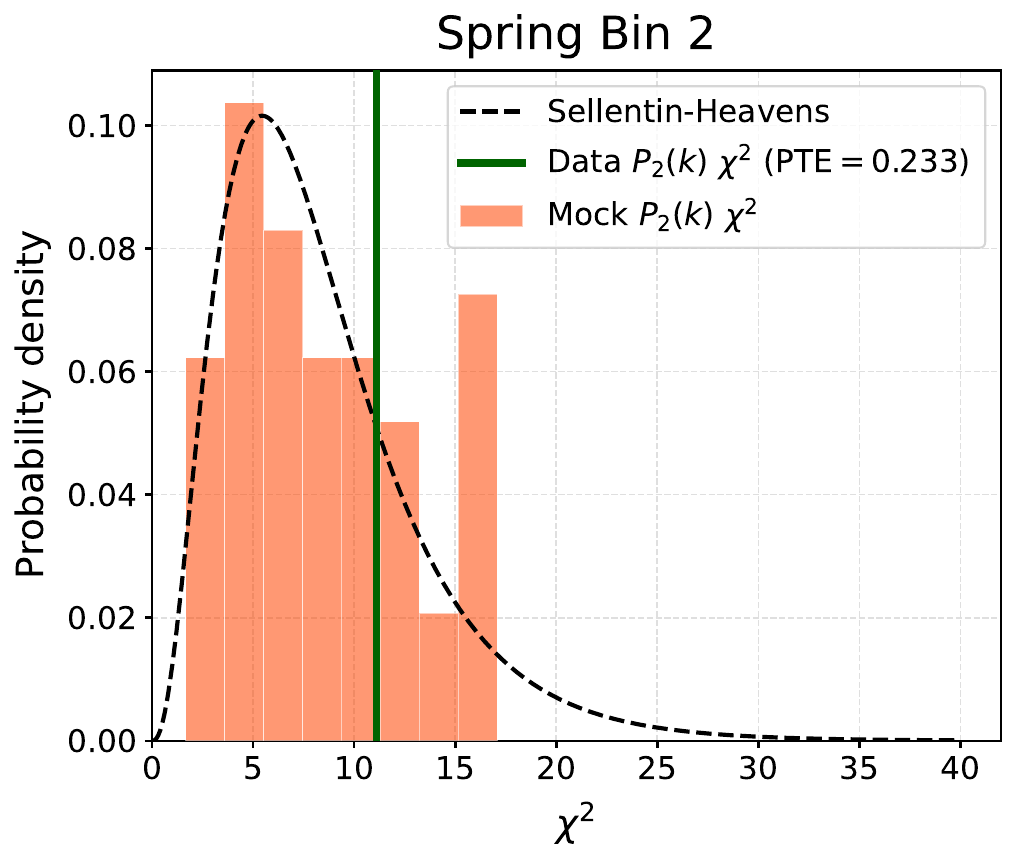}
    \includegraphics[width=0.69\columnwidth]{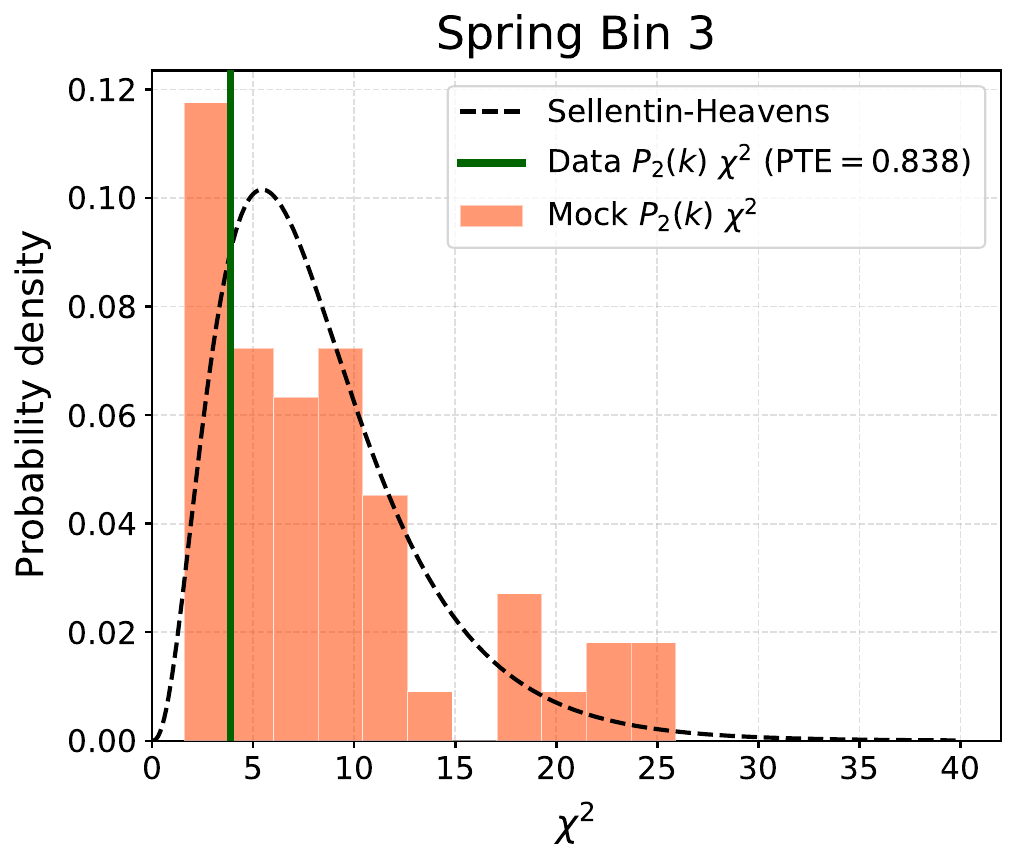}
    \includegraphics[width=0.69\columnwidth]{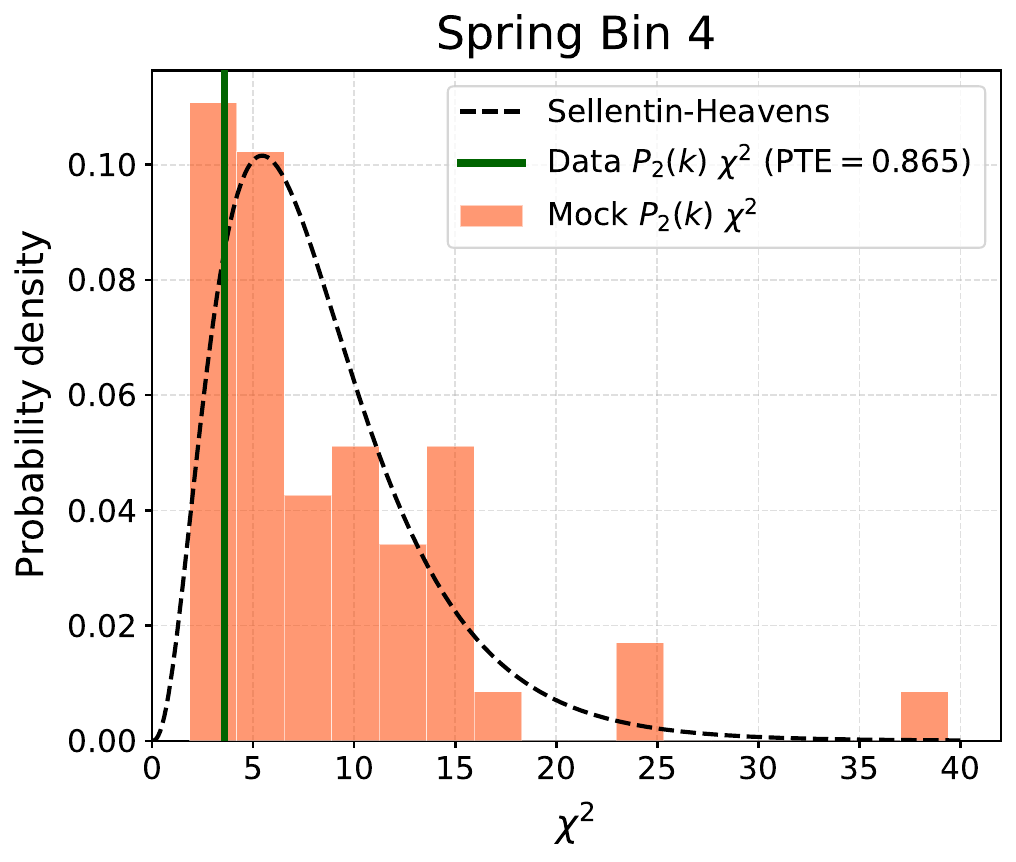}
    \includegraphics[width=0.69\columnwidth]{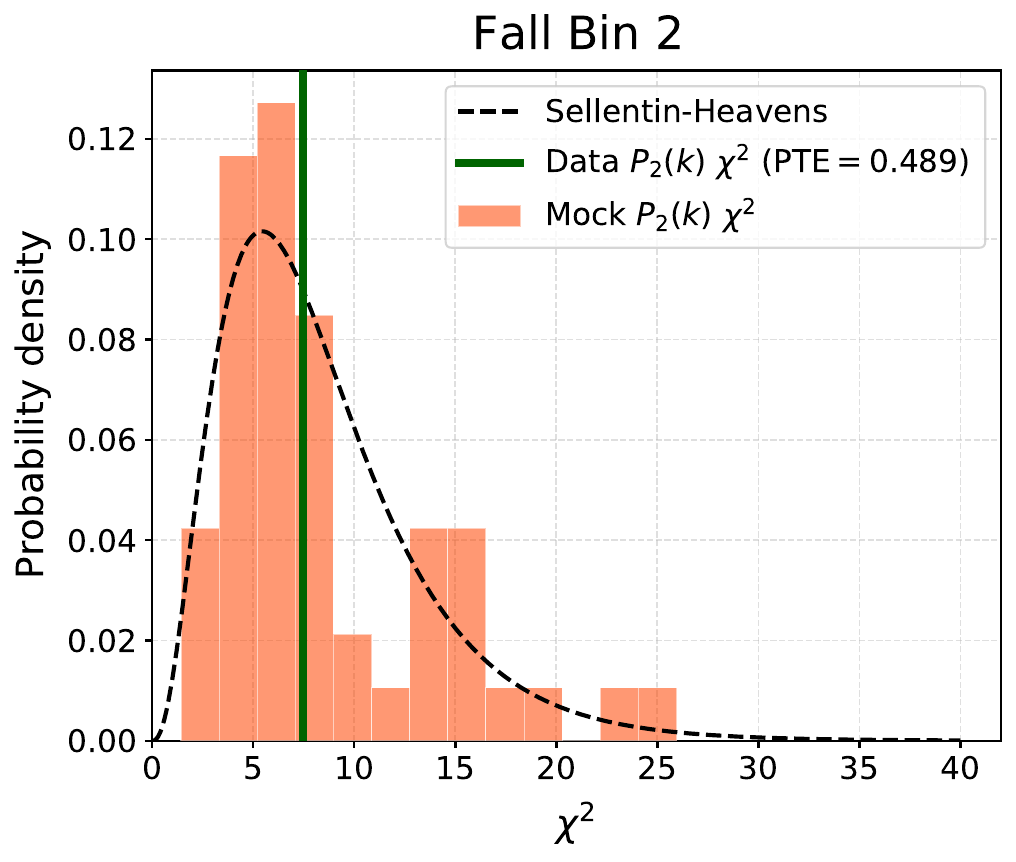}
    \includegraphics[width=0.69\columnwidth]{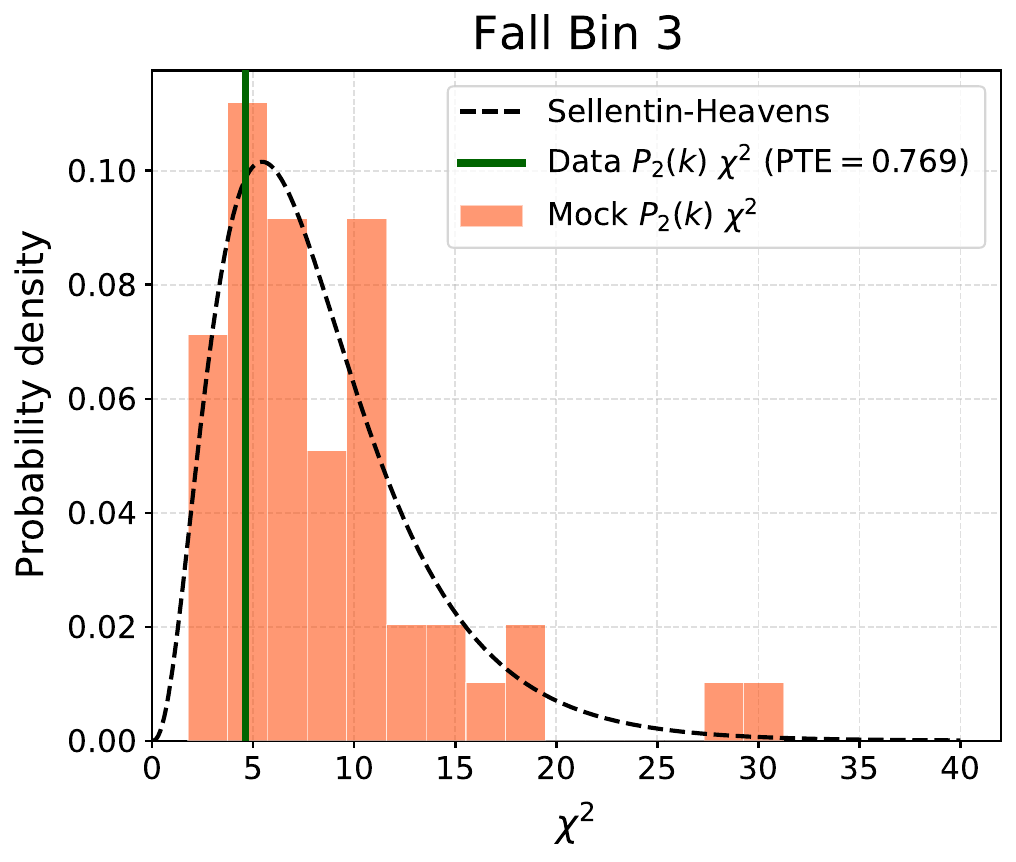}
    \includegraphics[width=0.69\columnwidth]{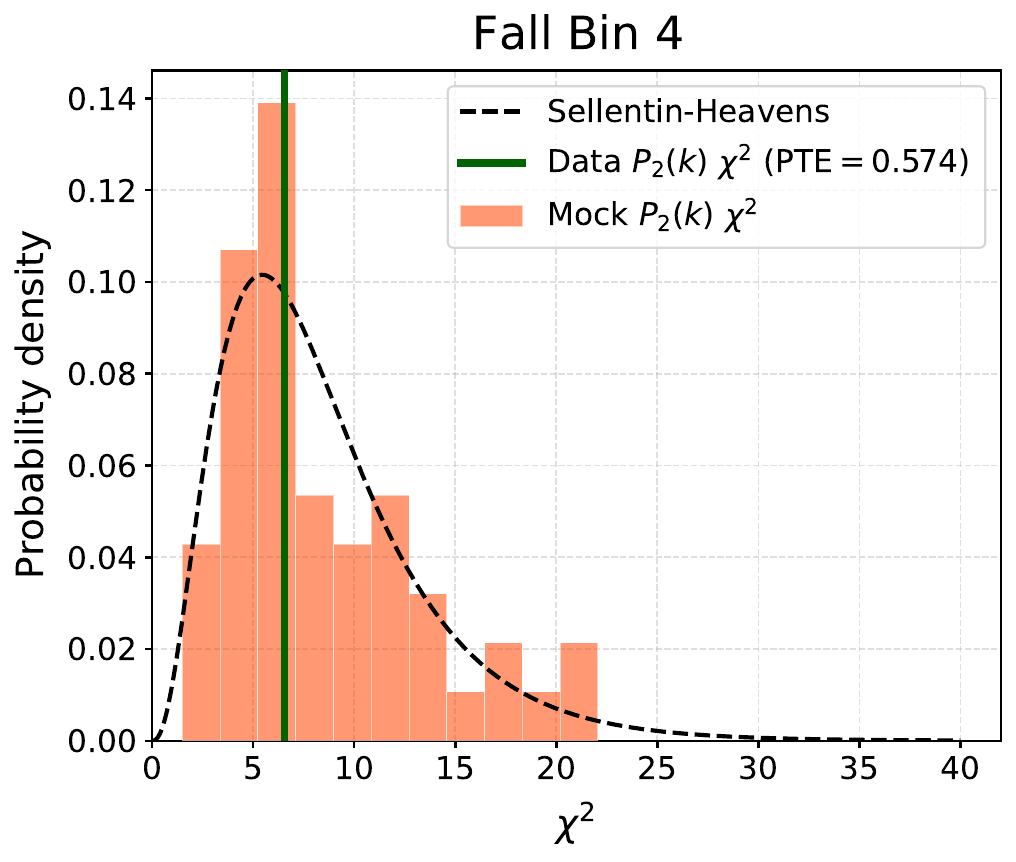}
    \caption{Same format as Figure~\ref{fig:plot_P0_chi2_distribution_best-fit_Mcut} but for the quadrupole power spectra. 
    The results are consistent with those of the monopole, confirming the goodness-of-fit across all bins and fields.}
    \label{fig:plot_P2_chi2_distribution_best-fit_Mcut}
\end{figure*}

The best-fit characteristic halo mass, $M_0$, is estimated by computing $\chi^2({\cal M})$ across a grid of ${\cal M}\equiv \log(M_0)$ values
\begin{equation}
\begin{split}
    \chi^2({\cal M}) = \sum_{i,j} \left[\hat{P}_0(k_i) - 
    \bar{P}_0^{\rm mock}(k_i, {\cal M})\right] \\
    \times~C_{ij}^{-1}({\cal M})
    \left[\hat{P}_0(k_j) - 
    \bar{P}_0^{\rm mock}(k_j, {\cal M})\right],
    \label{eq:chi2_full}
\end{split}
\end{equation}
where $\hat{P}_0(k_i)$ is the measured monopole power spectrum in the $i$-th wavenumber bin, and $\bar{P}_0^{\rm mock}(k_i, {\cal M})$ is the mean mock monopole power spectrum calculated from 50 realizations at a given value of ${\cal M}$. The inverse of the covariance matrix, $C_{ij}^{-1}$, is estimated from the 50 realizations as
\begin{equation}
\begin{split}
    C_{ij}({\cal M}) = \frac{1}{N_\mathrm{s} - 1}\sum_{n=1}^{N_\mathrm{s}}
    \left[P_0^{(n)}(k_i, {\cal M}) - \bar{P}_0^{\rm mock}(k_i, {\cal M})\right]\\
    \times~\left[P_0^{(n)}(k_j, {\cal M}) - \bar{P}_0^{\rm mock}(k_j, {\cal M})\right],
    \label{eq:cov}
\end{split}
\end{equation}
where $P_0^{(n)}(k_j, {\cal M})$ is the mock monopole power spectrum calculated from the $n$-th realization, and $N_\mathrm{s}=50$ is the number of mock realizations.

We use the Sellentin-Heavens (S\&H) likelihood \citep{Sellentin2016} to account for the finite number of mock realizations used to estimate the covariance matrix:
\begin{equation}
\label{eq:chisqSH}
    \chi_\mathrm{SH}^2({\cal M})\equiv-2\ln\mathcal{L}_{\rm SH}({\cal M}) = N_\mathrm{s} \ln\left(1 + 
    \frac{\chi^2({\cal M})}{N_\mathrm{s} - 1}\right).
\end{equation}
A parabola is fitted to $\chi_\mathrm{SH}^2({\cal M})$,
\begin{equation}
\label{eq:parabola_SH}
\chi_\mathrm{SH}^2({\cal M}) = a_{\mathrm SH}\,{\cal M}^2 + b_{\mathrm SH}\,{\cal M} + c_{\mathrm SH},
\end{equation}
to yield the best-fit $\mathcal{M}$ value at the minimum, ${\cal M}^{\rm best} = -b_{\mathrm SH}/(2a_{\mathrm SH})$, with a $1\sigma$ uncertainty of $\sigma_{{\cal M}} = 1/\sqrt{a_{\rm SH}}$ derived from the condition $\Delta \chi^2_\mathrm{SH} \equiv \chi_\mathrm{SH}^2-\chi_\mathrm{SH,min}^2 = 1$.

Equation~\eqref{eq:chisqSH} is derived using the Jeffreys prior, as described in \citet{Sellentin2016}, and we use this likelihood in the analysis.
We have also tested the modified likelihood using the frequentist-matching prior~\citep{Percival2022}, finding the same best-fit values with slightly more conservative uncertainties (approximately 5 percent).

The goodness-of-fit is assessed using the Hotelling $T^2$ test.
The measured $T^2$ distribution is the raw $\chi^2$ distribution evaluated at the ${\cal M}^{\rm best}$ following Equation~(17) of \citet{Sellentin2016}:
\begin{equation}
\label{eq:F_dist}
F_{\rm obs} \equiv \frac{\chi^2_\mathrm{min}\,(N_\mathrm{s} - N_\mathrm{d})}{(N_\mathrm{s}-1)\,N_\mathrm{d}} 
\sim F(N_\mathrm{d},\,N_\mathrm{s} - N_\mathrm{d}),
\end{equation}
where $N_\mathrm{d}$ is the number of $k$-bins and $F\!\left(N_\mathrm{d},\; N_\mathrm{s} - N_\mathrm{d}\right)$ is the $F$-distribution of $N_\mathrm{d}$ and $N_\mathrm{s}-N_\mathrm{d}$ degrees of freedom~\citep{Anderson2003}. A probability-to-exceed (PTE) is defined as $\mathrm{PTE} = P(F > F_{\mathrm{obs}})$.

For the likelihood analysis, we adopt a binning of $\Delta k = 0.1~h~\mathrm{Mpc}^{-1}$, which yields $N_\mathrm{d}=7$ $k$-bins for the monopole power spectra. This choice is validated by verifying the positive-definiteness of the covariance matrix, as well as the condition number and the PTE value.

Figure~\ref{fig:plot_chi2_interpolation} displays the parabolic fits to $\chi_\mathrm{SH}^2({\cal M})$, the best-fit value of ${\cal M}$, and its uncertainty.
Figures~\ref{fig:plot_P0_chi2_distribution_best-fit_Mcut} and 
\ref{fig:plot_P2_chi2_distribution_best-fit_Mcut} show the distribution of $\chi^2$ for the monopole and quadrupole power spectra, respectively. The mock $\chi^2$ values are computed via a leave-one-out approach, where each mock serves as pseudo-data and the covariance matrix is estimated from the remaining 49 mocks. 
The theoretical distribution of $\chi^2$ under the S\&H framework, indicated by dashed lines, is a scaled $F$ distribution as described in Equation~\eqref{eq:F_dist}, with $N_\mathrm{d} = 7$ and $N_\mathrm{s} = 50$. 
We verify that the empirical distribution of $\chi^2$ from the mock realizations is consistent with the theoretical expectation, and assess whether the data $\chi^2$ lies within the mock distribution.  
We find excellent PTE values in all cases and their $\chi^2$ values remain well within the mock distribution.

\subsection{Results}
\label{subsec:results}
\begin{figure*}
    \centering 
    \includegraphics[width=0.69\columnwidth]{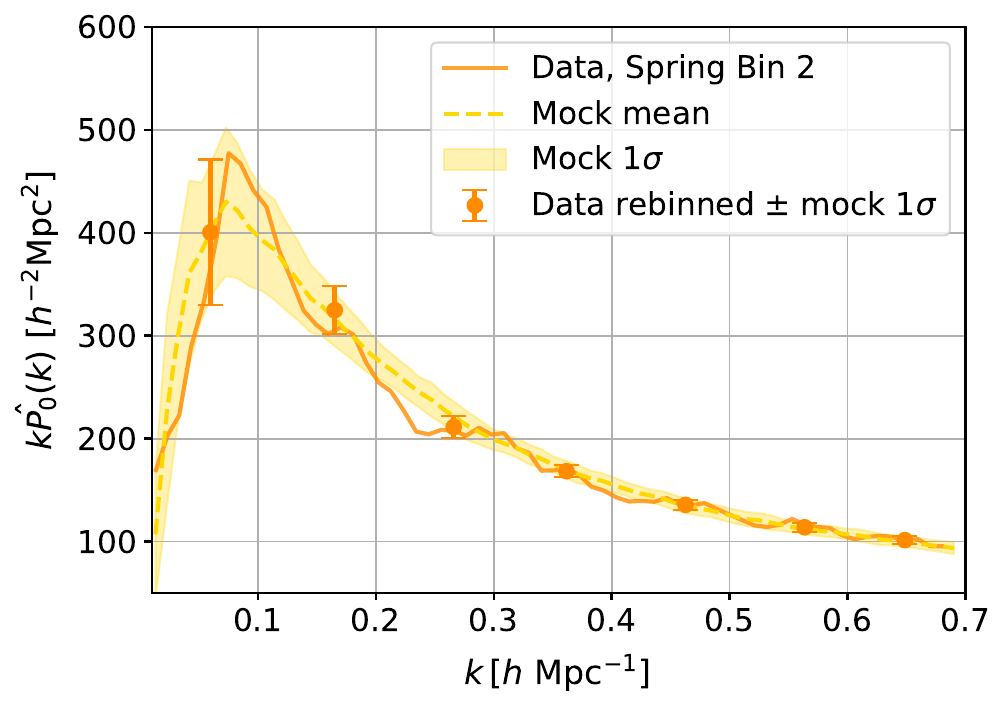}
    \includegraphics[width=0.69\columnwidth]{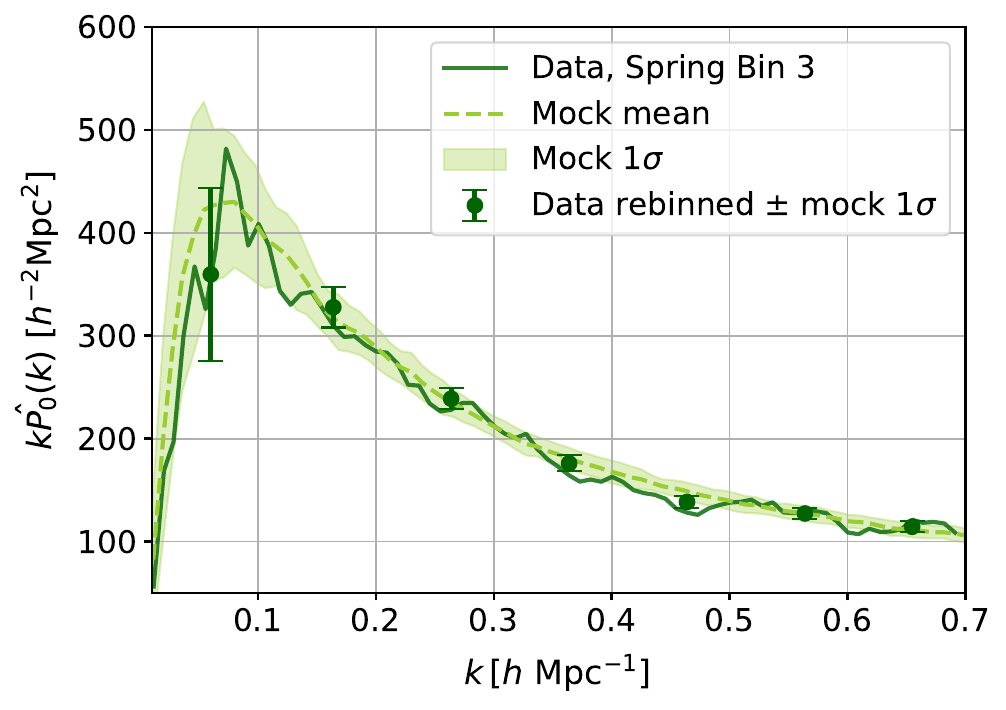}
    \includegraphics[width=0.69\columnwidth]{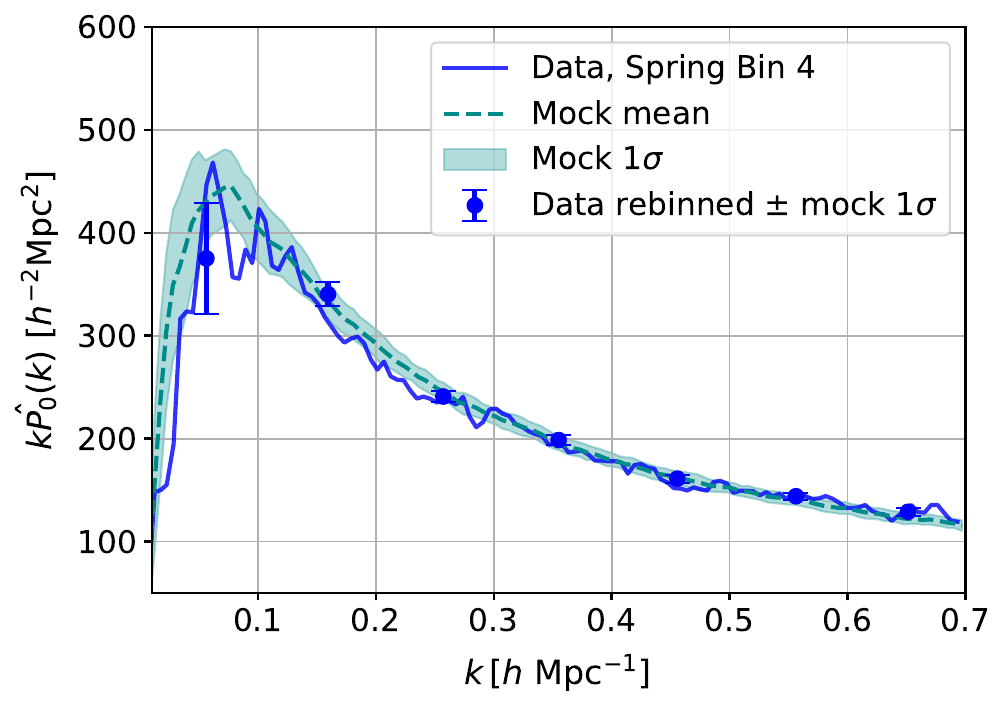}
    \includegraphics[width=0.69\columnwidth]{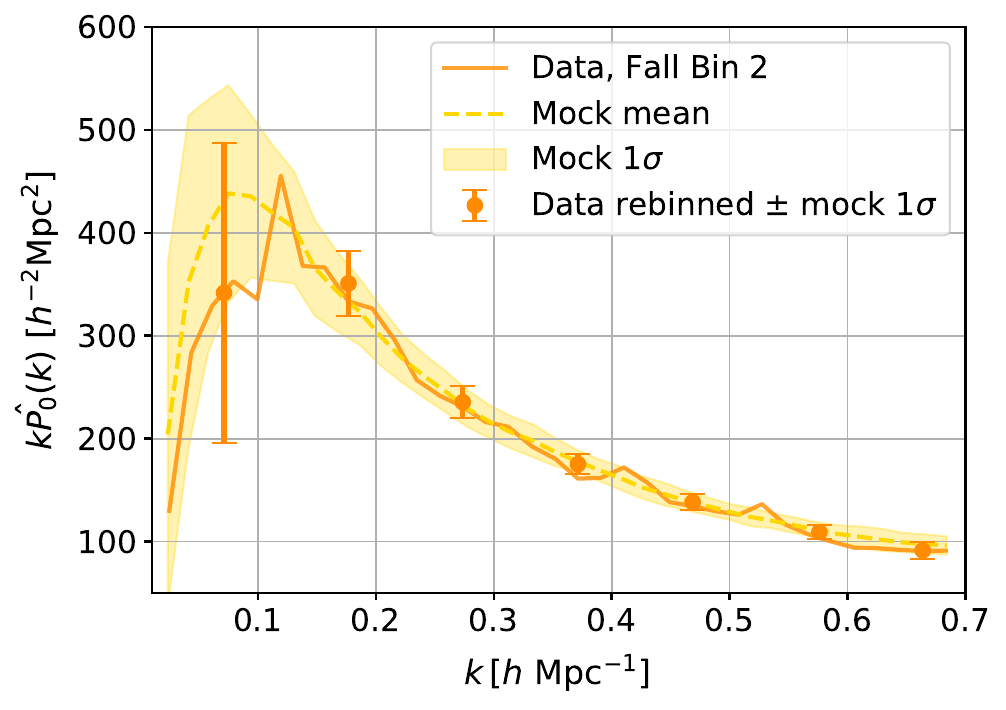}
    \includegraphics[width=0.69\columnwidth]{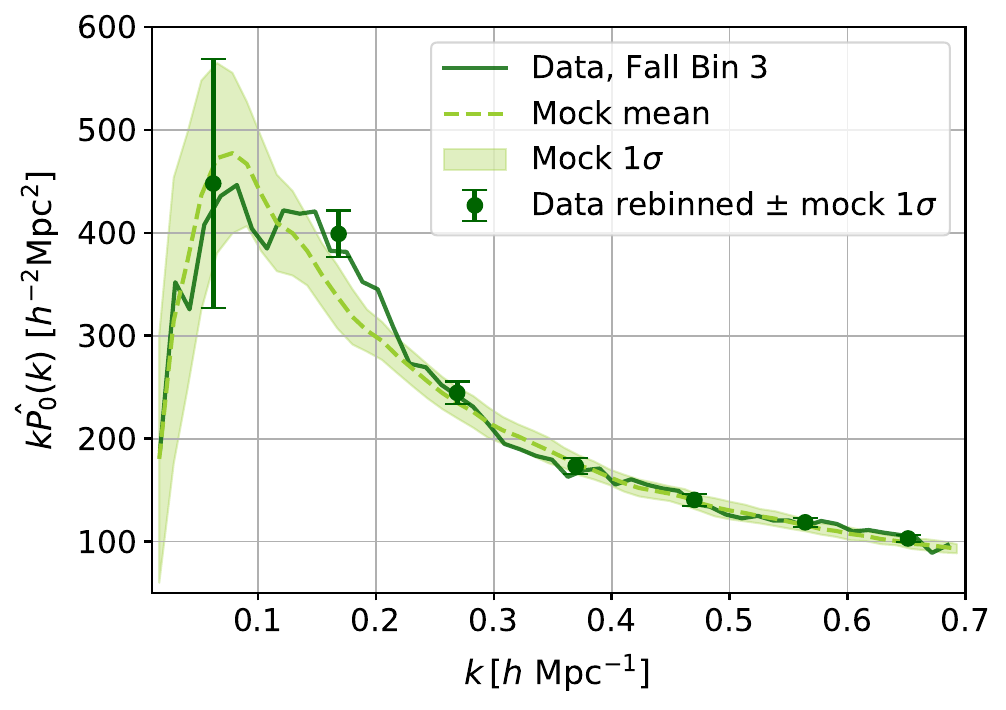}
    \includegraphics[width=0.69\columnwidth]{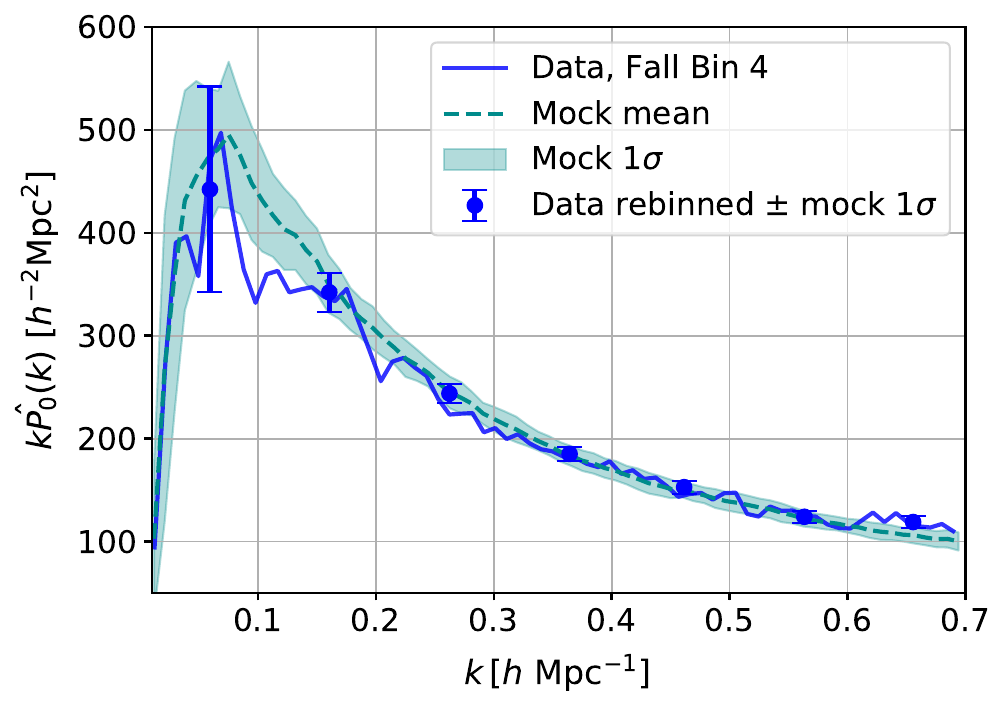}
    \caption{Monopole power spectra of the HETDEX [\ion{O}{2}] volume-limited samples in Bins 2 (left panels), 3 (middle panels), and 4 (right panels) in the Spring (top panels) and Fall (bottom panels) fields, compared with the best-fit Uchuu mock. The solid lines represent the HETDEX data with finer bins ($\Delta k\simeq0.01~h~\mathrm{Mpc}^{-1}$), while the points with the error bars show the rebinned data with $\Delta k=0.1~h~\mathrm{Mpc}^{-1}$. The error bars show $1\sigma$ uncertainties estimated from 50 mock realizations. The dashed lines show the mean of the best-fit mock, and the shaded areas indicate its $1\sigma$ uncertainties in the finer bins.
    The data and mock are in excellent agreement at all wavenumbers up to $k_\mathrm{max}= 0.7\,h\,\mathrm{Mpc}^{-1}$, which is in the non-linear regime.
    }
    \label{fig:plot_p0_Mcut_best-fit}
\end{figure*}

\begin{figure*}
    \centering 
    \includegraphics[width=0.69\columnwidth]{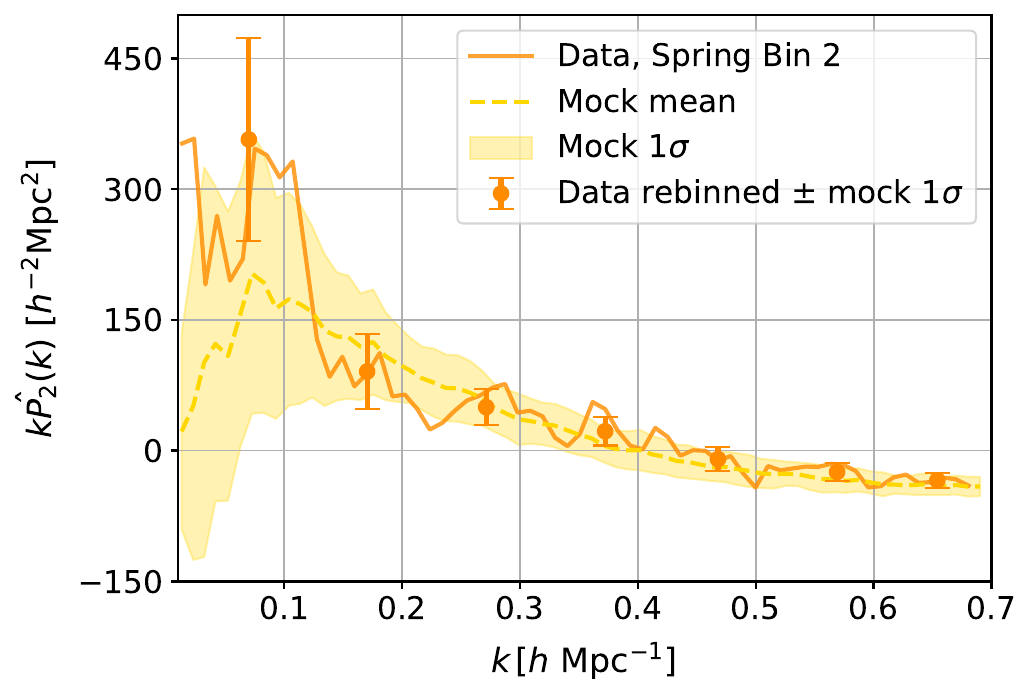}
    \includegraphics[width=0.69\columnwidth]{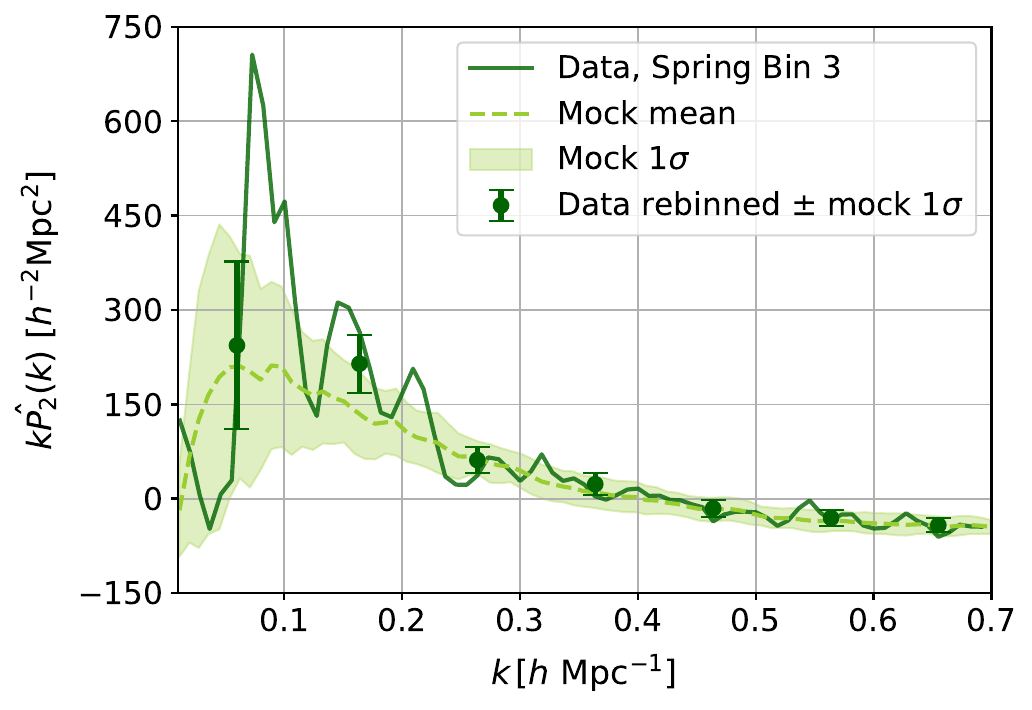}
    \includegraphics[width=0.69\columnwidth]{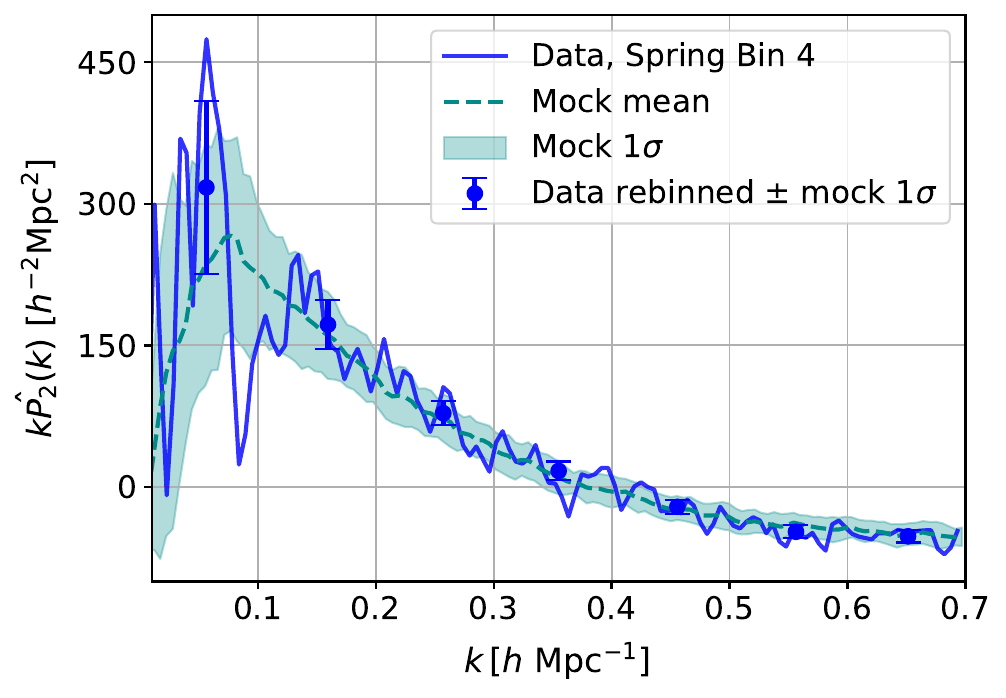}
    \includegraphics[width=0.69\columnwidth]{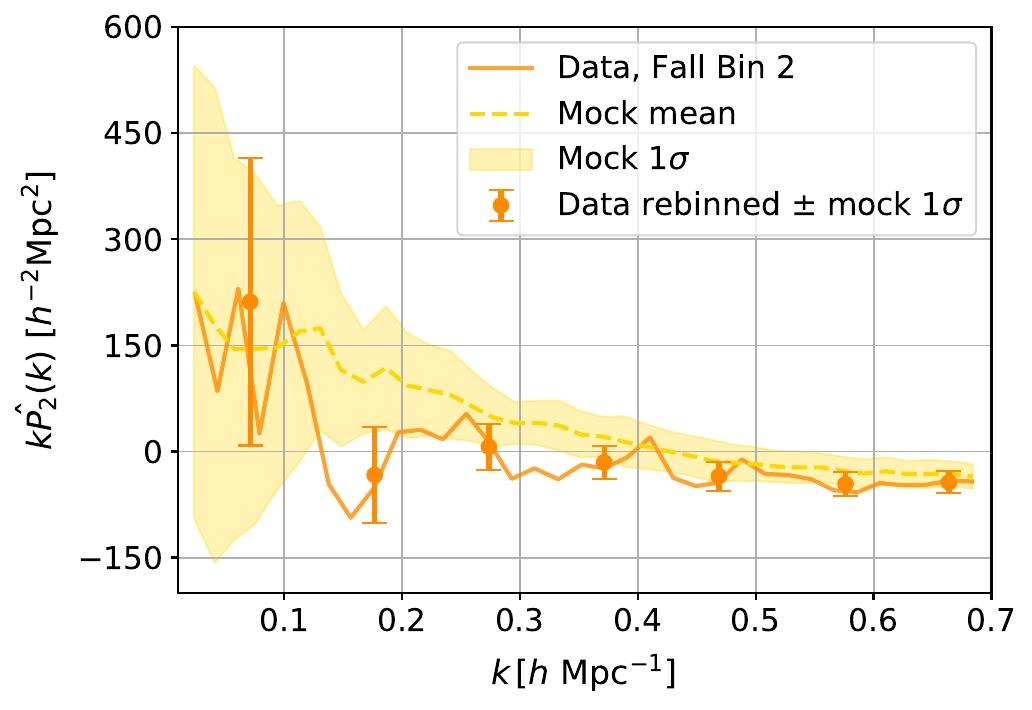}
    \includegraphics[width=0.69\columnwidth]{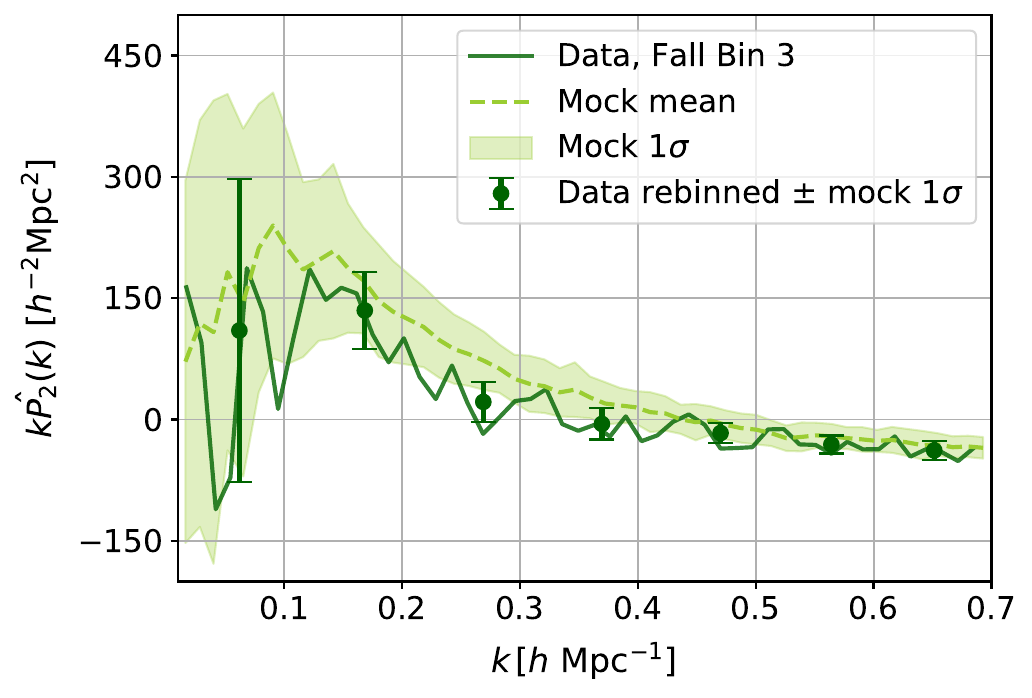}
    \includegraphics[width=0.69\columnwidth]{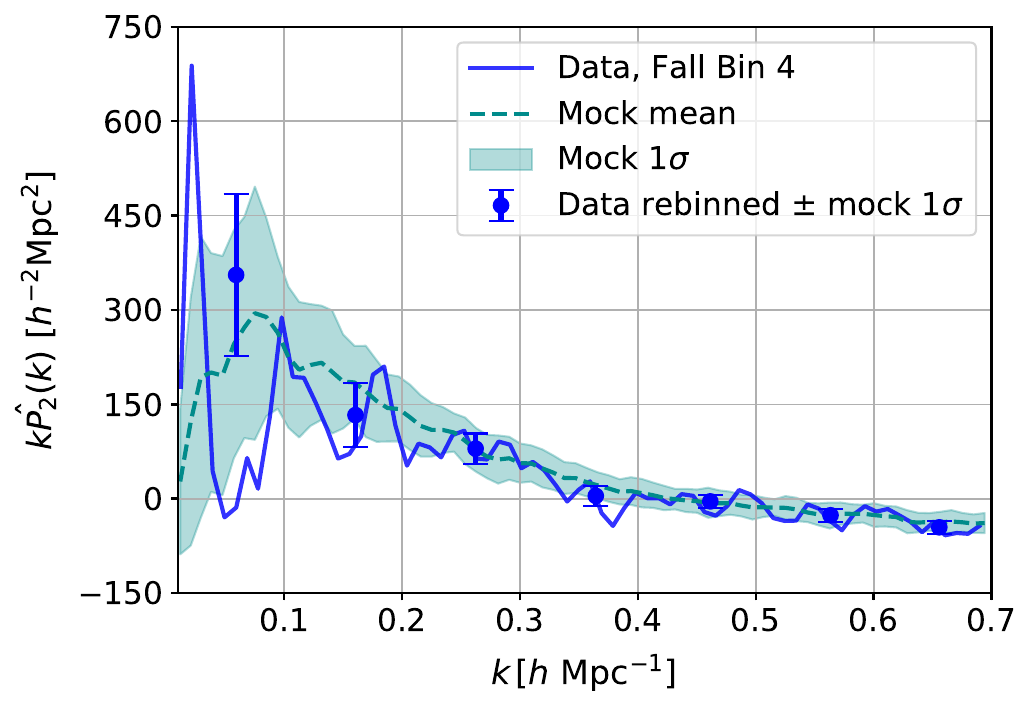}
    \caption{Same format as Figure~\ref{fig:plot_p0_Mcut_best-fit} but for the quadrupole power spectra. The data and mock remain in excellent agreement across all bins and fields, consistent with the monopole results.}
    \label{fig:plot_p2_Mcut_best-fit}
\end{figure*}

Figures~\ref{fig:plot_p0_Mcut_best-fit} and \ref{fig:plot_p2_Mcut_best-fit} compare the monopole and quadrupole power spectra of the HETDEX [\ion{O}{2}] volume-limited samples in the Spring and Fall fields to those of the best-fit mock.
The data and mock are in excellent agreement at all wavenumbers up to $k_\mathrm{max}= 0.7\,h\,\mathrm{Mpc}^{-1}$, which is in the non-linear regime. Note that the quadrupole power spectra are not used in the likelihood analysis, so the mock quadrupole power spectra are predictions rather than fits to the data. Nevertheless, they remain in great agreement with the data. 

It is remarkable and encouraging that such a simple HOD provides an excellent description of the data at all wavenumbers, making the results straightforward to interpret in terms of the characteristic host halo mass.
The best-fit mocks yield $F_\mathrm{g}= 0.52\pm0.09$, $0.63\pm0.12$, and $0.49\pm0.06$ for Spring Bins 2, 3, and 4, and $F_\mathrm{g}=0.64\pm0.14$, $0.49\pm0.07$, and $0.37\pm0.07$ for Fall Bins 2, 3, and 4, respectively (see Equation~\eqref{eq:HOD}), where the uncertainties are propagated from those in $\log(M_0)$. 

So far, we have fixed $\sigma_{\log M} = 0.6$. Varying $\sigma_{\log M} = 0.6\pm 0.1$, we find that the best-fit values of $F_\mathrm{g}$ change by $^{+0.17\sim0.31}_{-0.13\sim0.21}$ depending on the bins, still within the physically allowed range of $0<F_\mathrm{g}\le 1$. While $F_\mathrm{g}$ affects the number density of galaxies, we check that it does not significantly change the power spectrum amplitude. Therefore, the uncertainties in $F_\mathrm{g}$ arising from varying $\sigma_{\log M}$ do not affect our results on the galaxy clustering.

The mocks also suggest that approximately 13\% of the [\ion{O}{2}] galaxies in the volume-limited samples reside in subhalos, with a typical statistical uncertainty of $\pm0.2\%$ based on 50 realizations.

\begin{figure}
    \centering 
    \includegraphics[width=0.95\columnwidth]{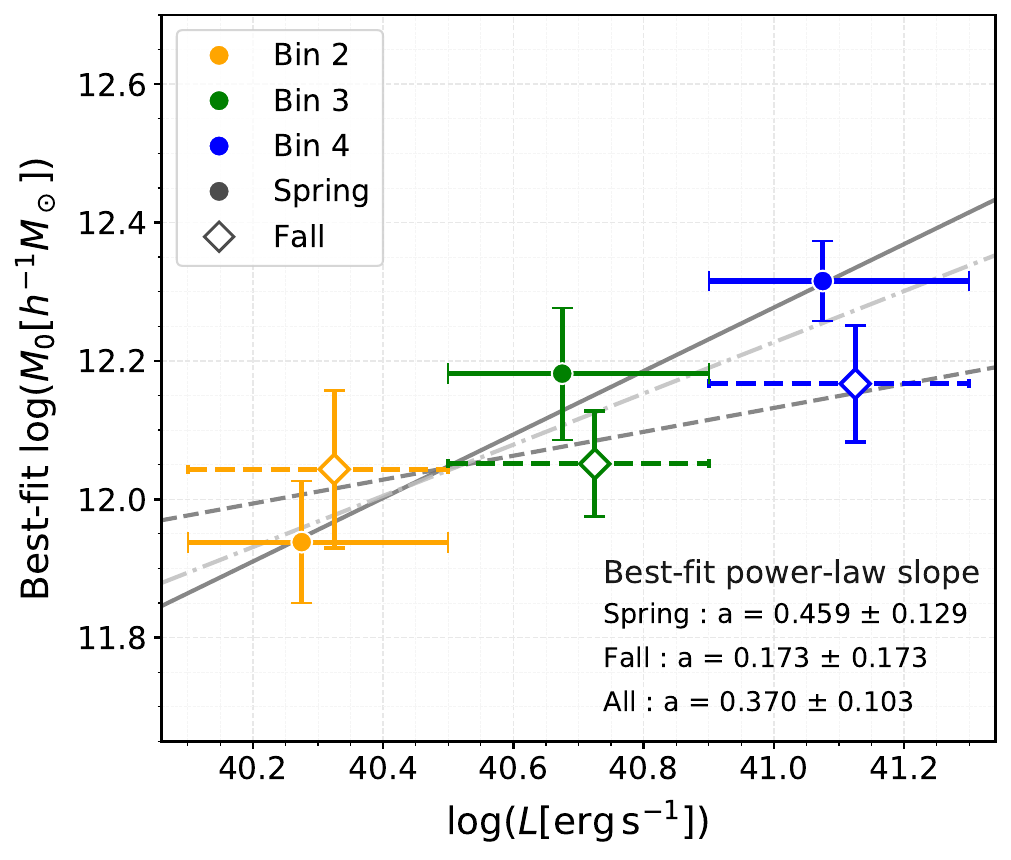}
    \caption{Best-fit values of $\log(M_0)$ as a function of $\log(L)$. The solid (dashed) horizontal bars indicate the luminosity ranges of the Spring (Fall) fields, and the vertical bars show the $1\sigma$ uncertainties. The Spring (Fall) field points are slightly shifted to the left (right) for clarity. The gray solid, dashed, and dash-dotted lines indicate the power-law fits, $M_0\propto L^a$, for the Spring, Fall, and all fields combined, respectively.}
    \label{fig:plot_log_L_vs_log_Mcut}
\end{figure}

Figure~\ref{fig:plot_log_L_vs_log_Mcut} presents the best-fit values of $\log(M_0)$ as a function of $\log(L)$ and their power-law fits, $M_0\propto L^a$. 
We find that the best-fit characteristic halo mass slightly increases with luminosity. Although the Spring field shows a steeper slope than the Fall field, both are statistically consistent with a value of $a=0.37\pm 0.10$. Since the [\ion{O}{2}] luminosity is proportional to the star formation rate~\citep{Kennicutt1998,Kennicutt2012}, our result can be explained by an increase in the star formation rate with stellar mass and a correlation between stellar mass and dark matter halo mass. 
For comparison, our $\log{M_0}$-$\log(L)$ result is in good agreement with previous studies \citep{Favole2017, Khostovan2018}, both of which demonstrate that [\ion{O}{2}] emission line luminosity correlates with galaxy clustering amplitude, bias, and dark matter halo mass.
\citet{GAMA_2021} found a stronger dependence on the $r$-band luminosity of magnitude-selected galaxies in the GAMA survey at $z\lesssim 0.5$. Since extinction affects emission lines more than continuum brightness, the [\ion{O}{2}] emission line dependence will naturally be different from the $r$-band dependence.

\begin{figure}
    \centering 
    \includegraphics[width=0.95\columnwidth]{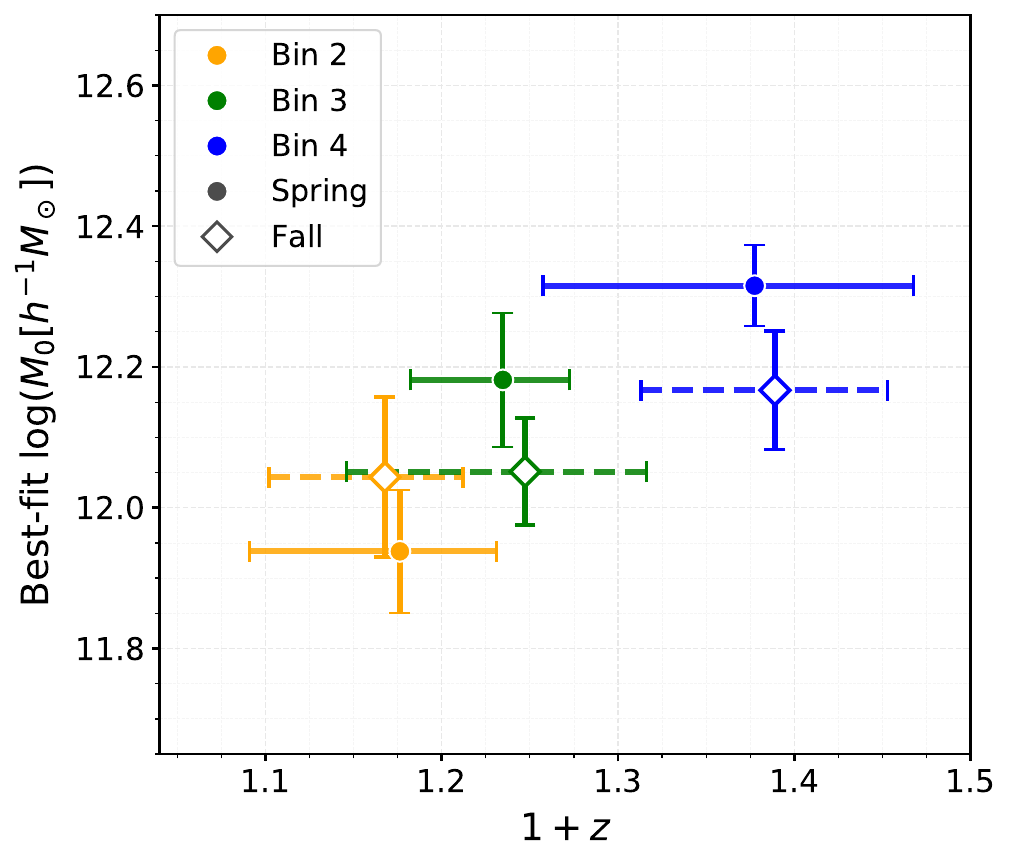}
    \caption{Same format as Figure~\ref{fig:plot_log_L_vs_log_Mcut} but for $\log(M_0)$ as a function of $1+z$.}
    \label{fig:plot_1+z_vs_log_Mcut}
\end{figure}

However, this explanation is not the only interpretation. Since our luminosity bins have different mean redshifts (see Table~\ref{tab:table_oii_sub}), our results could suggest time evolution rather than luminosity evolution.
Figure~\ref{fig:plot_1+z_vs_log_Mcut} shows the best-fit values of $\log(M_0)$ as a function of $1+z$.

These HOD results can be converted into linear galaxy bias parameters using a halo model~\citep[see][for reviews]{halomodel1,halomodel2}. We find $b_1\simeq 0.80$, 0.84, and 0.90 for Spring Bins 2, 3, and 4, and $b_1\simeq 0.81$, 0.83, and 0.88 for Fall Bins 2, 3, and 4, respectively, with a typical statistical uncertainty of $\pm 0.01$ derived from the uncertainty in $\log(M_0)$. Varying $\sigma_{\log M} = 0.6\pm 0.1$, we find that $b_1$ changes by $\pm 0.02$ for Bin 2, which increases to $\pm 0.03$ for Bin 4. These results are consistent with the well-established picture that galaxy bias increases with both halo mass and redshift~\citep{Mo1996, Tinker2010}, and are in line with expectations for low-mass, star-forming galaxies at $z\leq0.48$.

\section{Conclusions}
\label{sec:conclusions}
In this paper, we have presented the first measurement of the power spectrum multipoles of emission-line-selected [\ion{O}{2}] galaxies from the HETDEX Public Data Release 1~\citep[PDR1;][]{pdr1} in the low-redshift universe, $z \le 0.48$. The untargeted nature of the HETDEX survey, combined with its unprecedented spectroscopic depth, has enabled the construction of well-characterized volume-limited samples of [\ion{O}{2}] emitting galaxies. These samples provide a unique, independent probe of the growth of large-scale structures complementary to other contemporary cosmological surveys~\citep{Blake_2011, Takada_2014, Alam_2017, Alam_2021, DESIDR2_2025}.

We constructed volume-limited samples in three luminosity bins 
from the Spring and Fall fields of the HETDEX PDR1. 
Using two-sample KS tests to validate the volume-limited nature of each sample, we confirmed that the distributions of galaxy number densities are consistent with being roughly constant across their respective redshift ranges.

We measured the monopole and quadrupole power spectra of these volume-limited [\ion{O}{2}] samples using an FFT-based Yamamoto estimator~\citep{Yamamoto_2006,Hand_2017}, and modeled them using mock catalogs constructed from the Uchuu $N$-body simulation~\citep{Uchuu_dr1} with a simple lognormal HOD model. 
By fixing $\sigma_{\log M} = 0.6$ and the galaxy fraction, $F_\mathrm{g}$, to match the observed number density of each sample, we reduced the HOD to have a single free parameter: the characteristic halo mass $M_0$. 
The best-fit mocks provide an excellent description of both the monopole and quadrupole power spectra across all wavenumbers up to $k_\mathrm{max} = 0.7~h~\mathrm{Mpc}^{-1}$, demonstrating that this simple HOD model is sufficient to describe the clustering properties of the HETDEX [\ion{O}{2}] volume-limited samples without additional complexity. We find that approximately 13 percent of the [\ion{O}{2}] galaxies in the best-fit mock reside in subhalos.

The main results of this paper are summarized as follows:
\begin{itemize}
\item The best-fit characteristic halo mass, $\log(M_0)$, slightly increases with [\ion{O}{2}] luminosity following a power-law relation, $M_0 \propto L^{a}$, with a slope of $a = 0.37 \pm 0.10$, consistent between the Spring and Fall fields. This trend is expected, as [\ion{O}{2}] luminosity is a proxy for the star formation rate, which correlates with stellar mass, and stellar mass in turn correlates with dark matter halo mass~\citep{Vale2004, Behroozi2010,GAMA_2021}.

\item The best-fit $\log(M_0)$ values also show a slight increase with redshift, consistent with a scenario in which higher-luminosity, higher-redshift galaxies tend to reside in more massive dark matter halos~\citep{Mo1996, Tinker2010}. However, since luminosity and redshift are correlated in volume-limited samples by definition, the two effects --- luminosity and time evolution --- cannot be disentangled from the current analysis alone.

\item The inferred linear galaxy bias parameters increase from $b_1 \simeq 0.8$ to $0.9$ from Bin 2 to Bin 4 in both the Spring and Fall fields. These values are consistent with the well-established picture that galaxy bias increases with both halo mass and redshift~\citep{Mo1996, Tinker2010}, and are in line with expectations for low-mass, star-forming galaxies at $z \leq 0.48$.

\end{itemize}

The HETDEX [\ion{O}{2}] volume-limited samples offer several unique advantages over other spectroscopic galaxy surveys using [\ion{O}{2}] emitting galaxies at higher redshifts \citep{Takada_2014,Alam_2021, DESIDR2_2025}. 
In particular, the emission-line-selected and untargeted nature of the HETDEX survey eliminates the need to model complex photometric pre-selection effects, resulting in a cleaner and more straightforward interpretation of the galaxy--halo connection. 
The high number densities, $\bar{n} \simeq (2\text{--}5) \times 10^{-3}~h^3~\mathrm{Mpc}^{-3}$, which are five to ten times higher than those of typical cosmological spectroscopic surveys of emission-line galaxies~\citep{Blake_2011,Takada_2014,eBOSS_2021,Wang_2022,DESIDR2_2025,euclid_2025}, make these samples not only well-suited for cosmological applications, but also valuable for astrophysical studies of galaxy formation and evolution, as well as for cross-correlation analyses with other large-scale structure tracers in the low-redshift universe.

These results represent the first step in a broader program of cosmological analyses using the HETDEX [\ion{O}{2}] galaxy catalog. In future work, we will extend this analysis to constrain the growth of structure using the quadrupole and higher-order multipoles of the power spectrum, as well as to study the galaxy--halo connection.

These data will also be useful for void cosmology analyses~\citep{Contarini2026}.
The combination of these measurements with complementary probes --- such as weak gravitational lensing~\citep{HSCY3_shear, KIDSLegacy_shear,DESY6_shear}, the CMB~\citep{Planck2020cosmo, ACT2025}, and other spectroscopic surveys in a low-redshift universe~\citep{DESIBGS_2023, DESIPV_2025} --- will provide powerful independent constraints on the matter density parameter, $\Omega_\mathrm{m}$, and the $S_8$ parameter, helping to shed light on the current cosmological tensions between early- and late-universe probes~\citep[see][for a review]{intertwined_2022}.

\section*{Acknowledgements}

HETDEX is led by the University of Texas at Austin McDonald Observatory and Department of Astronomy with participation from the Ludwig-Maximilians-Universität München, Max-Planck-Institut für Extraterrestrische Physik (MPE), Leibniz-Institut für Astrophysik Potsdam (AIP), Texas A\&M University, Pennsylvania State University, Institut für Astrophysik Göttingen, The University of Oxford, Max-Planck-Institut für Astrophysik (MPA), The University of Tokyo and Missouri University of Science and Technology.

Observations for HETDEX were obtained with the Hobby-Eberly Telescope (HET), which is a joint project of the University of Texas at Austin, the Pennsylvania State University, Ludwig-Maximilians-Universität München, and Georg-August-Universität Göttingen. The HET is named in honor of its principal benefactors, William P. Hobby and Robert E. Eberly. The Visible Integral-field Replicable Unit Spectrograph (VIRUS) was used for HETDEX observations. VIRUS is a joint project of the University of Texas at Austin, Leibniz-Institut für Astrophysik Potsdam (AIP), Texas A\&M University, Max-Planck-Institut fürExtraterrestrische Physik (MPE), Ludwig-Maximilians-Universität München, Pennsylvania State University, Institut für Astrophysik Göttingen, University of Oxford, and the Max-Planck-Institut fur Astrophysik (MPA).

The authors acknowledge the Texas Advanced Computing Center (TACC) at The University of Texas at Austin for providing high performance computing, visualization, and storage resources that have contributed to the research results reported within this paper. URL: http://www.tacc.utexas.edu

Funding for HETDEX has been provided by the partner institutions, the National Science Foundation, the State of Texas, the US Air Force, and by generous support from private individuals and foundations.

This work was also supported in part by the Deutsche Forschungsgemeinschaft (DFG, German Research Foundation) under Germany's Excellence Strategy – EXC-2094/2 – 390783311.

The Institute for Gravitation and the Cosmos is supported by the Eberly College of Science and the Office of the Senior Vice President for Research at the Pennsylvania State University.

S.S. acknowledges support from the National Science Foundation under grants NSF-2219212 and NSF-2511145.

D.J. was supported by NSF grants AST-2307026 and AST-2407298 at PSU, as well as by a KIAS Individual Grant, PG088301.

We thank Instituto de Astrofisica de Andalucia (IAA-CSIC), Centro de Supercomputacion de Galicia (CESGA) and the Spanish academic and research network (RedIRIS) in Spain for hosting Uchuu DR1, DR2 and DR3 in the Skies $\&$ Universes site for cosmological simulations. The Uchuu simulations were carried out on Aterui II supercomputer at Center for Computational Astrophysics, CfCA, of National Astronomical Observatory of Japan, and the K computer at the RIKEN Advanced Institute for Computational Science. The Uchuu Data Releases efforts have made use of the skun@IAA\_RedIRIS and skun6@IAA computer facilities managed by the IAA-CSIC in Spain (MICINN EU-Feder grant EQC2018-004366-P).

\bibliography{sample701}{}

@ARTICLE{GAMADR1,
       author = {{Driver}, S.~P. and {Hill}, D.~T. and {Kelvin}, L.~S. and {Robotham}, A.~S.~G. and {Liske}, J. and {Norberg}, P. and {Baldry}, I.~K. and {Bamford}, S.~P. and {Hopkins}, A.~M. and {Loveday}, J. and {Peacock}, J.~A. and {Andrae}, E. and {Bland-Hawthorn}, J. and {Brough}, S. and {Brown}, M.~J.~I. and {Cameron}, E. and {Ching}, J.~H.~Y. and {Colless}, M. and {Conselice}, C.~J. and {Croom}, S.~M. and {Cross}, N.~J.~G. and {de Propris}, R. and {Dye}, S. and {Drinkwater}, M.~J. and {Ellis}, S. and {Graham}, Alister W. and {Grootes}, M.~W. and {Gunawardhana}, M. and {Jones}, D.~H. and {van Kampen}, E. and {Maraston}, C. and {Nichol}, R.~C. and {Parkinson}, H.~R. and {Phillipps}, S. and {Pimbblet}, K. and {Popescu}, C.~C. and {Prescott}, M. and {Roseboom}, I.~G. and {Sadler}, E.~M. and {Sansom}, A.~E. and {Sharp}, R.~G. and {Smith}, D.~J.~B. and {Taylor}, E. and {Thomas}, D. and {Tuffs}, R.~J. and {Wijesinghe}, D. and {Dunne}, L. and {Frenk}, C.~S. and {Jarvis}, M.~J. and {Madore}, B.~F. and {Meyer}, M.~J. and {Seibert}, M. and {Staveley-Smith}, L. and {Sutherland}, W.~J. and {Warren}, S.~J.},
        title = "{Galaxy and Mass Assembly (GAMA): survey diagnostics and core data release}",
      journal = {\mnras},
         year = 2011,
        month = may,
       volume = {413},
       number = {2},
        pages = {971-995},
          doi = {10.1111/j.1365-2966.2010.18188.x},
archivePrefix = {arXiv},
       eprint = {1009.0614},
 primaryClass = {astro-ph.CO},
       adsurl = {https://ui.adsabs.harvard.edu/abs/2011MNRAS.413..971D}
}

@ARTICLE{GAMADR2,
       author = {{Liske}, J. and {Baldry}, I.~K. and {Driver}, S.~P. and {Tuffs}, R.~J. and {Alpaslan}, M. and {Andrae}, E. and {Brough}, S. and {Cluver}, M.~E. and {Grootes}, M.~W. and {Gunawardhana}, M.~L.~P. and {Kelvin}, L.~S. and {Loveday}, J. and {Robotham}, A.~S.~G. and {Taylor}, E.~N. and {Bamford}, S.~P. and {Bland-Hawthorn}, J. and {Brown}, M.~J.~I. and {Drinkwater}, M.~J. and {Hopkins}, A.~M. and {Meyer}, M.~J. and {Norberg}, P. and {Peacock}, J.~A. and {Agius}, N.~K. and {Andrews}, S.~K. and {Bauer}, A.~E. and {Ching}, J.~H.~Y. and {Colless}, M. and {Conselice}, C.~J. and {Croom}, S.~M. and {Davies}, L.~J.~M. and {De Propris}, R. and {Dunne}, L. and {Eardley}, E.~M. and {Ellis}, S. and {Foster}, C. and {Frenk}, C.~S. and {H{\"a}u{\ss}ler}, B. and {Holwerda}, B.~W. and {Howlett}, C. and {Ibarra}, H. and {Jarvis}, M.~J. and {Jones}, D.~H. and {Kafle}, P.~R. and {Lacey}, C.~G. and {Lange}, R. and {Lara-L{\'o}pez}, M.~A. and {L{\'o}pez-S{\'a}nchez}, {\'A}. R. and {Maddox}, S. and {Madore}, B.~F. and {McNaught-Roberts}, T. and {Moffett}, A.~J. and {Nichol}, R.~C. and {Owers}, M.~S. and {Palamara}, D. and {Penny}, S.~J. and {Phillipps}, S. and {Pimbblet}, K.~A. and {Popescu}, C.~C. and {Prescott}, M. and {Proctor}, R. and {Sadler}, E.~M. and {Sansom}, A.~E. and {Seibert}, M. and {Sharp}, R. and {Sutherland}, W. and {V{\'a}zquez-Mata}, J.~A. and {van Kampen}, E. and {Wilkins}, S.~M. and {Williams}, R. and {Wright}, A.~H.},
        title = "{Galaxy And Mass Assembly (GAMA): end of survey report and data release 2}",
      journal = {\mnras},
         year = 2015,
        month = sep,
       volume = {452},
       number = {2},
        pages = {2087-2126},
          doi = {10.1093/mnras/stv1436},
archivePrefix = {arXiv},
       eprint = {1506.08222},
 primaryClass = {astro-ph.GA},
       adsurl = {https://ui.adsabs.harvard.edu/abs/2015MNRAS.452.2087L}
}

@ARTICLE{GAMA_2021,
       author = {{Alam}, Shadab and {Peacock}, John A. and {Farrow}, Daniel J. and {Loveday}, J. and {Hopkins}, A.~M.},
        title = "{Using GAMA to probe the impact of small-scale galaxy physics on nonlinear redshift-space distortions}",
      journal = {\mnras},
         year = 2021,
        month = may,
       volume = {503},
       number = {1},
        pages = {59-76},
          doi = {10.1093/mnras/stab409},
archivePrefix = {arXiv},
       eprint = {2006.05383},
 primaryClass = {astro-ph.CO},
       adsurl = {https://ui.adsabs.harvard.edu/abs/2021MNRAS.503...59A}
}

@ARTICLE{Farrow_2015,
       author = {{Farrow}, D.~J. and {Cole}, Shaun and {Norberg}, Peder and {Metcalfe}, N. and {Baldry}, I. and {Bland-Hawthorn}, Joss and {Brown}, Michael J.~I. and {Hopkins}, A.~M. and {Lacey}, Cedric G. and {Liske}, J. and {Loveday}, Jon and {Palamara}, David P. and {Robotham}, A.~S.~G. and {Sridhar}, Srivatsan},
        title = "{Galaxy and mass assembly (GAMA): projected galaxy clustering}",
      journal = {\mnras},
         year = 2015,
        month = dec,
       volume = {454},
       number = {2},
        pages = {2120-2145},
          doi = {10.1093/mnras/stv2075},
archivePrefix = {arXiv},
       eprint = {1509.02159},
 primaryClass = {astro-ph.GA},
       adsurl = {https://ui.adsabs.harvard.edu/abs/2015MNRAS.454.2120F}
}

@ARTICLE{intertwined_2022,
       author = {{Abdalla}, Elcio and {Abell{\'a}n}, Guillermo Franco and {Aboubrahim}, Amin and {Agnello}, Adriano and {Akarsu}, {\"O}zg{\"u}r and {Akrami}, Yashar and {Alestas}, George and {Aloni}, Daniel and {Amendola}, Luca and {Anchordoqui}, Luis A. and {Anderson}, Richard I. and {Arendse}, Nikki and {Asgari}, Marika and {Ballardini}, Mario and {Barger}, Vernon and {Basilakos}, Spyros and {Batista}, Ronaldo C. and {Battistelli}, Elia S. and {Battye}, Richard and {Benetti}, Micol and {Benisty}, David and {Berlin}, Asher and {de Bernardis}, Paolo and {Berti}, Emanuele and {Bidenko}, Bohdan and {Birrer}, Simon and {Blakeslee}, John P. and {Boddy}, Kimberly K. and {Bom}, Clecio R. and {Bonilla}, Alexander and {Borghi}, Nicola and {Bouchet}, Fran{\c{c}}ois R. and {Braglia}, Matteo and {Buchert}, Thomas and {Buckley-Geer}, Elizabeth and {Calabrese}, Erminia and {Caldwell}, Robert R. and {Camarena}, David and {Capozziello}, Salvatore and {Casertano}, Stefano and {Chen}, Geoff C.-F. and {Chluba}, Jens and {Chen}, Angela and {Chen}, Hsin-Yu and {Chudaykin}, Anton and {Cicoli}, Michele and {Copi}, Craig J. and {Courbin}, Fred and {Cyr-Racine}, Francis-Yan and {Czerny}, Bo{\.z}ena and {Dainotti}, Maria and {D'Amico}, Guido and {Davis}, Anne-Christine and {de Cruz P{\'e}rez}, Javier and {de Haro}, Jaume and {Delabrouille}, Jacques and {Denton}, Peter B. and {Dhawan}, Suhail and {Dienes}, Keith R. and {Di Valentino}, Eleonora and {Du}, Pu and {Eckert}, Dominique and {Escamilla-Rivera}, Celia and {Fert{\'e}}, Agn{\`e}s and {Finelli}, Fabio and {Fosalba}, Pablo and {Freedman}, Wendy L. and {Frusciante}, Noemi and {Gazta{\~n}aga}, Enrique and {Giar{\`e}}, William and {Giusarma}, Elena and {G{\'o}mez-Valent}, Adri{\`a} and {Handley}, Will and {Harrison}, Ian and {Hart}, Luke and {Hazra}, Dhiraj Kumar and {Heavens}, Alan and {Heinesen}, Asta and {Hildebrandt}, Hendrik and {Hill}, J. Colin and {Hogg}, Natalie B. and {Holz}, Daniel E. and {Hooper}, Deanna C. and {Hosseininejad}, Nikoo and {Huterer}, Dragan and {Ishak}, Mustapha and {Ivanov}, Mikhail M. and {Jaffe}, Andrew H. and {Jang}, In Sung and {Jedamzik}, Karsten and {Jimenez}, Raul and {Joseph}, Melissa and {Joudaki}, Shahab and {Kamionkowski}, Marc and {Karwal}, Tanvi and {Kazantzidis}, Lavrentios and {Keeley}, Ryan E. and {Klasen}, Michael and {Komatsu}, Eiichiro and {Koopmans}, L{\'e}on V.~E. and {Kumar}, Suresh and {Lamagna}, Luca and {Lazkoz}, Ruth and {Lee}, Chung-Chi and {Lesgourgues}, Julien and {Levi Said}, Jackson and {Lewis}, Tiffany R. and {L'Huillier}, Benjamin and {Lucca}, Matteo and {Maartens}, Roy and {Macri}, Lucas M. and {Marfatia}, Danny and {Marra}, Valerio and {Martins}, Carlos J.~A.~P. and {Masi}, Silvia and {Matarrese}, Sabino and {Mazumdar}, Arindam and {Melchiorri}, Alessandro and {Mena}, Olga and {Mersini-Houghton}, Laura and {Mertens}, James and {Milakovi{\'c}}, Dinko and {Minami}, Yuto and {Miranda}, Vivian and {Moreno-Pulido}, Cristian and {Moresco}, Michele and {Mota}, David F. and {Mottola}, Emil and {Mozzon}, Simone and {Muir}, Jessica and {Mukherjee}, Ankan and {Mukherjee}, Suvodip and {Naselsky}, Pavel and {Nath}, Pran and {Nesseris}, Savvas and {Niedermann}, Florian and {Notari}, Alessio and {Nunes}, Rafael C. and {{\'O} Colg{\'a}in}, Eoin and {Owens}, Kayla A. and {{\"O}z{\"u}lker}, Emre and {Pace}, Francesco and {Paliathanasis}, Andronikos and {Palmese}, Antonella and {Pan}, Supriya and {Paoletti}, Daniela and {Perez Bergliaffa}, Santiago E. and {Perivolaropoulos}, Leandros and {Pesce}, Dominic W. and {Pettorino}, Valeria and {Philcox}, Oliver H.~E. and {Pogosian}, Levon and {Poulin}, Vivian and {Poulot}, Gaspard and {Raveri}, Marco and {Reid}, Mark J. and {Renzi}, Fabrizio and {Riess}, Adam G. and {Sabla}, Vivian I. and {Salucci}, Paolo and {Salzano}, Vincenzo and {Saridakis}, Emmanuel N. and {Sathyaprakash}, Bangalore S. and {Schmaltz}, Martin and {Sch{\"o}neberg}, Nils and {Scolnic}, Dan and {Sen}, Anjan A. and {Sehgal}, Neelima and {Shafieloo}, Arman and {Sheikh-Jabbari}, M.~M. and {Silk}, Joseph and {Silvestri}, Alessandra and {Skara}, Foteini and {Sloth}, Martin S. and {Soares-Santos}, Marcelle and {Sol{\`a} Peracaula}, Joan and {Songsheng}, Yu-Yang and {Soriano}, Jorge F. and {Staicova}, Denitsa and {Starkman}, Glenn D. and {Szapudi}, Istv{\'a}n and {Teixeira}, Elsa M. and {Thomas}, Brooks and {Treu}, Tommaso and {Trott}, Emery and {van de Bruck}, Carsten and {Vazquez}, J. Alberto and {Verde}, Licia and {Visinelli}, Luca and {Wang}, Deng and {Wang}, Jian-Min and {Wang}, Shao-Jiang and {Watkins}, Richard and {Watson}, Scott and {Webb}, John K. and {Weiner}, Neal and {Weltman}, Amanda and {Witte}, Samuel J. and {Wojtak}, Rados{\l}aw and {Yadav}, Anil Kumar},
        title = "{Cosmology intertwined: A review of the particle physics, astrophysics, and cosmology associated with the cosmological tensions and anomalies}",
      journal = {Journal of High Energy Astrophysics},
         year = 2022,
        month = jun,
       volume = {34},
        pages = {49-211},
          doi = {10.1016/j.jheap.2022.04.002},
archivePrefix = {arXiv},
       eprint = {2203.06142},
 primaryClass = {astro-ph.CO},
       adsurl = {https://ui.adsabs.harvard.edu/abs/2022JHEAp..34...49A}
}

@ARTICLE{halomodel1,
       author = {{Cooray}, Asantha and {Sheth}, Ravi},
        title = "{Halo models of large scale structure}",
      journal = {\physrep},
         year = 2002,
        month = dec,
       volume = {372},
       number = {1},
        pages = {1-129},
          doi = {10.1016/S0370-1573(02)00276-4},
archivePrefix = {arXiv},
       eprint = {astro-ph/0206508},
 primaryClass = {astro-ph},
       adsurl = {https://ui.adsabs.harvard.edu/abs/2002PhR...372....1C}
}

@ARTICLE{halomodel2,
       author = {{Asgari}, Marika and {Mead}, Alexander J. and {Heymans}, Catherine},
        title = "{The halo model for cosmology: a pedagogical review}",
      journal = {The Open Journal of Astrophysics},
         year = 2023,
        month = nov,
       volume = {6},
          eid = {39},
        pages = {39},
          doi = {10.21105/astro.2303.08752},
archivePrefix = {arXiv},
       eprint = {2303.08752},
 primaryClass = {astro-ph.CO},
       adsurl = {https://ui.adsabs.harvard.edu/abs/2023OJAp....6E..39A}
}

@ARTICLE{planck2015cosmo,
       author = {{Planck Collaboration} and {Ade}, P.~A.~R. and {Aghanim}, N. and {Arnaud}, M. and {Ashdown}, M. and {Aumont}, J. and {Baccigalupi}, C. and {Banday}, A.~J. and {Barreiro}, R.~B. and {Bartlett}, J.~G. and {Bartolo}, N. and {Battaner}, E. and {Battye}, R. and {Benabed}, K. and {Beno{\^\i}t}, A. and {Benoit-L{\'e}vy}, A. and {Bernard}, J.-P. and {Bersanelli}, M. and {Bielewicz}, P. and {Bock}, J.~J. and {Bonaldi}, A. and {Bonavera}, L. and {Bond}, J.~R. and {Borrill}, J. and {Bouchet}, F.~R. and {Boulanger}, F. and {Bucher}, M. and {Burigana}, C. and {Butler}, R.~C. and {Calabrese}, E. and {Cardoso}, J.-F. and {Catalano}, A. and {Challinor}, A. and {Chamballu}, A. and {Chary}, R.-R. and {Chiang}, H.~C. and {Chluba}, J. and {Christensen}, P.~R. and {Church}, S. and {Clements}, D.~L. and {Colombi}, S. and {Colombo}, L.~P.~L. and {Combet}, C. and {Coulais}, A. and {Crill}, B.~P. and {Curto}, A. and {Cuttaia}, F. and {Danese}, L. and {Davies}, R.~D. and {Davis}, R.~J. and {de Bernardis}, P. and {de Rosa}, A. and {de Zotti}, G. and {Delabrouille}, J. and {D{\'e}sert}, F.-X. and {Di Valentino}, E. and {Dickinson}, C. and {Diego}, J.~M. and {Dolag}, K. and {Dole}, H. and {Donzelli}, S. and {Dor{\'e}}, O. and {Douspis}, M. and {Ducout}, A. and {Dunkley}, J. and {Dupac}, X. and {Efstathiou}, G. and {Elsner}, F. and {En{\ss}lin}, T.~A. and {Eriksen}, H.~K. and {Farhang}, M. and {Fergusson}, J. and {Finelli}, F. and {Forni}, O. and {Frailis}, M. and {Fraisse}, A.~A. and {Franceschi}, E. and {Frejsel}, A. and {Galeotta}, S. and {Galli}, S. and {Ganga}, K. and {Gauthier}, C. and {Gerbino}, M. and {Ghosh}, T. and {Giard}, M. and {Giraud-H{\'e}raud}, Y. and {Giusarma}, E. and {Gjerl{\o}w}, E. and {Gonz{\'a}lez-Nuevo}, J. and {G{\'o}rski}, K.~M. and {Gratton}, S. and {Gregorio}, A. and {Gruppuso}, A. and {Gudmundsson}, J.~E. and {Hamann}, J. and {Hansen}, F.~K. and {Hanson}, D. and {Harrison}, D.~L. and {Helou}, G. and {Henrot-Versill{\'e}}, S. and {Hern{\'a}ndez-Monteagudo}, C. and {Herranz}, D. and {Hildebrandt}, S.~R. and {Hivon}, E. and {Hobson}, M. and {Holmes}, W.~A. and {Hornstrup}, A. and {Hovest}, W. and {Huang}, Z. and {Huffenberger}, K.~M. and {Hurier}, G. and {Jaffe}, A.~H. and {Jaffe}, T.~R. and {Jones}, W.~C. and {Juvela}, M. and {Keih{\"a}nen}, E. and {Keskitalo}, R. and {Kisner}, T.~S. and {Kneissl}, R. and {Knoche}, J. and {Knox}, L. and {Kunz}, M. and {Kurki-Suonio}, H. and {Lagache}, G. and {L{\"a}hteenm{\"a}ki}, A. and {Lamarre}, J.-M. and {Lasenby}, A. and {Lattanzi}, M. and {Lawrence}, C.~R. and {Leahy}, J.~P. and {Leonardi}, R. and {Lesgourgues}, J. and {Levrier}, F. and {Lewis}, A. and {Liguori}, M. and {Lilje}, P.~B. and {Linden-V{\o}rnle}, M. and {L{\'o}pez-Caniego}, M. and {Lubin}, P.~M. and {Mac{\'\i}as-P{\'e}rez}, J.~F. and {Maggio}, G. and {Maino}, D. and {Mandolesi}, N. and {Mangilli}, A. and {Marchini}, A. and {Maris}, M. and {Martin}, P.~G. and {Martinelli}, M. and {Mart{\'\i}nez-Gonz{\'a}lez}, E. and {Masi}, S. and {Matarrese}, S. and {McGehee}, P. and {Meinhold}, P.~R. and {Melchiorri}, A. and {Melin}, J.-B. and {Mendes}, L. and {Mennella}, A. and {Migliaccio}, M. and {Millea}, M. and {Mitra}, S. and {Miville-Desch{\^e}nes}, M.-A. and {Moneti}, A. and {Montier}, L. and {Morgante}, G. and {Mortlock}, D. and {Moss}, A. and {Munshi}, D. and {Murphy}, J.~A. and {Naselsky}, P. and {Nati}, F. and {Natoli}, P. and {Netterfield}, C.~B. and {N{\o}rgaard-Nielsen}, H.~U. and {Noviello}, F. and {Novikov}, D. and {Novikov}, I. and {Oxborrow}, C.~A. and {Paci}, F. and {Pagano}, L. and {Pajot}, F. and {Paladini}, R. and {Paoletti}, D. and {Partridge}, B. and {Pasian}, F. and {Patanchon}, G. and {Pearson}, T.~J. and {Perdereau}, O. and {Perotto}, L. and {Perrotta}, F. and {Pettorino}, V. and {Piacentini}, F. and {Piat}, M. and {Pierpaoli}, E. and {Pietrobon}, D. and {Plaszczynski}, S. and {Pointecouteau}, E. and {Polenta}, G. and {Popa}, L. and {Pratt}, G.~W. and {Pr{\'e}zeau}, G.},
        title = "{Planck 2015 results. XIII. Cosmological parameters}",
      journal = {\aap},
         year = 2016,
        month = sep,
       volume = {594},
          eid = {A13},
        pages = {A13},
          doi = {10.1051/0004-6361/201525830},
archivePrefix = {arXiv},
       eprint = {1502.01589},
 primaryClass = {astro-ph.CO},
       adsurl = {https://ui.adsabs.harvard.edu/abs/2016A&A...594A..13P}
}

@ARTICLE{DESIPV_2025,
       author = {{Said}, Khaled and {Howlett}, Cullan and {Davis}, Tamara and {Lucey}, John and {Saulder}, Christoph and {Douglass}, Kelly and {Kim}, Alex G. and {Kremin}, Anthony and {Ross}, Caitlin and {Aldering}, Greg and {Aguilar}, Jessica Nicole and {Ahlen}, Steven and {BenZvi}, Segev and {Bianchi}, Davide and {Brooks}, David and {Claybaugh}, Todd and {Dawson}, Kyle and {de la Macorra}, Axel and {Dey}, Biprateep and {Doel}, Peter and {Fanning}, Kevin and {Ferraro}, Simone and {Font-Ribera}, Andreu and {Forero-Romero}, Jaime E. and {Gazta{\~n}aga}, Enrique and {Gontcho}, Satya Gontcho A. and {Guy}, Julien and {Honscheid}, Klaus and {Kehoe}, Robert and {Kisner}, Theodore and {Lambert}, Andrew and {Landriau}, Martin and {Le Guillou}, Laurent and {Manera}, Marc and {Meisner}, Aaron and {Miquel}, Ramon and {Moustakas}, John and {Mu{\~n}oz-Guti{\'e}rrez}, Andrea and {Myers}, Adam and {Nie}, Jundan and {Palanque-Delabrouille}, Nathalie and {Percival}, Will and {Prada}, Francisco and {Rossi}, Graziano and {Sanchez}, Eusebio and {Schlegel}, David and {Schubnell}, Michael and {Silber}, Joseph Harry and {Sprayberry}, David and {Tarl{\'e}}, Gregory and {Magana}, Mariana Vargas and {Weaver}, Benjamin Alan and {Wechsler}, Risa and {Zhou}, Zhimin and {Zou}, Hu},
        title = "{DESI peculiar velocity survey ─ Fundamental Plane}",
      journal = {\mnras},
         year = 2025,
        month = jun,
       volume = {539},
       number = {4},
        pages = {3627-3644},
          doi = {10.1093/mnras/staf700},
archivePrefix = {arXiv},
       eprint = {2408.13842},
 primaryClass = {astro-ph.CO},
       adsurl = {https://ui.adsabs.harvard.edu/abs/2025MNRAS.539.3627S}
}

@ARTICLE{DESIBGS_2023,
       author = {{Hahn}, ChangHoon and {Wilson}, Michael J. and {Ruiz-Macias}, Omar and {Cole}, Shaun and {Weinberg}, David H. and {Moustakas}, John and {Kremin}, Anthony and {Tinker}, Jeremy L. and {Smith}, Alex and {Wechsler}, Risa H. and {Ahlen}, Steven and {Alam}, Shadab and {Bailey}, Stephen and {Brooks}, David and {Cooper}, Andrew P. and {Davis}, Tamara M. and {Dawson}, Kyle and {Dey}, Arjun and {Dey}, Biprateep and {Eftekharzadeh}, Sarah and {Eisenstein}, Daniel J. and {Fanning}, Kevin and {Forero-Romero}, Jaime E. and {Frenk}, Carlos S. and {Gazta{\~n}aga}, Enrique and {A Gontcho}, Satya Gontcho and {Guy}, Julien and {Honscheid}, Klaus and {Ishak}, Mustapha and {Juneau}, St{\'e}phanie and {Kehoe}, Robert and {Kisner}, Theodore and {Lan}, Ting-Wen and {Landriau}, Martin and {Le Guillou}, Laurent and {Levi}, Michael E. and {Magneville}, Christophe and {Martini}, Paul and {Meisner}, Aaron and {Myers}, Adam D. and {Nie}, Jundan and {Norberg}, Peder and {Palanque-Delabrouille}, Nathalie and {Percival}, Will J. and {Poppett}, Claire and {Prada}, Francisco and {Raichoor}, Anand and {Ross}, Ashley J. and {Gaines}, Sasha and {Saulder}, Christoph and {Schlafly}, Eddie and {Schlegel}, David and {Sierra-Porta}, David and {Tarle}, Gregory and {Weaver}, Benjamin A. and {Y{\`e}che}, Christophe and {Zarrouk}, Pauline and {Zhou}, Rongpu and {Zhou}, Zhimin and {Zou}, Hu},
        title = "{The DESI Bright Galaxy Survey: Final Target Selection, Design, and Validation}",
      journal = {\aj},
         year = 2023,
        month = jun,
       volume = {165},
       number = {6},
          eid = {253},
        pages = {253},
          doi = {10.3847/1538-3881/accff8},
archivePrefix = {arXiv},
       eprint = {2208.08512},
 primaryClass = {astro-ph.CO},
       adsurl = {https://ui.adsabs.harvard.edu/abs/2023AJ....165..253H}
}

@ARTICLE{KIDSLegacy_shear,
       author = {{Wright}, Angus H. and {St{\"o}lzner}, Benjamin and {Asgari}, Marika and {Bilicki}, Maciej and {Giblin}, Benjamin and {Heymans}, Catherine and {Hildebrandt}, Hendrik and {Hoekstra}, Henk and {Joachimi}, Benjamin and {Kuijken}, Konrad and {Li}, Shun-Sheng and {Reischke}, Robert and {von Wietersheim-Kramsta}, Maximilian and {Yoon}, Mijin and {Burger}, Pierre and {Chisari}, Nora Elisa and {de Jong}, Jelte and {Dvornik}, Andrej and {Georgiou}, Christos and {Harnois-D{\'e}raps}, Joachim and {Jalan}, Priyanka and {William}, Anjitha John and {Joudaki}, Shahab and {Lesci}, Giorgio Francesco and {Linke}, Laila and {Loureiro}, Arthur and {Mahony}, Constance and {Maturi}, Matteo and {Miller}, Lance and {Moscardini}, Lauro and {Napolitano}, Nicola R. and {Porth}, Lucas and {Radovich}, Mario and {Schneider}, Peter and {Tr{\"o}ster}, Tilman and {Valentijn}, Edwin and {Wittje}, Anna and {Yan}, Ziang and {Zhang}, Yun-Hao},
        title = "{KiDS-Legacy: Cosmological constraints from cosmic shear with the complete Kilo-Degree Survey}",
      journal = {\aap},
         year = 2025,
        month = nov,
       volume = {703},
          eid = {A158},
        pages = {A158},
          doi = {10.1051/0004-6361/202554908},
archivePrefix = {arXiv},
       eprint = {2503.19441},
 primaryClass = {astro-ph.CO},
       adsurl = {https://ui.adsabs.harvard.edu/abs/2025A&A...703A.158W}
}

@ARTICLE{HSCY3_shear,
       author = {{Dalal}, Roohi and {Li}, Xiangchong and {Nicola}, Andrina and {Zuntz}, Joe and {Strauss}, Michael A. and {Sugiyama}, Sunao and {Zhang}, Tianqing and {Rau}, Markus M. and {Mandelbaum}, Rachel and {Takada}, Masahiro and {More}, Surhud and {Miyatake}, Hironao and {Kannawadi}, Arun and {Shirasaki}, Masato and {Taniguchi}, Takanori and {Takahashi}, Ryuichi and {Osato}, Ken and {Hamana}, Takashi and {Oguri}, Masamune and {Nishizawa}, Atsushi J. and {Malag{\'o}n}, Andr{\'e}s A. Plazas and {Sunayama}, Tomomi and {Alonso}, David and {Slosar}, An{\v{z}}e and {Luo}, Wentao and {Armstrong}, Robert and {Bosch}, James and {Hsieh}, Bau-Ching and {Komiyama}, Yutaka and {Lupton}, Robert H. and {Lust}, Nate B. and {MacArthur}, Lauren A. and {Miyazaki}, Satoshi and {Murayama}, Hitoshi and {Nishimichi}, Takahiro and {Okura}, Yuki and {Price}, Paul A. and {Tait}, Philip J. and {Tanaka}, Masayuki and {Wang}, Shiang-Yu},
        title = "{Hyper Suprime-Cam Year 3 results: Cosmology from cosmic shear power spectra}",
      journal = {\prd},
         year = 2023,
        month = dec,
       volume = {108},
       number = {12},
          eid = {123519},
        pages = {123519},
          doi = {10.1103/PhysRevD.108.123519},
archivePrefix = {arXiv},
       eprint = {2304.00701},
 primaryClass = {astro-ph.CO},
       adsurl = {https://ui.adsabs.harvard.edu/abs/2023PhRvD.108l3519D}
}

@ARTICLE{DESY6_shear,
       author = {{DES Collaboration} and {Abbott}, T.~M.~C. and {Aguena}, M. and {Alarcon}, A. and {Alves}, O. and {Amon}, A. and {Anbajagane}, D. and {Andrade-Oliveira}, F. and {d'Assignies}, W. and {Avila}, S. and {Bacon}, D. and {Beas-Gonzalez}, J. and {Bechtol}, K. and {Becker}, M.~R. and {Bernstein}, G.~M. and {Blazek}, J. and {Bocquet}, S. and {Brooks}, D. and {Camacho}, H. and {Camacho-Ciurana}, G. and {Camilleri}, R. and {Campailla}, G. and {Campos}, A. and {Carnero Rosell}, A. and {Carrasco Kind}, M. and {Carretero}, J. and {Castander}, F.~J. and {Cawthon}, R. and {Chang}, C. and {Choi}, A. and {Coloma-Nadal}, J.~M. and {Conselice}, C. and {da Costa}, L.~N. and {Costanzi}, M. and {Crocce}, M. and {Davis}, T.~M. and {De Vicente}, J. and {DePoy}, D.~L. and {DeRose}, J. and {Desai}, S. and {Diehl}, H.~T. and {Doel}, P. and {Doux}, C. and {Drlica-Wagner}, A. and {Eifler}, T.~F. and {Everett}, S. and {Evrard}, A.~E. and {Fert{\'e}}, A. and {Flaugher}, B. and {Fosalba}, P. and {Friedrich}, O. and {Frieman}, J. and {Garc{\'\i}a-Bellido}, J. and {Gatti}, M. and {Giannini}, G. and {Giles}, P. and {Glazebrook}, K. and {Gruen}, D. and {Gruendl}, R.~A. and {Gutierrez}, G. and {Harrison}, I. and {Hartley}, W.~G. and {Herner}, K. and {Hinton}, S.~R. and {Hollowood}, D.~L. and {Honscheid}, K. and {Huterer}, D. and {Jain}, B. and {James}, D.~J. and {Jarvis}, M. and {Jeffrey}, N. and {Jeltema}, T. and {Kacprzak}, T. and {Kent}, S. and {Krause}, E. and {Lahav}, O. and {Lee}, S. and {Legnani}, E. and {Lin}, H. and {Marshall}, J.~L. and {Mau}, S. and {Mena-Fern{\'a}ndez}, J. and {Menanteau}, F. and {Miquel}, R. and {Mohr}, J.~J. and {Muir}, J. and {Myles}, J. and {Nichol}, R.~C. and {Ogando}, R.~L.~C. and {Palmese}, A. and {Paterno}, M. and {Percival}, W.~J. and {Petravick}, D. and {Plazas Malag{\'o}n}, A.~A. and {Porredon}, A. and {Prat}, J. and {Preston}, C. and {Raveri}, M. and {Rodriguez-Monroy}, M. and {Romer}, A.~K. and {Roodman}, A. and {Rykoff}, E.~S. and {Samuroff}, S. and {S{\'a}nchez}, C. and {Sanchez}, E. and {Sanchez Cid}, D. and {Schutt}, T. and {Sevilla-Noarbe}, I. and {Sheldon}, E. and {Shin}, T. and {da Silva Pereira}, M.~E. and {Smith}, M. and {Soares-Santos}, M. and {Suchyta}, E. and {Swanson}, M.~E.~C. and {Tabbutt}, M. and {Tarle}, G. and {Thomas}, D. and {To}, C. and {Troxel}, M.~A. and {Vikram}, V. and {Vincenzi}, M. and {Weaverdyck}, N. and {Weller}, J. and {Wiseman}, P. and {Yamamoto}, M. and {Yanny}, B. and {Yin}, B. and {Zuntz}, J.},
        title = "{Dark Energy Survey Year 6 Results: Cosmological Constraints from Cosmic Shear}",
      journal = {arXiv e-prints},
         year = 2026,
        month = feb,
          eid = {arXiv:2602.10065},
        pages = {arXiv:2602.10065},
          doi = {10.48550/arXiv.2602.10065},
archivePrefix = {arXiv},
       eprint = {2602.10065},
 primaryClass = {astro-ph.CO},
       adsurl = {https://ui.adsabs.harvard.edu/abs/2026arXiv260210065D}
}

@ARTICLE{Takada_2014,
       author = {{Takada}, Masahiro and {Ellis}, Richard S. and {Chiba}, Masashi and {Greene}, Jenny E. and {Aihara}, Hiroaki and {Arimoto}, Nobuo and {Bundy}, Kevin and {Cohen}, Judith and {Dor{\'e}}, Olivier and {Graves}, Genevieve and {Gunn}, James E. and {Heckman}, Timothy and {Hirata}, Christopher M. and {Ho}, Paul and {Kneib}, Jean-Paul and {Le F{\`e}vre}, Olivier and {Lin}, Lihwai and {More}, Surhud and {Murayama}, Hitoshi and {Nagao}, Tohru and {Ouchi}, Masami and {Seiffert}, Michael and {Silverman}, John D. and {Sodr{\'e}}, Laerte and {Spergel}, David N. and {Strauss}, Michael A. and {Sugai}, Hajime and {Suto}, Yasushi and {Takami}, Hideki and {Wyse}, Rosemary},
        title = "{Extragalactic science, cosmology, and Galactic archaeology with the Subaru Prime Focus Spectrograph}",
      journal = {\pasj},
         year = 2014,
        month = feb,
       volume = {66},
       number = {1},
          eid = {R1},
        pages = {R1},
          doi = {10.1093/pasj/pst019},
archivePrefix = {arXiv},
       eprint = {1206.0737},
 primaryClass = {astro-ph.CO},
       adsurl = {https://ui.adsabs.harvard.edu/abs/2014PASJ...66R...1T}
}

@ARTICLE{Wang_2022,
       author = {{Wang}, Yun and {Zhai}, Zhongxu and {Alavi}, Anahita and {Massara}, Elena and {Pisani}, Alice and {Benson}, Andrew and {Hirata}, Christopher M. and {Samushia}, Lado and {Weinberg}, David H. and {Colbert}, James and {Dor{\'e}}, Olivier and {Eifler}, Tim and {Heinrich}, Chen and {Ho}, Shirley and {Krause}, Elisabeth and {Padmanabhan}, Nikhil and {Spergel}, David and {Teplitz}, Harry I.},
        title = "{The High Latitude Spectroscopic Survey on the Nancy Grace Roman Space Telescope}",
      journal = {\apj},
         year = 2022,
        month = mar,
       volume = {928},
       number = {1},
          eid = {1},
        pages = {1},
          doi = {10.3847/1538-4357/ac4973},
archivePrefix = {arXiv},
       eprint = {2110.01829},
 primaryClass = {astro-ph.CO},
       adsurl = {https://ui.adsabs.harvard.edu/abs/2022ApJ...928....1W}
}

@ARTICLE{Alam_2021,
       author = {{Alam}, Shadab and {Aubert}, Marie and {Avila}, Santiago and {Balland}, Christophe and {Bautista}, Julian E. and {Bershady}, Matthew A. and {Bizyaev}, Dmitry and {Blanton}, Michael R. and {Bolton}, Adam S. and {Bovy}, Jo and {Brinkmann}, Jonathan and {Brownstein}, Joel R. and {Burtin}, Etienne and {Chabanier}, Sol{\`e}ne and {Chapman}, Michael J. and {Choi}, Peter Doohyun and {Chuang}, Chia-Hsun and {Comparat}, Johan and {Cousinou}, Marie-Claude and {Cuceu}, Andrei and {Dawson}, Kyle S. and {de la Torre}, Sylvain and {de Mattia}, Arnaud and {Agathe}, Victoria de Sainte and {des Bourboux}, H{\'e}lion du Mas and {Escoffier}, Stephanie and {Etourneau}, Thomas and {Farr}, James and {Font-Ribera}, Andreu and {Frinchaboy}, Peter M. and {Fromenteau}, Sebastien and {Gil-Mar{\'\i}n}, H{\'e}ctor and {Le Goff}, Jean-Marc and {Gonzalez-Morales}, Alma X. and {Gonzalez-Perez}, Violeta and {Grabowski}, Kathleen and {Guy}, Julien and {Hawken}, Adam J. and {Hou}, Jiamin and {Kong}, Hui and {Parker}, James and {Klaene}, Mark and {Kneib}, Jean-Paul and {Lin}, Sicheng and {Long}, Daniel and {Lyke}, Brad W. and {de la Macorra}, Axel and {Martini}, Paul and {Masters}, Karen and {Mohammad}, Faizan G. and {Moon}, Jeongin and {Mueller}, Eva-Maria and {Mu{\~n}oz-Guti{\'e}rrez}, Andrea and {Myers}, Adam D. and {Nadathur}, Seshadri and {Neveux}, Richard and {Newman}, Jeffrey A. and {Noterdaeme}, Pasquier and {Oravetz}, Audrey and {Oravetz}, Daniel and {Palanque-Delabrouille}, Nathalie and {Pan}, Kaike and {Paviot}, Romain and {Percival}, Will J. and {P{\'e}rez-R{\`a}fols}, Ignasi and {Petitjean}, Patrick and {Pieri}, Matthew M. and {Prakash}, Abhishek and {Raichoor}, Anand and {Ravoux}, Corentin and {Rezaie}, Mehdi and {Rich}, James and {Ross}, Ashley J. and {Rossi}, Graziano and {Ruggeri}, Rossana and {Ruhlmann-Kleider}, Vanina and {S{\'a}nchez}, Ariel G. and {S{\'a}nchez}, F. Javier and {S{\'a}nchez-Gallego}, Jos{\'e} R. and {Sayres}, Conor and {Schneider}, Donald P. and {Seo}, Hee-Jong and {Shafieloo}, Arman and {Slosar}, An{\v{z}}e and {Smith}, Alex and {Stermer}, Julianna and {Tamone}, Amelie and {Tinker}, Jeremy L. and {Tojeiro}, Rita and {Vargas-Maga{\~n}a}, Mariana and {Variu}, Andrei and {Wang}, Yuting and {Weaver}, Benjamin A. and {Weijmans}, Anne-Marie and {Y{\`e}che}, Christophe and {Zarrouk}, Pauline and {Zhao}, Cheng and {Zhao}, Gong-Bo and {Zheng}, Zheng},
        title = "{Completed SDSS-IV extended Baryon Oscillation Spectroscopic Survey: Cosmological implications from two decades of spectroscopic surveys at the Apache Point Observatory}",
      journal = {\prd},
         year = 2021,
        month = apr,
       volume = {103},
       number = {8},
          eid = {083533},
        pages = {083533},
          doi = {10.1103/PhysRevD.103.083533},
archivePrefix = {arXiv},
       eprint = {2007.08991},
 primaryClass = {astro-ph.CO},
       adsurl = {https://ui.adsabs.harvard.edu/abs/2021PhRvD.103h3533A}
}

@ARTICLE{Alam_2017,
       author = {{Alam}, Shadab and {Ata}, Metin and {Bailey}, Stephen and {Beutler}, Florian and {Bizyaev}, Dmitry and {Blazek}, Jonathan A. and {Bolton}, Adam S. and {Brownstein}, Joel R. and {Burden}, Angela and {Chuang}, Chia-Hsun and {Comparat}, Johan and {Cuesta}, Antonio J. and {Dawson}, Kyle S. and {Eisenstein}, Daniel J. and {Escoffier}, Stephanie and {Gil-Mar{\'\i}n}, H{\'e}ctor and {Grieb}, Jan Niklas and {Hand}, Nick and {Ho}, Shirley and {Kinemuchi}, Karen and {Kirkby}, David and {Kitaura}, Francisco and {Malanushenko}, Elena and {Malanushenko}, Viktor and {Maraston}, Claudia and {McBride}, Cameron K. and {Nichol}, Robert C. and {Olmstead}, Matthew D. and {Oravetz}, Daniel and {Padmanabhan}, Nikhil and {Palanque-Delabrouille}, Nathalie and {Pan}, Kaike and {Pellejero-Ibanez}, Marcos and {Percival}, Will J. and {Petitjean}, Patrick and {Prada}, Francisco and {Price-Whelan}, Adrian M. and {Reid}, Beth A. and {Rodr{\'\i}guez-Torres}, Sergio A. and {Roe}, Natalie A. and {Ross}, Ashley J. and {Ross}, Nicholas P. and {Rossi}, Graziano and {Rubi{\~n}o-Mart{\'\i}n}, Jose Alberto and {Saito}, Shun and {Salazar-Albornoz}, Salvador and {Samushia}, Lado and {S{\'a}nchez}, Ariel G. and {Satpathy}, Siddharth and {Schlegel}, David J. and {Schneider}, Donald P. and {Sc{\'o}ccola}, Claudia G. and {Seo}, Hee-Jong and {Sheldon}, Erin S. and {Simmons}, Audrey and {Slosar}, An{\v{z}}e and {Strauss}, Michael A. and {Swanson}, Molly E.~C. and {Thomas}, Daniel and {Tinker}, Jeremy L. and {Tojeiro}, Rita and {Maga{\~n}a}, Mariana Vargas and {Vazquez}, Jose Alberto and {Verde}, Licia and {Wake}, David A. and {Wang}, Yuting and {Weinberg}, David H. and {White}, Martin and {Wood-Vasey}, W. Michael and {Y{\`e}che}, Christophe and {Zehavi}, Idit and {Zhai}, Zhongxu and {Zhao}, Gong-Bo},
        title = "{The clustering of galaxies in the completed SDSS-III Baryon Oscillation Spectroscopic Survey: cosmological analysis of the DR12 galaxy sample}",
      journal = {\mnras},
         year = 2017,
        month = sep,
       volume = {470},
       number = {3},
        pages = {2617-2652},
          doi = {10.1093/mnras/stx721},
archivePrefix = {arXiv},
       eprint = {1607.03155},
 primaryClass = {astro-ph.CO},
       adsurl = {https://ui.adsabs.harvard.edu/abs/2017MNRAS.470.2617A}
}

@ARTICLE{eBOSS_2021,
       author = {{de Mattia}, Arnaud and {Ruhlmann-Kleider}, Vanina and {Raichoor}, Anand and {Ross}, Ashley J. and {Tamone}, Am{\'e}lie and {Zhao}, Cheng and {Alam}, Shadab and {Avila}, Santiago and {Burtin}, Etienne and {Bautista}, Julian and {Beutler}, Florian and {Brinkmann}, Jonathan and {Brownstein}, Joel R. and {Chapman}, Michael J. and {Chuang}, Chia-Hsun and {Comparat}, Johan and {du Mas des Bourboux}, H{\'e}lion and {Dawson}, Kyle S. and {de la Macorra}, Axel and {Gil-Mar{\'\i}n}, H{\'e}ctor and {Gonzalez-Perez}, Violeta and {Gorgoni}, Claudio and {Hou}, Jiamin and {Kong}, Hui and {Lin}, Sicheng and {Nadathur}, Seshadri and {Newman}, Jeffrey A. and {Mueller}, Eva-Maria and {Percival}, Will J. and {Rezaie}, Mehdi and {Rossi}, Graziano and {Schneider}, Donald P. and {Tiwari}, Prabhakar and {Vivek}, M. and {Wang}, Yuting and {Zhao}, Gong-Bo},
        title = "{The completed SDSS-IV extended Baryon Oscillation Spectroscopic Survey: measurement of the BAO and growth rate of structure of the emission line galaxy sample from the anisotropic power spectrum between redshift 0.6 and 1.1}",
      journal = {\mnras},
         year = 2021,
        month = mar,
       volume = {501},
       number = {4},
        pages = {5616-5645},
          doi = {10.1093/mnras/staa3891},
archivePrefix = {arXiv},
       eprint = {2007.09008},
 primaryClass = {astro-ph.CO},
       adsurl = {https://ui.adsabs.harvard.edu/abs/2021MNRAS.501.5616D}
}

@ARTICLE{Blake_2011,
       author = {{Blake}, Chris and {Kazin}, Eyal A. and {Beutler}, Florian and {Davis}, Tamara M. and {Parkinson}, David and {Brough}, Sarah and {Colless}, Matthew and {Contreras}, Carlos and {Couch}, Warrick and {Croom}, Scott and {Croton}, Darren and {Drinkwater}, Michael J. and {Forster}, Karl and {Gilbank}, David and {Gladders}, Mike and {Glazebrook}, Karl and {Jelliffe}, Ben and {Jurek}, Russell J. and {Li}, I.-Hui and {Madore}, Barry and {Martin}, D. Christopher and {Pimbblet}, Kevin and {Poole}, Gregory B. and {Pracy}, Michael and {Sharp}, Rob and {Wisnioski}, Emily and {Woods}, David and {Wyder}, Ted K. and {Yee}, H.~K.~C.},
        title = "{The WiggleZ Dark Energy Survey: mapping the distance-redshift relation with baryon acoustic oscillations}",
      journal = {\mnras},
         year = 2011,
        month = dec,
       volume = {418},
       number = {3},
        pages = {1707-1724},
          doi = {10.1111/j.1365-2966.2011.19592.x},
archivePrefix = {arXiv},
       eprint = {1108.2635},
 primaryClass = {astro-ph.CO},
       adsurl = {https://ui.adsabs.harvard.edu/abs/2011MNRAS.418.1707B}
}

@ARTICLE{euclid_2025,
       author = {{Euclid Collaboration} and {Mellier}, Y. and {Abdurro'uf} and {Acevedo Barroso}, J.~A. and {Ach{\'u}carro}, A. and {Adamek}, J. and {Adam}, R. and {Addison}, G.~E. and {Aghanim}, N. and {Aguena}, M. and {Ajani}, V. and {Akrami}, Y. and {Al-Bahlawan}, A. and {Alavi}, A. and {Albuquerque}, I.~S. and {Alestas}, G. and {Alguero}, G. and {Allaoui}, A. and {Allen}, S.~W. and {Allevato}, V. and {Alonso-Tetilla}, A.~V. and {Altieri}, B. and {Alvarez-Candal}, A. and {Alvi}, S. and {Amara}, A. and {Amendola}, L. and {Amiaux}, J. and {Andika}, I.~T. and {Andreon}, S. and {Andrews}, A. and {Angora}, G. and {Angulo}, R.~E. and {Annibali}, F. and {Anselmi}, A. and {Anselmi}, S. and {Arcari}, S. and {Archidiacono}, M. and {Aric{\`o}}, G. and {Arnaud}, M. and {Arnouts}, S. and {Asgari}, M. and {Asorey}, J. and {Atayde}, L. and {Atek}, H. and {Atrio-Barandela}, F. and {Aubert}, M. and {Aubourg}, E. and {Auphan}, T. and {Auricchio}, N. and {Aussel}, B. and {Aussel}, H. and {Avelino}, P.~P. and {Avgoustidis}, A. and {Avila}, S. and {Awan}, S. and {Azzollini}, R. and {Baccigalupi}, C. and {Bachelet}, E. and {Bacon}, D. and {Baes}, M. and {Bagley}, M.~B. and {Bahr-Kalus}, B. and {Balaguera-Antolinez}, A. and {Balbinot}, E. and {Balcells}, M. and {Baldi}, M. and {Baldry}, I. and {Balestra}, A. and {Ballardini}, M. and {Ballester}, O. and {Balogh}, M. and {Ba{\~n}ados}, E. and {Barbier}, R. and {Bardelli}, S. and {Baron}, M. and {Barreiro}, T. and {Barrena}, R. and {Barriere}, J.-C. and {Barros}, B.~J. and {Barthelemy}, A. and {Bartolo}, N. and {Basset}, A. and {Battaglia}, P. and {Battisti}, A.~J. and {Baugh}, C.~M. and {Baumont}, L. and {Bazzanini}, L. and {Beaulieu}, J.-P. and {Beckmann}, V. and {Belikov}, A.~N. and {Bel}, J. and {Bellagamba}, F. and {Bella}, M. and {Bellini}, E. and {Benabed}, K. and {Bender}, R. and {Benevento}, G. and {Bennett}, C.~L. and {Benson}, K. and {Bergamini}, P. and {Bermejo-Climent}, J.~R. and {Bernardeau}, F. and {Bertacca}, D. and {Berthe}, M. and {Berthier}, J. and {Bethermin}, M. and {Beutler}, F. and {Bevillon}, C. and {Bhargava}, S. and {Bhatawdekar}, R. and {Bianchi}, D. and {Bisigello}, L. and {Biviano}, A. and {Blake}, R.~P. and {Blanchard}, A. and {Blazek}, J. and {Blot}, L. and {Bosco}, A. and {Bodendorf}, C. and {Boenke}, T. and {B{\"o}hringer}, H. and {Boldrini}, P. and {Bolzonella}, M. and {Bonchi}, A. and {Bonici}, M. and {Bonino}, D. and {Bonino}, L. and {Bonvin}, C. and {Bon}, W. and {Booth}, J.~T. and {Borgani}, S. and {Borlaff}, A.~S. and {Borsato}, E. and {Bose}, B. and {Botticella}, M.~T. and {Boucaud}, A. and {Bouche}, F. and {Boucher}, J.~S. and {Boutigny}, D. and {Bouvard}, T. and {Bouwens}, R. and {Bouy}, H. and {Bowler}, R.~A.~A. and {Bozza}, V. and {Bozzo}, E. and {Branchini}, E. and {Brando}, G. and {Brau-Nogue}, S. and {Brekke}, P. and {Bremer}, M.~N. and {Brescia}, M. and {Breton}, M.-A. and {Brinchmann}, J. and {Brinckmann}, T. and {Brockley-Blatt}, C. and {Brodwin}, M. and {Brouard}, L. and {Brown}, M.~L. and {Bruton}, S. and {Bucko}, J. and {Buddelmeijer}, H. and {Buenadicha}, G. and {Buitrago}, F. and {Burger}, P. and {Burigana}, C. and {Busillo}, V. and {Busonero}, D. and {Cabanac}, R. and {Cabayol-Garcia}, L. and {Cagliari}, M.~S. and {Caillat}, A. and {Caillat}, L. and {Calabrese}, M. and {Calabro}, A. and {Calderone}, G. and {Calura}, F. and {Camacho Quevedo}, B. and {Camera}, S. and {Campos}, L. and {Ca{\~n}as-Herrera}, G. and {Candini}, G.~P. and {Cantiello}, M. and {Capobianco}, V. and {Cappellaro}, E. and {Cappelluti}, N. and {Cappi}, A. and {Caputi}, K.~I. and {Cara}, C. and {Carbone}, C. and {Cardone}, V.~F. and {Carella}, E. and {Carlberg}, R.~G. and {Carle}, M. and {Carminati}, L. and {Caro}, F. and {Carrasco}, J.~M. and {Carretero}, J. and {Carrilho}, P. and {Carron Duque}, J. and {Carry}, B.},
        title = "{Euclid: I. Overview of the Euclid mission}",
      journal = {\aap},
         year = 2025,
        month = may,
       volume = {697},
          eid = {A1},
        pages = {A1},
          doi = {10.1051/0004-6361/202450810},
archivePrefix = {arXiv},
       eprint = {2405.13491},
 primaryClass = {astro-ph.CO},
       adsurl = {https://ui.adsabs.harvard.edu/abs/2025A&A...697A...1E}
}

@ARTICLE{pdr1,
author = {Mentuch Cooper, Erin and Gebhardt, Karl and Davis, Dustin and Liu, Chenxu and Castanheira, Barbara G. and Chase, Owen and Chávez Ortiz, Óscar A. and Ciardullo, Robin and Curtis, Olivia and Dunne, Delaney A. and Evans, Neal J. and Farrow, Daniel J. and Fabricius, Maximilian and Finkelstein, Steven L. and Gronwall, Caryl and Hamme, Nathaniel J. and Hill, Gary J. and House, Lindsay R. and Jarvis, Matt J. and Jeong, Donghui and Kelz, Andreas and Komatsu, Eiichiro and Mirza Khanlari, Mahan and Khoraminezhad, Hasti and Kollatschny, Wolfram and Lujan Niemeyer, Maja and Lee, Hanshin and MacQueen, Phillip and Mitra, Deeshani and Mukae, Shiro and Ouchi, Masami and Poppe, Jennifer and Powell, Meredith C. and Qezlou, Mahdi and Saito, Shun and Schneider, Donald P. and Weiss, Laurel and Wisotzki, Lutz and Zeimann, Gregory R.},
title = "{HETDEX Public Data Release 1: Source Catalog 2 and Datacubes
                 from $\sim$90 deg$^2$ of Integral-Field Optical Spectroscopy}",
      journal = {\apjs},
         year = 2026,
        month = jun,
       volume = {284},
       number = {2},
        pages = {67},
        doi = {10.3847/1538-4365/ae6068},
url = {https://doi.org/10.3847/1538-4365/ae6068},
}

@ARTICLE{Hand_2017,
       author = {{Hand}, Nick and {Li}, Yin and {Slepian}, Zachary and {Seljak}, Uro{\v{s}}},
        title = "{An optimal FFT-based anisotropic power spectrum estimator}",
      journal = {\jcap},
         year = 2017,
        month = jul,
       volume = {2017},
       number = {07},
        pages = {002},
          doi = {10.1088/1475-7516/2017/07/002},
archivePrefix = {arXiv},
       eprint = {1704.02357},
 primaryClass = {astro-ph.CO},
}

@ARTICLE{Yamamoto_2006,
       author = {{Yamamoto}, Kazuhiro and {Nakamichi}, Masashi and {Kamino}, Akinari
                 and {Bassett}, Bruce A. and {Nishioka}, Hiroaki},
        title = "{A Measurement of the Quadrupole Power Spectrum in the Clustering
                 of the 2dF QSO Survey}",
      journal = {\pasj},
         year = 2006,
        month = feb,
       volume = {58},
       number = {1},
        pages = {93-102},
          doi = {10.1093/pasj/58.1.93},
archivePrefix = {arXiv},
       eprint = {astro-ph/0603004},
 primaryClass = {astro-ph},
}

@ARTICLE{FKP_1994,
       author = {{Feldman}, Hume A. and {Kaiser}, Nick and {Peacock}, John A.},
        title = "{Power-Spectrum Analysis of Three-dimensional Redshift Surveys}",
      journal = {\apj},
         year = 1994,
        month = may,
       volume = {426},
        pages = {23},
          doi = {10.1086/174036},
archivePrefix = {arXiv},
       eprint = {astro-ph/9304022},
 primaryClass = {astro-ph},
       adsurl = {https://ui.adsabs.harvard.edu/abs/1994ApJ...426...23F}
}

@ARTICLE{Jing_2005,
       author = {{Jing}, Y.~P.},
        title = "{Correcting for the Alias Effect When Measuring the Power Spectrum
                 Using a Fast Fourier Transform}",
      journal = {\apj},
         year = 2005,
        month = feb,
       volume = {620},
       number = {2},
        pages = {559},
          doi = {10.1086/427087},
archivePrefix = {arXiv},
       eprint = {astro-ph/0409240},
 primaryClass = {astro-ph},
}

@ARTICLE{Wang_2024,
       author = {{Wang}, Yipeng and {Yu}, Yu},
        title = "{Accurate power spectrum estimation toward Nyquist limit}",
      journal = {\jcap},
         year = 2024,
        month = sep,
       volume = {2024},
       number = {09},
        pages = {044},
          doi = {10.1088/1475-7516/2024/09/044},
archivePrefix = {arXiv},
       eprint = {2309.00052},
 primaryClass = {astro-ph.CO},
}

@ARTICLE{Tegmark_2004,
       author = {{Tegmark}, Max and {Blanton}, Michael R. and {Strauss}, Michael A.
                 and {Hoyle}, Fiona and {Schlegel}, David and {Scoccimarro}, Roman
                 and {Vogeley}, Michael S. and {Weinberg}, David H. and {Zehavi}, Idit
                 and {Berlind}, Andreas and {Budavari}, Tam{\'a}s and {Connolly}, Andrew
                 and {Eisenstein}, Daniel J. and {Finkbeiner}, Douglas and {Frieman}, Joshua A.
                 and {Gunn}, James E. and {Hamilton}, Andrew J.~S. and {Hui}, Lam
                 and {Jain}, Bhuvnesh and {Johnston}, David and {Kent}, Stephen
                 and {Lin}, Huan and {Nakajima}, Reiko and {Nichol}, Robert C.
                 and {Ostriker}, Jeremiah P. and {Pope}, Adrian and {Scranton}, Ryan
                 and {Seljak}, Uro{\v{s}} and {Sheth}, Ravi K. and {Stebbins}, Albert
                 and {Szalay}, Alexander S. and {Szapudi}, Istv{\'a}n and {Verde}, Licia
                 and {Xu}, Yongzhong and {Annis}, James and {Bahcall}, Neta A.
                 and {Brinkmann}, J. and {Burles}, Scott and {Castander}, Francisco J.
                 and {Csabai}, Istvan and {Loveday}, Jon and {Doi}, Mamoru
                 and {Fukugita}, Masataka and {Gott}, J.~Richard and {Hennessy}, Greg
                 and {Hogg}, David W. and {Ivezi{\'c}}, {\v{Z}}eljko and {Knapp}, Gillian R.
                 and {Lamb}, Don Q. and {Lee}, Brian C. and {Lupton}, Robert H.
                 and {McKay}, Timothy A. and {Kunszt}, Peter and {Munn}, Jeffrey A.
                 and {O'Connell}, Liam and {Peoples}, John and {Pier}, Jeffrey R.
                 and {Richmond}, Michael and {Rockosi}, Constance and {Schneider}, Donald P.
                 and {Stoughton}, Christopher and {Tucker}, Douglas L.
                 and {Vanden Berk}, Daniel E. and {Yanny}, Brian and {York}, Donald G.},
        title = "{The Three-Dimensional Power Spectrum of Galaxies from the Sloan
                 Digital Sky Survey}",
      journal = {\apj},
         year = 2004,
        month = may,
       volume = {606},
       number = {2},
        pages = {702-740},
          doi = {10.1086/382125},
archivePrefix = {arXiv},
       eprint = {astro-ph/0310725},
 primaryClass = {astro-ph},
}

@ARTICLE{Deng_2007,
       author = {{Deng}, Xin-Fa and {He}, Ji-Zhou and {Jiang}, Peng},
        title = "{The Dependence of Galaxy Luminosity on Environment in a
                 Volume-limited Sample of the Main Galaxy Sample of SDSS DR6}",
      journal = {\apjl},
         year = 2007,
        month = nov,
       volume = {671},
       number = {2},
        pages = {L101-L104},
          doi = {10.1086/524950},
}

@ARTICLE{Norberg_2001,
       author = {{Norberg}, Peder and {Baugh}, Carlton M. and {Hawkins}, Ed and {Maddox}, Steve and {Peacock}, John A. and {Cole}, Shaun and {Frenk}, Carlos S. and {Bland-Hawthorn}, Joss and {Bridges}, Terry and {Cannon}, Russell and {Colless}, Matthew and {Collins}, Chris and {Couch}, Warrick and {Dalton}, Gavin and {De Propris}, Roberto and {Driver}, Simon P. and {Efstathiou}, George and {Ellis}, Richard S. and {Glazebrook}, Karl and {Jackson}, Carole and {Lahav}, Ofer and {Lewis}, Ian and {Lumsden}, Stuart and {Madgwick}, Darren and {Peterson}, Bruce A. and {Sutherland}, Will and {Taylor}, Keith},
        title = "{The 2dF Galaxy Redshift Survey: luminosity dependence of galaxy clustering}",
      journal = {\mnras},
         year = 2001,
        month = nov,
       volume = {328},
       number = {1},
        pages = {64-70},
          doi = {10.1046/j.1365-8711.2001.04839.x},
archivePrefix = {arXiv},
       eprint = {astro-ph/0105500},
 primaryClass = {astro-ph},
       adsurl = {https://ui.adsabs.harvard.edu/abs/2001MNRAS.328...64N}
}

@ARTICLE{Tempel_2014,
       author = {{Tempel}, E. and {Tamm}, A. and {Gramann}, M. and {Tuvikene}, T. and {Liivam{\"a}gi}, L.~J. and {Suhhonenko}, I. and {Kipper}, R. and {Einasto}, M. and {Saar}, E.},
        title = "{Flux- and volume-limited groups/clusters for the SDSS galaxies: catalogues and mass estimation}",
      journal = {\aap},
         year = 2014,
        month = jun,
       volume = {566},
          eid = {A1},
        pages = {A1},
          doi = {10.1051/0004-6361/201423585},
archivePrefix = {arXiv},
       eprint = {1402.1350},
 primaryClass = {astro-ph.CO},
       adsurl = {https://ui.adsabs.harvard.edu/abs/2014A&A...566A...1T}
}

@ARTICLE{Zehavi_2002,
       author = {{Zehavi}, Idit and {Blanton}, Michael R. and {Frieman}, Joshua A. and {Weinberg}, David H. and {Mo}, Houjun J. and {Strauss}, Michael A. and {Anderson}, Scott F. and {Annis}, James and {Bahcall}, Neta A. and {Bernardi}, Mariangela and {Briggs}, John W. and {Brinkmann}, Jon and {Burles}, Scott and {Carey}, Larry and {Castander}, Francisco J. and {Connolly}, Andrew J. and {Csabai}, Istvan and {Dalcanton}, Julianne J. and {Dodelson}, Scott and {Doi}, Mamoru and {Eisenstein}, Daniel and {Evans}, Michael L. and {Finkbeiner}, Douglas P. and {Friedman}, Scott and {Fukugita}, Masataka and {Gunn}, James E. and {Hennessy}, Greg S. and {Hindsley}, Robert B. and {Ivezi{\'c}}, {\v{Z}}eljko and {Kent}, Stephen and {Knapp}, Gillian R. and {Kron}, Richard and {Kunszt}, Peter and {Lamb}, Donald Q. and {Leger}, R. French and {Long}, Daniel C. and {Loveday}, Jon and {Lupton}, Robert H. and {McKay}, Timothy and {Meiksin}, Avery and {Merrelli}, Aronne and {Munn}, Jeffrey A. and {Narayanan}, Vijay and {Newcomb}, Matt and {Nichol}, Robert C. and {Owen}, Russell and {Peoples}, John and {Pope}, Adrian and {Rockosi}, Constance M. and {Schlegel}, David and {Schneider}, Donald P. and {Scoccimarro}, Roman and {Sheth}, Ravi K. and {Siegmund}, Walter and {Smee}, Stephen and {Snir}, Yehuda and {Stebbins}, Albert and {Stoughton}, Christopher and {SubbaRao}, Mark and {Szalay}, Alexander S. and {Szapudi}, Istvan and {Tegmark}, Max and {Tucker}, Douglas L. and {Uomoto}, Alan and {Vanden Berk}, Dan and {Vogeley}, Michael S. and {Waddell}, Patrick and {Yanny}, Brian and {York}, Donald G.},
        title = "{Galaxy Clustering in Early Sloan Digital Sky Survey Redshift Data}",
      journal = {\apj},
         year = 2002,
        month = may,
       volume = {571},
       number = {1},
        pages = {172-190},
          doi = {10.1086/339893},
archivePrefix = {arXiv},
       eprint = {astro-ph/0106476},
 primaryClass = {astro-ph},
       adsurl = {https://ui.adsabs.harvard.edu/abs/2002ApJ...571..172Z}
}

@ARTICLE{Hill_2021,
       author = {{Hill}, Gary J. and {Lee}, Hanshin and {MacQueen}, Phillip J.
                 and {Kelz}, Andreas and {Drory}, Niv and {Vattiat}, Brian L.
                 and {Good}, John M. and {Ramsey}, Jason and {Kriel}, Herman
                 and {Peterson}, Trent and {DePoy}, D.~L. and {Gebhardt}, Karl
                 and {Marshall}, J.~L. and {Tuttle}, Sarah E. and {Bauer}, Svend M.
                 and {Chonis}, Taylor S. and {Fabricius}, Maximilian H.
                 and {Froning}, Cynthia and {H{\"a}user}, Marco and {Indahl}, Briana L.
                 and {Jahn}, Thomas and {Landriau}, Martin and {Leck}, Ron
                 and {Montesano}, Francesco and {Prochaska}, Travis and {Snigula}, Jan M.
                 and {Zeimann}, Greg and {Bryant}, Randy and {Damm}, George
                 and {Fowler}, J.~R. and {Janowiecki}, Steven and {Martin}, Jerry
                 and {Mrozinski}, Emily and {Odewahn}, Stephen and {Rostopchin}, Sergey
                 and {Shetrone}, Matthew and {Spencer}, Renny and {Mentuch Cooper}, Erin
                 and {Armandroff}, Taft and {Bender}, Ralf and {Dalton}, Gavin
                 and {Hopp}, Ulrich and {Komatsu}, Eiichiro and {Nicklas}, Harald
                 and {Ramsey}, Lawrence W. and {Roth}, Martin M. and {Schneider}, Donald P.
                 and {Sneden}, Chris and {Steinmetz}, Matthias},
        title = "{The HETDEX Instrumentation: Hobby--Eberly Telescope Wide-field
                 Upgrade and VIRUS}",
      journal = {\aj},
         year = 2021,
        month = dec,
       volume = {162},
       number = {6},
          eid = {298},
        pages = {298},
          doi = {10.3847/1538-3881/ac2c02},
archivePrefix = {arXiv},
       eprint = {2110.10273},
 primaryClass = {astro-ph.IM},
       adsurl = {https://ui.adsabs.harvard.edu/abs/2021AJ....162..298H}
}

@INPROCEEDINGS{Ramsey_1998,
       author = {{Ramsey}, Lawrence W. and {Adams}, M.~T. and {Barnes}, Thomas G. and {Booth}, John A. and {Cornell}, Mark E. and {Fowler}, James R. and {Gaffney}, Niall I. and {Glaspey}, John W. and {Good}, John M. and {Hill}, Gary J. and {Kelton}, Philip W. and {Krabbendam}, Victor L. and {Long}, L. and {MacQueen}, Phillip J. and {Ray}, Frank B. and {Ricklefs}, Randall L. and {Sage}, J. and {Sebring}, Thomas A. and {Spiesman}, W.~J. and {Steiner}, M.},
        title = "{Early performance and present status of the Hobby-Eberly Telescope}",
    booktitle = {Advanced Technology Optical/IR Telescopes VI},
         year = 1998,
       editor = {{Stepp}, Larry M.},
       series = {Society of Photo-Optical Instrumentation Engineers (SPIE) Conference Series},
       volume = {3352},
        month = aug,
        pages = {34-42},
          doi = {10.1117/12.319287},
       adsurl = {https://ui.adsabs.harvard.edu/abs/1998SPIE.3352...34R}
}

@ARTICLE{Dustin_2023,
       author = {{Davis}, Dustin and {Gebhardt}, Karl and {Cooper}, Erin Mentuch and {Ciardullo}, Robin and {Fabricius}, Maximilian and {Farrow}, Daniel J. and {Feldmeier}, John J. and {Finkelstein}, Steven L. and {Gawiser}, Eric and {Gronwall}, Caryl and {Hill}, Gary J. and {Hopp}, Ulrich and {House}, Lindsay R. and {Jeong}, Donghui and {Kollatschny}, Wolfram and {Komatsu}, Eiichiro and {Landriau}, Martin and {Liu}, Chenxu and {Saito}, Shun and {Tuttle}, Sarah and {Wold}, Isak G.~B. and {Zeimann}, Gregory R. and {Zhang}, Yechi},
        title = "{The HETDEX Survey Emission-line Exploration and Source Classification}",
      journal = {\apj},
         year = 2023,
        month = apr,
       volume = {946},
       number = {2},
          eid = {86},
        pages = {86},
          doi = {10.3847/1538-4357/acb0ca},
archivePrefix = {arXiv},
       eprint = {2301.01799},
 primaryClass = {astro-ph.GA},
       adsurl = {https://ui.adsabs.harvard.edu/abs/2023ApJ...946...86D}
}

@ARTICLE{Gebhardt_2021,
       author = {{Gebhardt}, Karl and {Mentuch Cooper}, Erin and {Ciardullo}, Robin and {Acquaviva}, Viviana and {Bender}, Ralf and {Bowman}, William P. and {Castanheira}, Barbara G. and {Dalton}, Gavin and {Davis}, Dustin and {de Jong}, Roelof S. and {DePoy}, D.~L. and {Devarakonda}, Yaswant and {Dongsheng}, Sun and {Drory}, Niv and {Fabricius}, Maximilian and {Farrow}, Daniel J. and {Feldmeier}, John and {Finkelstein}, Steven L. and {Froning}, Cynthia S. and {Gawiser}, Eric and {Gronwall}, Caryl and {Herold}, Laura and {Hill}, Gary J. and {Hopp}, Ulrich and {House}, Lindsay R. and {Janowiecki}, Steven and {Jarvis}, Matthew and {Jeong}, Donghui and {Jogee}, Shardha and {Kakuma}, Ryota and {Kelz}, Andreas and {Kollatschny}, W. and {Komatsu}, Eiichiro and {Krumpe}, Mirko and {Landriau}, Martin and {Liu}, Chenxu and {Niemeyer}, Maja Lujan and {MacQueen}, Phillip and {Marshall}, Jennifer and {Mawatari}, Ken and {McLinden}, Emily M. and {Mukae}, Shiro and {Nagaraj}, Gautam and {Ono}, Yoshiaki and {Ouchi}, Masami and {Papovich}, Casey and {Sakai}, Nao and {Saito}, Shun and {Schneider}, Donald P. and {Schulze}, Andreas and {Shanmugasundararaj}, Khavvia and {Shetrone}, Matthew and {Sneden}, Chris and {Snigula}, Jan and {Steinmetz}, Matthias and {Thomas}, Benjamin P. and {Thomas}, Brianna and {Tuttle}, Sarah and {Urrutia}, Tanya and {Wisotzki}, Lutz and {Wold}, Isak and {Zeimann}, Gregory and {Zhang}, Yechi},
        title = "{The Hobby-Eberly Telescope Dark Energy Experiment (HETDEX) Survey Design, Reductions, and Detections}",
      journal = {\apj},
         year = 2021,
        month = dec,
       volume = {923},
       number = {2},
          eid = {217},
        pages = {217},
          doi = {10.3847/1538-4357/ac2e03},
archivePrefix = {arXiv},
       eprint = {2110.04298},
 primaryClass = {astro-ph.IM},
       adsurl = {https://ui.adsabs.harvard.edu/abs/2021ApJ...923..217G}
}

@ARTICLE{Mentuch_Cooper_2023,
       author = {{Mentuch Cooper}, Erin and {Gebhardt}, Karl and {Davis}, Dustin and {Farrow}, Daniel J. and {Liu}, Chenxu and {Zeimann}, Gregory and {Ciardullo}, Robin and {Feldmeier}, John J. and {Drory}, Niv and {Jeong}, Donghui and {Benda}, Barbara and {Bowman}, William P. and {Boylan-Kolchin}, Michael and {Ch{\'a}vez Ortiz}, {\'O}scar A. and {Debski}, Maya H. and {Dentler}, Mona and {Fabricius}, Maximilian and {Farooq}, Rameen and {Finkelstein}, Steven L. and {Gawiser}, Eric and {Gronwall}, Caryl and {Hill}, Gary J. and {Hopp}, Ulrich and {House}, Lindsay R. and {Janowiecki}, Steven and {Khoraminezhad}, Hasti and {Kollatschny}, Wolfram and {Komatsu}, Eiichiro and {Landriau}, Martin and {Niemeyer}, Maja Lujan and {Lee}, Hanshin and {MacQueen}, Phillip and {Mawatari}, Ken and {McKay}, Brianna and {Ouchi}, Masami and {Poppe}, Jennifer and {Saito}, Shun and {Schneider}, Donald P. and {Snigula}, Jan and {Thomas}, Benjamin P. and {Tuttle}, Sarah and {Urrutia}, Tanya and {Weiss}, Laurel and {Wisotzki}, Lutz and {Zhang}, Yechi and {HETDEX Collaboration}},
        title = "{HETDEX Public Source Catalog 1: 220 K Sources Including Over 50 K Ly{\ensuremath{\alpha}} Emitters from an Untargeted Wide-area Spectroscopic Survey}",
      journal = {\apj},
         year = 2023,
        month = feb,
       volume = {943},
       number = {2},
          eid = {177},
        pages = {177},
          doi = {10.3847/1538-4357/aca962},
archivePrefix = {arXiv},
       eprint = {2301.01826},
 primaryClass = {astro-ph.GA},
       adsurl = {https://ui.adsabs.harvard.edu/abs/2023ApJ...943..177M}
}

@article{Chiang2013,
    author  = {{Chiang}, Chi-Ting and {Wullstein}, Philipp and {Jeong}, Donghui
               and {Komatsu}, Eiichiro and {Blanc}, Guillermo A. and {Ciardullo}, Robin
               and {Drory}, Niv and {Fabricius}, Maximilian and {Finkelstein}, Steven
               and {Gebhardt}, Karl and {Gronwall}, Caryl and {Hagen}, Alex
               and {Hill}, Gary J. and {Jee}, Inh and {Jogee}, Shardha
               and {Landriau}, Martin and {Mentuch Cooper}, Erin
               and {Schneider}, Donald P. and {Tuttle}, Sarah},
    title   = {Galaxy redshift surveys with sparse sampling},
    journal = {\jcap},
    year    = 2013,
    volume  = {2013},
    number  = {12},
    pages   = {030},
    doi     = {10.1088/1475-7516/2013/12/030}
}

@ARTICLE{Nguyen_2023,
       author = {{Nguyen}, Nhat-Minh and {Huterer}, Dragan and {Wen}, Yuewei},
        title = "{Evidence for Suppression of Structure Growth in the Concordance
                 Cosmological Model}",
      journal = {\prl},
         year = 2023,
        month = sep,
       volume = {131},
       number = {11},
        pages = {111001},
          doi = {10.1103/PhysRevLett.131.111001},
archivePrefix = {arXiv},
       eprint = {2302.14435},
 primaryClass = {astro-ph.CO},
       adsurl = {https://ui.adsabs.harvard.edu/abs/2023PhRvL.131k1001N}
}

@ARTICLE{Planck2020cosmo,
       author = {{Planck Collaboration} and {Aghanim}, N. and {Akrami}, Y. and {Ashdown}, M. and {Aumont}, J. and {Baccigalupi}, C. and {Ballardini}, M. and {Banday}, A.~J. and {Barreiro}, R.~B. and {Bartolo}, N. and {Basak}, S. and {Battye}, R. and {Benabed}, K. and {Bernard}, J. -P. and {Bersanelli}, M. and {Bielewicz}, P. and {Bock}, J.~J. and {Bond}, J.~R. and {Borrill}, J. and {Bouchet}, F.~R. and {Boulanger}, F. and {Bucher}, M. and {Burigana}, C. and {Butler}, R.~C. and {Calabrese}, E. and {Cardoso}, J. -F. and {Carron}, J. and {Challinor}, A. and {Chiang}, H.~C. and {Chluba}, J. and {Colombo}, L.~P.~L. and {Combet}, C. and {Contreras}, D. and {Crill}, B.~P. and {Cuttaia}, F. and {de Bernardis}, P. and {de Zotti}, G. and {Delabrouille}, J. and {Delouis}, J. -M. and {Di Valentino}, E. and {Diego}, J.~M. and {Dor{\'e}}, O. and {Douspis}, M. and {Ducout}, A. and {Dupac}, X. and {Dusini}, S. and {Efstathiou}, G. and {Elsner}, F. and {En{\ss}lin}, T.~A. and {Eriksen}, H.~K. and {Fantaye}, Y. and {Farhang}, M. and {Fergusson}, J. and {Fernandez-Cobos}, R. and {Finelli}, F. and {Forastieri}, F. and {Frailis}, M. and {Fraisse}, A.~A. and {Franceschi}, E. and {Frolov}, A. and {Galeotta}, S. and {Galli}, S. and {Ganga}, K. and {G{\'e}nova-Santos}, R.~T. and {Gerbino}, M. and {Ghosh}, T. and {Gonz{\'a}lez-Nuevo}, J. and {G{\'o}rski}, K.~M. and {Gratton}, S. and {Gruppuso}, A. and {Gudmundsson}, J.~E. and {Hamann}, J. and {Handley}, W. and {Hansen}, F.~K. and {Herranz}, D. and {Hildebrandt}, S.~R. and {Hivon}, E. and {Huang}, Z. and {Jaffe}, A.~H. and {Jones}, W.~C. and {Karakci}, A. and {Keih{\"a}nen}, E. and {Keskitalo}, R. and {Kiiveri}, K. and {Kim}, J. and {Kisner}, T.~S. and {Knox}, L. and {Krachmalnicoff}, N. and {Kunz}, M. and {Kurki-Suonio}, H. and {Lagache}, G. and {Lamarre}, J. -M. and {Lasenby}, A. and {Lattanzi}, M. and {Lawrence}, C.~R. and {Le Jeune}, M. and {Lemos}, P. and {Lesgourgues}, J. and {Levrier}, F. and {Lewis}, A. and {Liguori}, M. and {Lilje}, P.~B. and {Lilley}, M. and {Lindholm}, V. and {L{\'o}pez-Caniego}, M. and {Lubin}, P.~M. and {Ma}, Y. -Z. and {Mac{\'\i}as-P{\'e}rez}, J.~F. and {Maggio}, G. and {Maino}, D. and {Mandolesi}, N. and {Mangilli}, A. and {Marcos-Caballero}, A. and {Maris}, M. and {Martin}, P.~G. and {Martinelli}, M. and {Mart{\'\i}nez-Gonz{\'a}lez}, E. and {Matarrese}, S. and {Mauri}, N. and {McEwen}, J.~D. and {Meinhold}, P.~R. and {Melchiorri}, A. and {Mennella}, A. and {Migliaccio}, M. and {Millea}, M. and {Mitra}, S. and {Miville-Desch{\^e}nes}, M. -A. and {Molinari}, D. and {Montier}, L. and {Morgante}, G. and {Moss}, A. and {Natoli}, P. and {N{\o}rgaard-Nielsen}, H.~U. and {Pagano}, L. and {Paoletti}, D. and {Partridge}, B. and {Patanchon}, G. and {Peiris}, H.~V. and {Perrotta}, F. and {Pettorino}, V. and {Piacentini}, F. and {Polastri}, L. and {Polenta}, G. and {Puget}, J. -L. and {Rachen}, J.~P. and {Reinecke}, M. and {Remazeilles}, M. and {Renzi}, A. and {Rocha}, G. and {Rosset}, C. and {Roudier}, G. and {Rubi{\~n}o-Mart{\'\i}n}, J.~A. and {Ruiz-Granados}, B. and {Salvati}, L. and {Sandri}, M. and {Savelainen}, M. and {Scott}, D. and {Shellard}, E.~P.~S. and {Sirignano}, C. and {Sirri}, G. and {Spencer}, L.~D. and {Sunyaev}, R. and {Suur-Uski}, A. -S. and {Tauber}, J.~A. and {Tavagnacco}, D. and {Tenti}, M. and {Toffolatti}, L. and {Tomasi}, M. and {Trombetti}, T. and {Valenziano}, L. and {Valiviita}, J. and {Van Tent}, B. and {Vibert}, L. and {Vielva}, P. and {Villa}, F. and {Vittorio}, N. and {Wandelt}, B.~D. and {Wehus}, I.~K. and {White}, M. and {White}, S.~D.~M. and {Zacchei}, A. and {Zonca}, A.},
        title = "{Planck 2018 results. VI. Cosmological parameters}",
     journal = {\aap},
         year = 2020,
        month = sep,
       volume = {641},
          eid = {A6},
        pages = {A6},
          doi = {10.1051/0004-6361/201833910},
archivePrefix = {arXiv},
       eprint = {1807.06209},
 primaryClass = {astro-ph.CO},
       adsurl = {https://ui.adsabs.harvard.edu/abs/2020A&A...641A...6P}
}

@ARTICLE{ACT2025,
       author = {{Qu}, Frank J. and {Hang}, Qianjun and {Farren}, Gerrit and {Bolliet}, Boris and {Aguilar}, Jessica Nicole and {Ahlen}, Steven and {Alam}, Shadab and {Brooks}, David and {Cai}, Yan-Chuan and {Calabrese}, Erminia and {Claybaugh}, Todd and {de la Macorra}, Axel and {Devlin}, Mark J. and {Doel}, Peter and {Embil-Villagra}, Carmen and {Ferraro}, Simone and {Font-Ribera}, Andreu and {Forero-Romero}, Jaime E. and {Gazta{\~n}aga}, Enrique and {Gluscevic}, Vera and {Gontcho}, Satya Gontcho A. and {Gutierrez}, Gaston and {Howlett}, Cullan and {Kehoe}, Robert and {Kim}, Joshua and {Kremin}, Anthony and {Lambert}, Andrew and {Landriau}, Martin and {Le Guillou}, Laurent and {Levi}, Michael and {Louis}, Thibaut and {Meisner}, Aaron and {Miquel}, Ramon and {Moustakas}, John and {Newman}, Jeffrey A. and {Niz}, Gustavo and {Peacock}, John A. and {Percival}, Will and {Poppett}, Claire and {Prada}, Francisco and {P{\'e}rez-R{\`a}fols}, Ignasi and {Rossi}, Graziano and {Sanchez}, Eusebio and {Schlegel}, David and {Sehgal}, Neelima and {Shaikh}, Shabbir and {Sherwin}, Blake and {Sif{\'o}n}, Crist{\'o}bal and {Schubnell}, Michael and {Sprayberry}, David and {Tarl{\'e}}, Gregory and {Weaver}, Benjamin Alan and {Wollack}, Edward J. and {Zou}, Hu},
        title = "{Atacama Cosmology Telescope DR6 and DESI: Structure growth measurements from the cross-correlation of DESI legacy imaging galaxies and CMB lensing from ACT DR6 and Planck PR4}",
      journal = {\prd},
         year = 2025,
        month = may,
       volume = {111},
       number = {10},
          eid = {103503},
        pages = {103503},
          doi = {10.1103/PhysRevD.111.103503},
archivePrefix = {arXiv},
       eprint = {2410.10808},
 primaryClass = {astro-ph.CO},
       adsurl = {https://ui.adsabs.harvard.edu/abs/2025PhRvD.111j3503Q}
}

@ARTICLE{DESIDR2_2025,
       author = {{Abdul Karim}, M. and {Aguilar}, J. and {Ahlen}, S. and {Alam}, S. and {Allen}, L. and {Allende Prieto}, C. and {Alves}, O. and {Anand}, A. and {Andrade}, U. and {Armengaud}, E. and {Aviles}, A. and {Bailey}, S. and {Baltay}, C. and {Bansal}, P. and {Bault}, A. and {Behera}, J. and {BenZvi}, S. and {Bianchi}, D. and {Blake}, C. and {Brieden}, S. and {Brodzeller}, A. and {Brooks}, D. and {Buckley-Geer}, E. and {Burtin}, E. and {Calderon}, R. and {Canning}, R. and {Rosell}, A. Carnero and {Carrilho}, P. and {Casas}, L. and {Castander}, F.~J. and {Charles}, M. and {Chaussidon}, E. and {Chaves-Montero}, J. and {Chebat}, D. and {Chen}, X. and {Claybaugh}, T. and {Cole}, S. and {Cooper}, A.~P. and {Cuceu}, A. and {Dawson}, K.~S. and {de la Macorra}, A. and {de Mattia}, A. and {Deiosso}, N. and {Della Costa}, J. and {Demina}, R. and {Dey}, A. and {Dey}, B. and {Ding}, Z. and {Doel}, P. and {Edelstein}, J. and {Eisenstein}, D.~J. and {Elbers}, W. and {Fagrelius}, P. and {Fanning}, K. and {Fern{\'a}ndez-Garc{\'\i}a}, E. and {Ferraro}, S. and {Font-Ribera}, A. and {Forero-Romero}, J.~E. and {Frenk}, C.~S. and {Garcia-Quintero}, C. and {Garrison}, L.~H. and {Gazta{\~n}aga}, E. and {Gil-Mar{\'\i}n}, H. and {Gontcho A Gontcho}, S. and {Gonzalez}, D. and {Gonzalez-Morales}, A.~X. and {Gordon}, C. and {Green}, D. and {Gutierrez}, G. and {Guy}, J. and {Hadzhiyska}, B. and {Hahn}, C. and {He}, S. and {Herbold}, M. and {Herrera-Alcantar}, H.~K. and {Ho}, M.-F. and {Honscheid}, K. and {Howlett}, C. and {Huterer}, D. and {Ishak}, M. and {Juneau}, S. and {Kamble}, N.~V. and {Kara{\c{c}}ayl{\i}}, N.~G. and {Kehoe}, R. and {Kent}, S. and {Kim}, A.~G. and {Kirkby}, D. and {Kisner}, T. and {Koposov}, S.~E. and {Kremin}, A. and {Krolewski}, A. and {Lahav}, O. and {Lamman}, C. and {Landriau}, M. and {Lang}, D. and {Lasker}, J. and {Le Goff}, J.~M. and {Le Guillou}, L. and {Leauthaud}, A. and {Levi}, M.~E. and {Li}, Q. and {Li}, T.~S. and {Lodha}, K. and {Lokken}, M. and {Lozano-Rodr{\'\i}guez}, F. and {Magneville}, C. and {Manera}, M. and {Martini}, P. and {Matthewson}, W.~L. and {Meisner}, A. and {Mena-Fern{\'a}ndez}, J. and {Menegas}, A. and {Mergulh{\~a}o}, T. and {Miquel}, R. and {Moustakas}, J. and {Mu{\~n}oz-Guti{\'e}rrez}, A. and {Mu{\~n}oz-Santos}, D. and {Myers}, A.~D. and {Nadathur}, S. and {Naidoo}, K. and {Napolitano}, L. and {Newman}, J.~A. and {Niz}, G. and {Noriega}, H.~E. and {Paillas}, E. and {Palanque-Delabrouille}, N. and {Pan}, J. and {Peacock}, J.~A. and {Pellejero Ibanez}, M. and {Percival}, W.~J. and {P{\'e}rez-Fern{\'a}ndez}, A. and {P{\'e}rez-R{\`a}fols}, I. and {Pieri}, M.~M. and {Poppett}, C. and {Prada}, F. and {Rabinowitz}, D. and {Raichoor}, A. and {Ram{\'\i}rez-P{\'e}rez}, C. and {Rashkovetskyi}, M. and {Ravoux}, C. and {Rich}, J. and {Rocher}, A. and {Rockosi}, C. and {Rohlf}, J. and {Rom{\'a}n-Herrera}, J.~O. and {Ross}, A.~J. and {Rossi}, G. and {Ruggeri}, R. and {Ruhlmann-Kleider}, V. and {Samushia}, L. and {Sanchez}, E. and {Sanders}, N. and {Schlegel}, D. and {Schubnell}, M. and {Seo}, H. and {Shafieloo}, A. and {Sharples}, R. and {Silber}, J. and {Sinigaglia}, F. and {Sprayberry}, D. and {Tan}, T. and {Tarl{\'e}}, G. and {Taylor}, P. and {Turner}, W. and {Ure{\~n}a-L{\'o}pez}, L.~A. and {Vaisakh}, R. and {Valdes}, F. and {Valogiannis}, G. and {Vargas-Maga{\~n}a}, M. and {Verde}, L. and {Walther}, M. and {Weaver}, B.~A. and {Weinberg}, D.~H. and {White}, M. and {Wolfson}, M. and {Y{\`e}che}, C. and {Yu}, J. and {Zaborowski}, E.~A. and {Zarrouk}, P. and {Zhai}, Z. and {Zhang}, H. and {Zhao}, C. and {Zhao}, G.~B. and {Zhou}, R. and {Zou}, H. and {DESI Collaboration}},
        title = "{DESI DR2 results. II. Measurements of baryon acoustic oscillations and cosmological constraints}",
      journal = {\prd},
         year = 2025,
        month = oct,
       volume = {112},
       number = {8},
          eid = {083515},
        pages = {083515},
          doi = {10.1103/tr6y-kpc6},
archivePrefix = {arXiv},
       eprint = {2503.14738},
 primaryClass = {astro-ph.CO},
       adsurl = {https://ui.adsabs.harvard.edu/abs/2025PhRvD.112h3515A}
}

@article{Desjacques2018,
    author  = {{Desjacques}, Vincent and {Jeong}, Donghui and {Schmidt}, Fabian},
    title   = {Large-Scale Galaxy Bias},
    journal = {Physics Reports},
    year    = 2018,
    volume  = {733},
    pages   = {1--193},
    doi     = {10.1016/j.physrep.2017.12.002}
}

@article{Philcox2022,
    author  = {{Philcox}, Oliver H.~E. and {Ivanov}, Mikhail M.},
    title   = {BOSS DR12 full-shape cosmology: {$\Lambda$CDM} constraints from
               the large-scale galaxy power spectrum and bispectrum monopole},
    journal = {\prd},
    year    = 2022,
    volume  = {105},
    pages   = {043517},
    doi     = {10.1103/PhysRevD.105.043517}
}

@article{Zheng2005,
    author  = {{Zheng}, Zheng and {Berlind}, Andreas A. and {Weinberg}, David H.
               and {Benson}, Andrew J. and {Baugh}, Carlton M. and {Cole}, Shaun
               and {Dav{\'e}}, Romeel and {Frenk}, Carlos S. and {Katz}, Neal
               and {Lacey}, Cedric G.},
    title   = {Theoretical Models of the Halo Occupation Distribution:
               Separating the Physics of Galaxy Formation},
    journal = {\apj},
    year    = 2005,
    volume  = {633},
    pages   = {791--809},
    doi     = {10.1086/466510}
}

@article{Behroozi2019,
    author  = {{Behroozi}, Peter and {Wechsler}, Risa H. and {Hearin}, Andrew P.
               and {Conroy}, Charlie},
    title   = {UniverseMachine: The Correlation between Galaxy Growth and
               Dark Matter Halo Assembly from z = 0--10},
    journal = {\mnras},
    year    = 2019,
    volume  = {488},
    pages   = {3143--3194},
    doi     = {10.1093/mnras/stz1182}
}

@article{Springel2005,
    author  = {{Springel}, Volker and {White}, Simon D.~M. and {Jenkins}, Adrian
               and {Frenk}, Carlos S. and {Yoshida}, Naoki and {Gao}, Liang
               and {Navarro}, Julio and {Thacker}, Robert and {Croton}, Darren
               and {Helly}, John and {Peacock}, John A. and {Cole}, Shaun
               and {Thomas}, Peter and {Couchman}, Hugh and {Evrard}, August
               and {Colberg}, J{\"o}rg and {Pearce}, Frazer},
    title   = {Simulations of the formation, evolution and clustering
               of galaxies and quasars},
    journal = {\nat},
    year    = 2005,
    volume  = {435},
    pages   = {629--636},
    doi     = {10.1038/nature03597}
}

@article{Pakmor2023,
    author  = {{Pakmor}, R{\"u}diger and {Springel}, Volker and {Coles}, Jon P.
               and {Guillet}, Thomas and {Bauer}, Andreas and {Delgado}, Ana Maria
               and {Ferlito}, Fulvio and {Hadzhiyska}, Boryana and {Hernandez-Aguayo},
               Cesar and {Kannan}, Rahul and {Kimmig}, Leonard and {Barrera}, Monica
               and {Bose}, Sownak and {Contreras}, Sergio and {Davies}, Christopher T.
               and {Frenk}, Carlos S. and {Garaldi}, Enrico and {Hernquist}, Lars
               and {Hours}, Borja and {Jasche}, Jens and {Kannan}, Rahul
               and {Lazeyras}, Tristan and {Littek}, Christian and {Lobato}, Roisin
               and {Marinacci}, Federico and {Mc{C}ullagh}, Niall and {Millea}, Meir
               and {Natarajan}, Priyamvada and {Ni}, Yueying and {Pillepich}, Annalisa
               and {Pinon}, Martine and {Rodriguez-Gomez}, Vicente and {Rosdahl}, Joakim
               and {Wanderman}, David and {Weinberger}, Reinhard and {White}, Simon D.~M.},
    title   = {The MillenniumTNG Project: the hydrodynamical full physics simulation},
    journal = {\mnras},
    year    = 2023,
    volume  = {524},
    pages   = {2539--2560},
    doi     = {10.1093/mnras/stad2027}
}

@ARTICLE{Uchuu_dr1,
       author = {{Ishiyama}, Tomoaki and {Prada}, Francisco and {Klypin}, Anatoly A. and {Sinha}, Manodeep and {Metcalf}, R. Benton and {Jullo}, Eric and {Altieri}, Bruno and {Cora}, Sof{\'\i}a A. and {Croton}, Darren and {de la Torre}, Sylvain and {Mill{\'a}n-Calero}, David E. and {Oogi}, Taira and {Ruedas}, Jos{\'e} and {Vega-Mart{\'\i}nez}, Cristian A.},
        title = "{The Uchuu simulations: Data Release 1 and dark matter halo concentrations}",
      journal = {\mnras},
         year = 2021,
        month = sep,
       volume = {506},
       number = {3},
        pages = {4210-4231},
          doi = {10.1093/mnras/stab1755},
archivePrefix = {arXiv},
       eprint = {2007.14720},
 primaryClass = {astro-ph.CO},
       adsurl = {https://ui.adsabs.harvard.edu/abs/2021MNRAS.506.4210I}
}

@ARTICLE{Uchuu_nu2GCC,
       author = {{Oogi}, Taira and {Ishiyama}, Tomoaki and {Prada}, Francisco and {Sinha}, Manodeep and {Croton}, Darren and {Cora}, Sof{\'\i}a A. and {Jullo}, Eric and {Klypin}, Anatoly A. and {Nagashima}, Masahiro and {L{\'o}pez Cacheiro}, J. and {Ruedas}, Jos{\'e} and {Kobayashi}, Masakazu A.~R. and {Makiya}, Ryu},
        title = "{Uchuu-{\ensuremath{\nu}}$^{2}$GC galaxies and AGN: cosmic variance forecasts of high-redshift AGN for JWST, Euclid, and LSST}",
      journal = {\mnras},
         year = 2023,
        month = nov,
       volume = {525},
       number = {3},
        pages = {3879-3895},
          doi = {10.1093/mnras/stad2401},
archivePrefix = {arXiv},
       eprint = {2207.14689},
 primaryClass = {astro-ph.GA},
       adsurl = {https://ui.adsabs.harvard.edu/abs/2023MNRAS.525.3879O}
}

@ARTICLE{Uchuu_SDSS,
       author = {{Dong-P{\'a}ez}, C.~A. and {Smith}, A. and {Szewciw}, A.~O. and {Ereza}, J. and {Abdullah}, M.~H. and {Hern{\'a}ndez-Aguayo}, C. and {Trusov}, S. and {Prada}, F. and {Klypin}, A. and {Ishiyama}, T. and {Berlind}, A. and {Zarrouk}, P. and {L{\'o}pez Cacheiro}, J. and {Ruedas}, J.},
        title = "{The Uchuu-SDSS galaxy light-cones: a clustering, redshift space distortion and baryonic acoustic oscillation study}",
      journal = {\mnras},
         year = 2024,
        month = mar,
       volume = {528},
       number = {4},
        pages = {7236-7255},
          doi = {10.1093/mnras/stae062},
archivePrefix = {arXiv},
       eprint = {2208.00540},
 primaryClass = {astro-ph.CO},
       adsurl = {https://ui.adsabs.harvard.edu/abs/2024MNRAS.528.7236D}
}

@ARTICLE{Uchuu_UM,
       author = {{Aung}, Han and {Nagai}, Daisuke and {Klypin}, Anatoly and {Behroozi}, Peter and {Abdullah}, Mohamed H. and {Ishiyama}, Tomoaki and {Prada}, Francisco and {P{\'e}rez}, Enrique and {L{\'o}pez Cacheiro}, Javier and {Ruedas}, Jos{\'e}},
        title = "{The Uchuu-universe machine data set: galaxies in and around clusters}",
      journal = {\mnras},
         year = 2023,
        month = feb,
       volume = {519},
       number = {2},
        pages = {1648-1656},
          doi = {10.1093/mnras/stac3514},
archivePrefix = {arXiv},
       eprint = {2209.12918},
 primaryClass = {astro-ph.GA},
       adsurl = {https://ui.adsabs.harvard.edu/abs/2023MNRAS.519.1648A}
}

@ARTICLE{Uchuu_RMG,
       author = {{Prada}, Francisco and {Behroozi}, Peter and {Ishiyama}, Tomoaki and {Klypin}, Anatoly and {P{\'e}rez}, Enrique},
        title = "{Confirmation of the standard cosmological model from red massive galaxies $\sim600$ Myr after the Big Bang}",
      journal = {arXiv e-prints},
         year = 2023,
        month = apr,
          eid = {arXiv:2304.11911},
        pages = {arXiv:2304.11911},
          doi = {10.48550/arXiv.2304.11911},
archivePrefix = {arXiv},
       eprint = {2304.11911},
 primaryClass = {astro-ph.GA},
       adsurl = {https://ui.adsabs.harvard.edu/abs/2023arXiv230411911P}
}

@ARTICLE{Geach2012,
       author = {{Geach}, J.~E. and {Sobral}, D. and {Hickox}, R.~C. and {Wake}, D.~A. and {Smail}, Ian and {Best}, P.~N. and {Baugh}, C.~M. and {Stott}, J.~P.},
        title = "{The clustering of H{\ensuremath{\alpha}} emitters at z=2.23 from HiZELS}",
      journal = {\mnras},
         year = 2012,
        month = oct,
       volume = {426},
       number = {1},
        pages = {679-689},
          doi = {10.1111/j.1365-2966.2012.21725.x},
archivePrefix = {arXiv},
       eprint = {1206.4052},
 primaryClass = {astro-ph.CO},
       adsurl = {https://ui.adsabs.harvard.edu/abs/2012MNRAS.426..679G}
}

@ARTICLE{Sellentin2016,
       author = {{Sellentin}, Elena and {Heavens}, Alan F.},
        title = "{Parameter inference with estimated covariance matrices}",
      journal = {\mnras},
         year = 2016,
        month = feb,
       volume = {456},
       number = {1},
        pages = {L132-L136},
          doi = {10.1093/mnrasl/slv190},
archivePrefix = {arXiv},
       eprint = {1511.05969},
 primaryClass = {astro-ph.CO},
}

@book{Anderson2003,
    author    = {{Anderson}, T.~W.},
    title     = {An Introduction to Multivariate Statistical Analysis},
    year      = 2003,
    edition   = {3rd},
    publisher = {Wiley-Interscience},
    address   = {Hoboken, NJ}
}

@ARTICLE{Mo1996,
       author = {{Mo}, H.~J. and {White}, S.~D.~M.},
        title = "{An analytic model for the spatial clustering of dark matter haloes}",
      journal = {\mnras},
         year = 1996,
        month = sep,
       volume = {282},
       number = {2},
        pages = {347-361},
          doi = {10.1093/mnras/282.2.347},
archivePrefix = {arXiv},
       eprint = {astro-ph/9512127},
 primaryClass = {astro-ph},
       adsurl = {https://ui.adsabs.harvard.edu/abs/1996MNRAS.282..347M}
}

@ARTICLE{Tinker2010,
       author = {{Tinker}, Jeremy L. and {Robertson}, Brant E. and {Kravtsov}, Andrey V. and {Klypin}, Anatoly and {Warren}, Michael S. and {Yepes}, Gustavo and {Gottl{\"o}ber}, Stefan},
        title = "{The Large-scale Bias of Dark Matter Halos: Numerical Calibration and Model Tests}",
      journal = {\apj},
         year = 2010,
        month = dec,
       volume = {724},
       number = {2},
        pages = {878-886},
          doi = {10.1088/0004-637X/724/2/878},
archivePrefix = {arXiv},
       eprint = {1001.3162},
 primaryClass = {astro-ph.CO},
       adsurl = {https://ui.adsabs.harvard.edu/abs/2010ApJ...724..878T}
}

@ARTICLE{Vale2004,
       author = {{Vale}, A. and {Ostriker}, J.~P.},
        title = "{Linking halo mass to galaxy luminosity}",
      journal = {\mnras},
         year = 2004,
        month = sep,
       volume = {353},
       number = {1},
        pages = {189-200},
          doi = {10.1111/j.1365-2966.2004.08059.x},
archivePrefix = {arXiv},
       eprint = {astro-ph/0402500},
 primaryClass = {astro-ph},
       adsurl = {https://ui.adsabs.harvard.edu/abs/2004MNRAS.353..189V}
}

@ARTICLE{Behroozi2010,
       author = {{Behroozi}, Peter S. and {Conroy}, Charlie and {Wechsler}, Risa H.},
        title = "{A Comprehensive Analysis of Uncertainties Affecting the Stellar Mass-Halo Mass Relation for 0 < z < 4}",
      journal = {\apj},
         year = 2010,
        month = jul,
       volume = {717},
       number = {1},
        pages = {379-403},
          doi = {10.1088/0004-637X/717/1/379},
archivePrefix = {arXiv},
       eprint = {1001.0015},
 primaryClass = {astro-ph.CO},
       adsurl = {https://ui.adsabs.harvard.edu/abs/2010ApJ...717..379B}
}

@ARTICLE{Percival2022,
       author = {{Percival}, Will J. and {Friedrich}, Oliver and {Sellentin}, Elena and {Heavens}, Alan},
        title = "{Matching Bayesian and frequentist coverage probabilities when using an approximate data covariance matrix}",
      journal = {\mnras},
         year = 2022,
        month = mar,
       volume = {510},
       number = {3},
        pages = {3207-3221},
          doi = {10.1093/mnras/stab3540},
archivePrefix = {arXiv},
       eprint = {2108.10402},
 primaryClass = {astro-ph.IM},
       adsurl = {https://ui.adsabs.harvard.edu/abs/2022MNRAS.510.3207P}
}

@ARTICLE{Jain1997,
       author = {{Jain}, Bhuvnesh and {Seljak}, Uro{\v{s}}},
        title = "{Cosmological Model Predictions for Weak Lensing: Linear and Nonlinear Regimes}",
      journal = {\apj},
         year = 1997,
        month = jul,
       volume = {484},
       number = {2},
        pages = {560-573},
          doi = {10.1086/304372},
archivePrefix = {arXiv},
       eprint = {astro-ph/9611077},
 primaryClass = {astro-ph},
       adsurl = {https://ui.adsabs.harvard.edu/abs/1997ApJ...484..560J}
}

@BOOK{Hockney1988,
       author = {{Hockney}, R.~W. and {Eastwood}, J.~W.},
        title = "{Computer simulation using particles}",
         year = 1988,
         publisher = "CRC Press",
       adsurl = {https://ui.adsabs.harvard.edu/abs/1988csup.book.....H}
}

@ARTICLE{Jimenez2020,
       author = {{Jim{\'e}nez}, Esteban and {Padilla}, Nelson and {Contreras}, Sergio and {Zehavi}, Idit and {Baugh}, Carlton M. and {Orsi}, {\'A}lvaro},
        title = "{The assembly bias of emission-line galaxies}",
      journal = {\mnras},
         year = 2021,
        month = sep,
       volume = {506},
       number = {3},
        pages = {3155-3168},
          doi = {10.1093/mnras/stab1819},
archivePrefix = {arXiv},
       eprint = {2010.08500},
 primaryClass = {astro-ph.GA},
       adsurl = {https://ui.adsabs.harvard.edu/abs/2021MNRAS.506.3155J}
}

@ARTICLE{Kajisawa2013,
       author = {{Kajisawa}, M. and {Shioya}, Y. and {Aida}, Y. and {Ideue}, Y. and {Taniguchi}, Y. and {Nagao}, T. and {Murayama}, T. and {Matsubayashi}, K. and {Riguccini}, L.},
        title = "{Environmental Effects on Star Formation Activity at z \raisebox{-0.5ex}\textasciitilde 0.9 in the COSMOS Field}",
      journal = {\apj},
         year = 2013,
        month = may,
       volume = {768},
       number = {1},
          eid = {51},
        pages = {51},
          doi = {10.1088/0004-637X/768/1/51},
archivePrefix = {arXiv},
       eprint = {1302.7148},
 primaryClass = {astro-ph.CO},
       adsurl = {https://ui.adsabs.harvard.edu/abs/2013ApJ...768...51K}
}
\bibliographystyle{aasjournalv7}

\appendix
\section{Effect of \texorpdfstring{$N_\mathrm{overlap}$}{Noverlap} weight}
\label{appendix:N_overlap}

In this appendix section, we check the effect of the weight, $w(\mathbf{r}) = N_\mathrm{overlap}^{-1}$, on the power spectrum measurements described in Section~\ref{subsec:pk_estimator}. 

\begin{figure}
    \centering 
    \includegraphics[width=0.48\columnwidth]{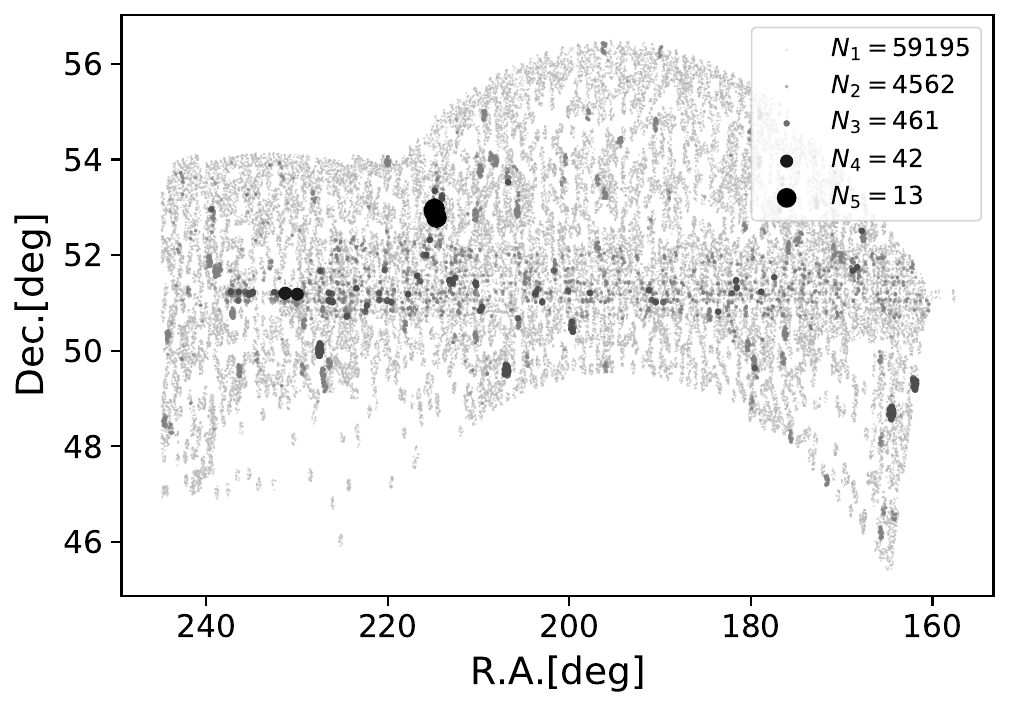}
    \includegraphics[width=0.48\columnwidth]{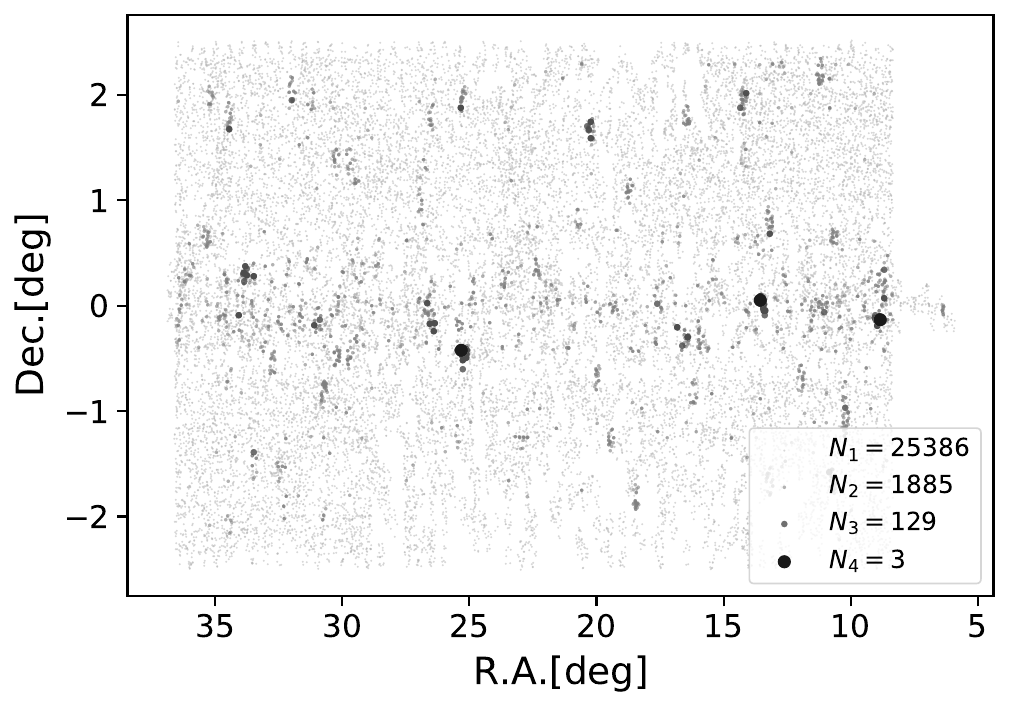}
    \caption{Sky distribution of $N_\mathrm{overlap}$ for Spring Bin 4 and Fall Bin 4 in the panel of right ascension and declination, where the size of the circles reflects the $N_\mathrm{overlap}$ value. The total number of galaxies corresponding to each $N_\mathrm{overlap}$, $N_{N_\mathrm{overlap}}$, is indicated in the legend.}
    \label{fig:plot_Noverlap_distribution}
\end{figure}

Figure~\ref{fig:plot_Noverlap_distribution} shows the sky distribution of $N_\mathrm{overlap}$ for Spring Bin 4 and Fall Bin 4, with the size of the circles reflecting the $N_\mathrm{overlap}$ value. We adopt Bin 4 from both fields as representative of all bins for this check, since these bins span the largest volume and contain the most galaxies. We find that the distribution of $N_\mathrm{overlap}$ is fairly random across the sky.

\begin{figure}
    \centering 
    \includegraphics[width=0.48\columnwidth]{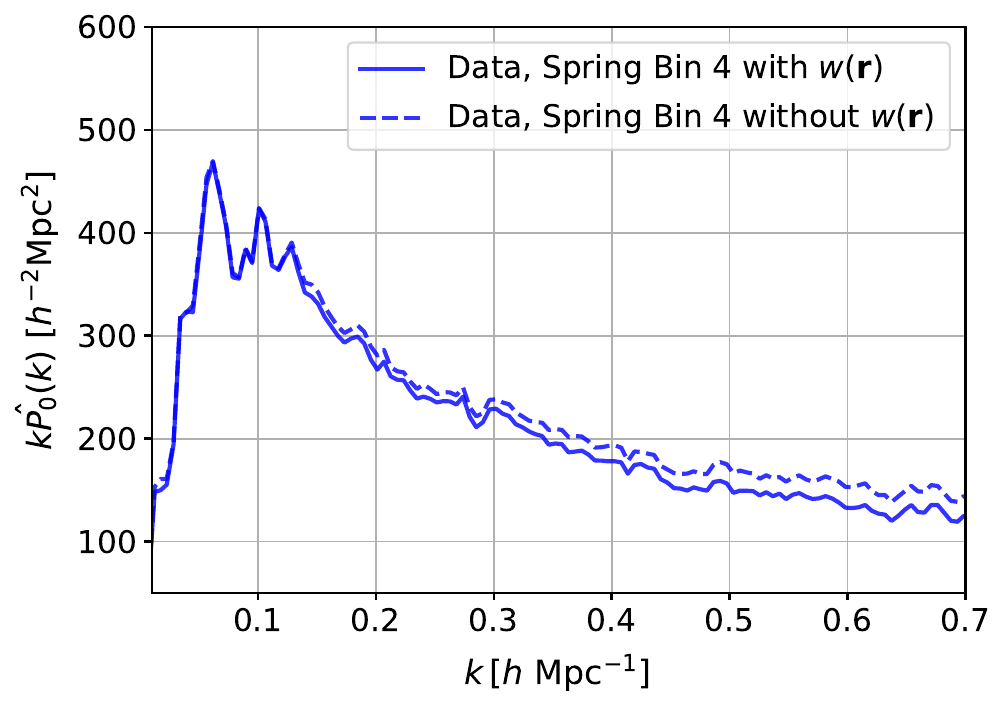}
    \includegraphics[width=0.48\columnwidth]{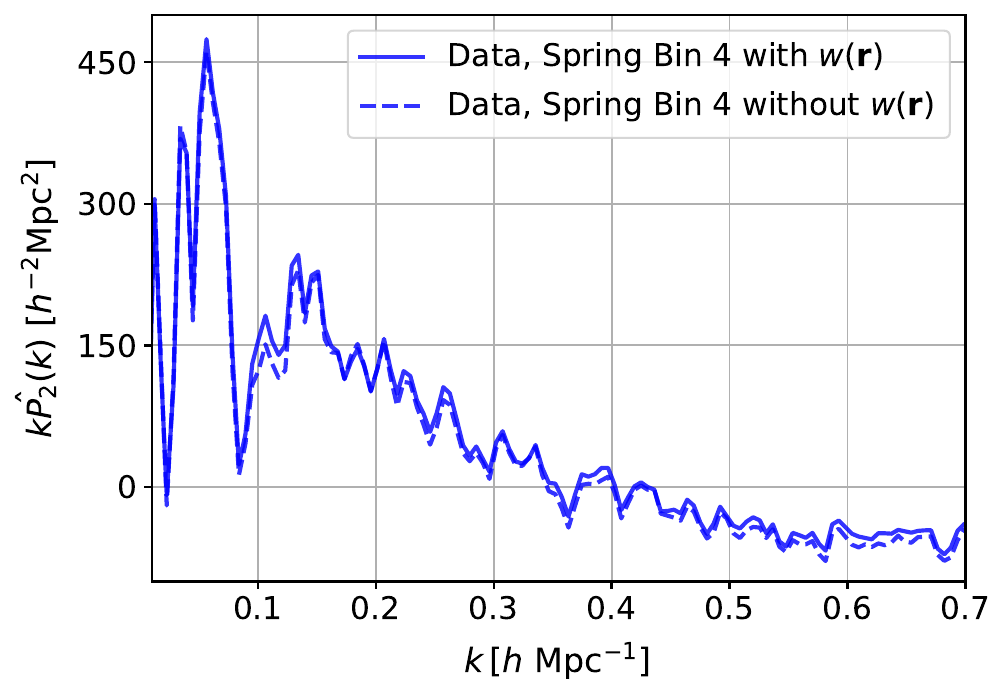}
    \includegraphics[width=0.48\columnwidth]{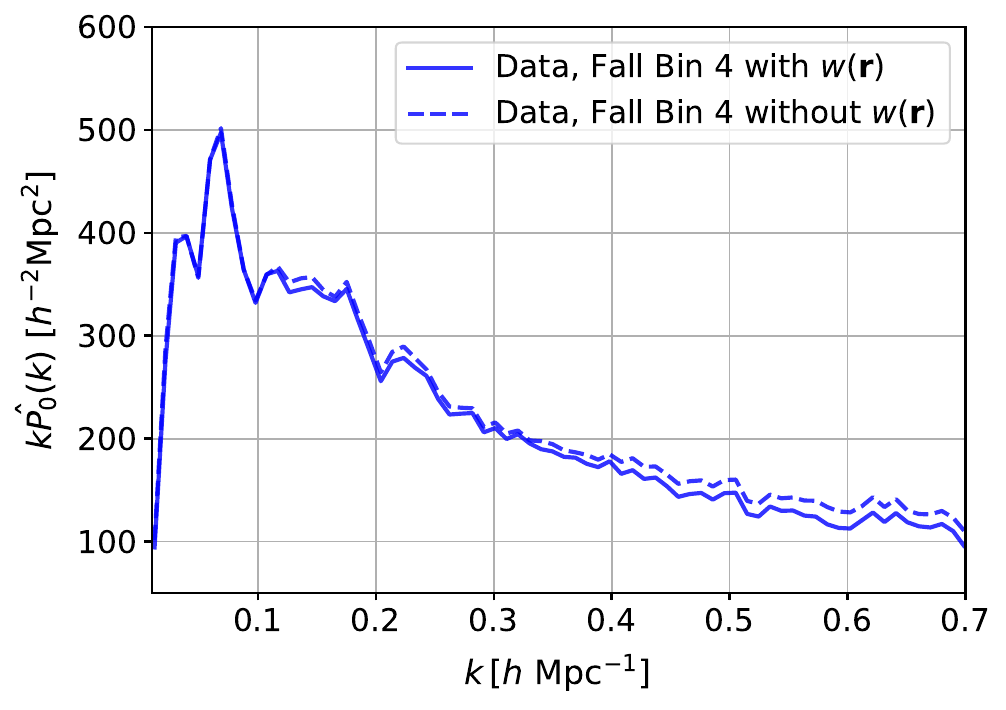}
    \includegraphics[width=0.48\columnwidth]{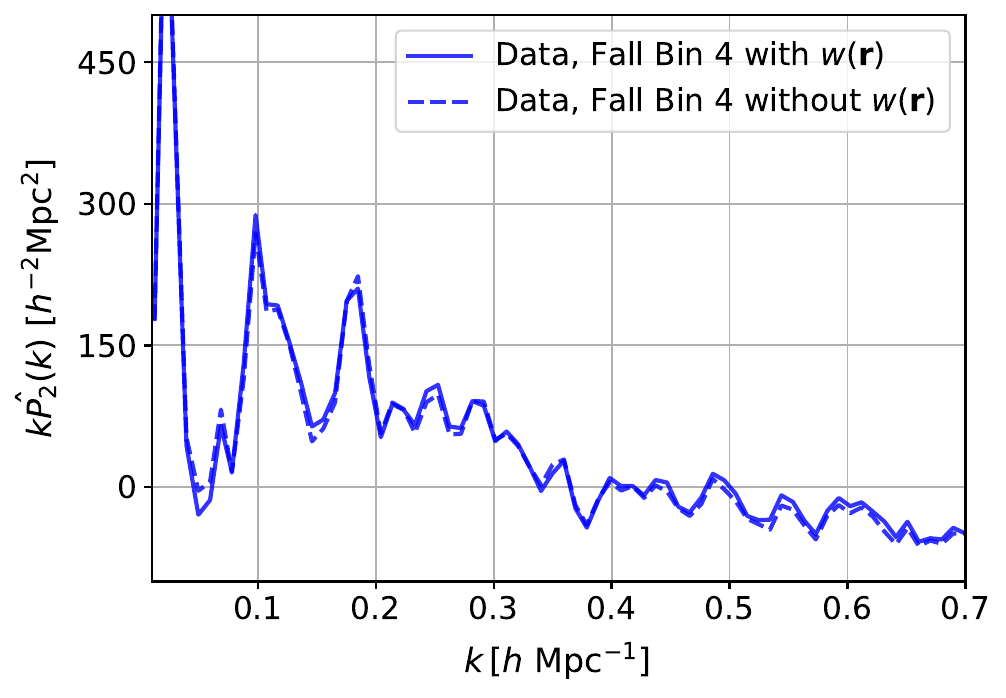}
    \caption{Monopole (left) and quadrupole (right) power spectra for Spring Bin 4 (top) and Fall Bin 4 (bottom), comparing results with (solid lines) and without (dashed lines) the weight, $w(\mathbf{r}) = N_\mathrm{overlap}^{-1}$.}
    \label{fig:plot_Noverlap_pk}
\end{figure}

Figure~\ref{fig:plot_Noverlap_pk} shows the monopole and quadrupole power spectra for Spring Bin 4 and Fall Bin 4 with and without the weight. This weight is either applied to both the data and random catalogs (solid lines), or omitted from both (dashed lines) for the comparison. Applying the weight lowers the monopole power spectrum amplitude at high $k$, since it prevents galaxies with $N_\mathrm{overlap} > 1$ from being counted multiple times. The effect on the quadrupole power spectra is minimal.

\section{Effect of satellite galaxies in power spectrum measurements}
\label{appendix:subhalo_exclusion}

This appendix describes how satellite galaxies, i.e., galaxies populated in dark matter subhalos, affect the power spectrum measurements, following the mock modeling described in Section~\ref{subsec:mock}. 

\begin{figure}
    \centering 
    \includegraphics[width=0.48\columnwidth]{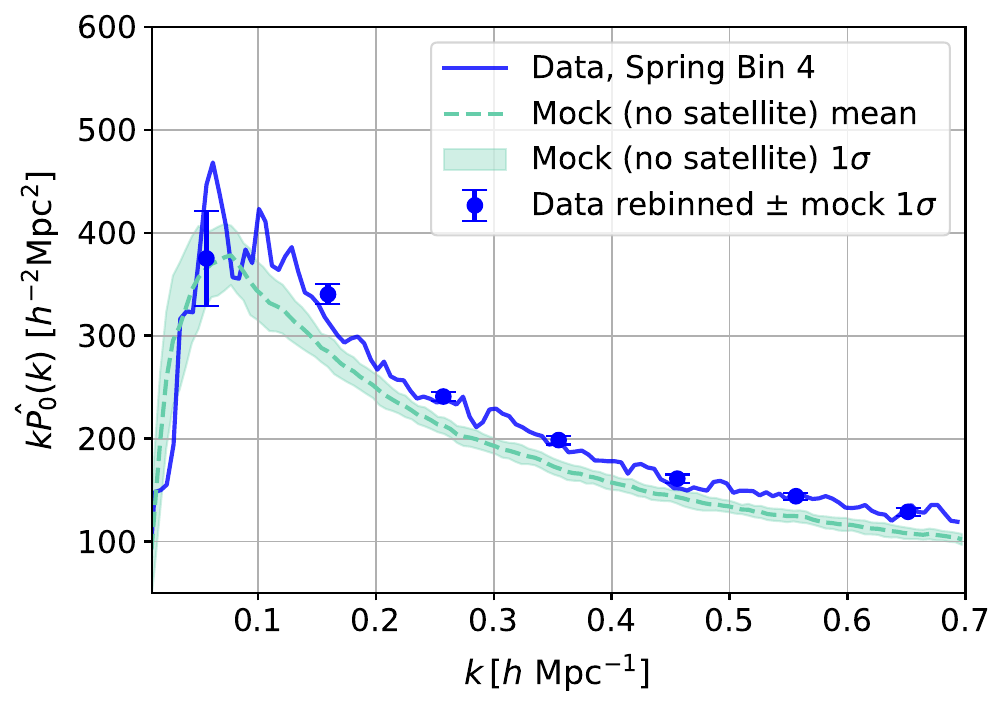}
    \includegraphics[width=0.48\columnwidth]{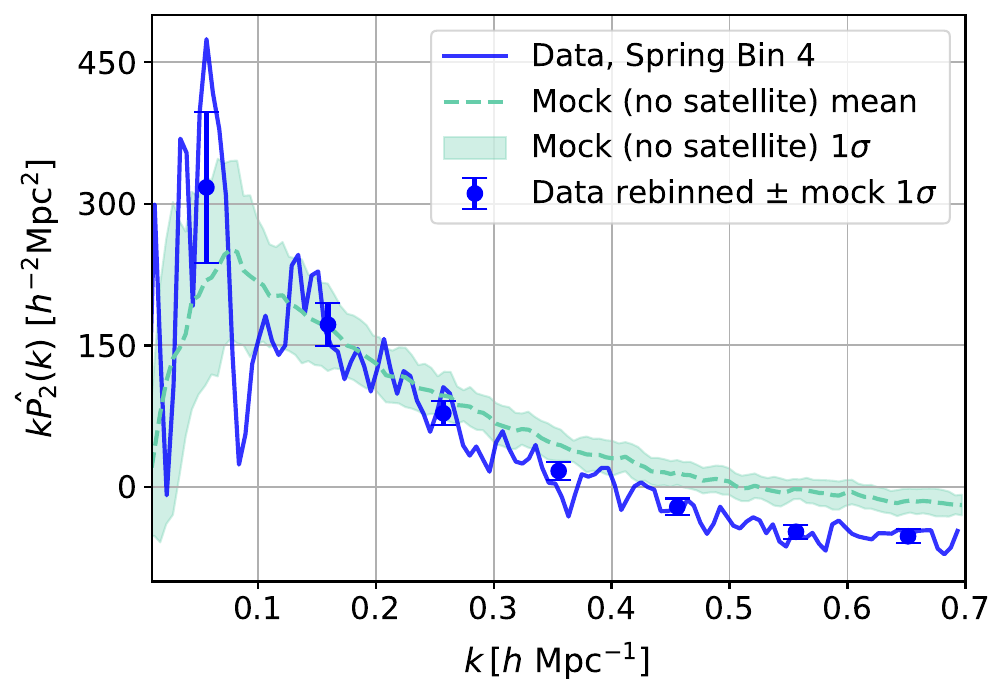}
    \caption{Monopole (left) and quadrupole (right) power spectra for the best-fit HOD mocks of Spring Bin 4, excluding galaxies populated in dark matter subhalos (dashed lines with shaded areas), compared to the data (points with error bars).}
    \label{fig:plot_pk_without_subhalos}
\end{figure}

Figure~\ref{fig:plot_pk_without_subhalos} compares the monopole and quadrupole power spectra  from the best-fit HOD mocks of Spring Bin 4 to those from the data, with satellite galaxies excluded from the mocks. 
The mock measurements show poor agreement with the data: the mock monopole amplitude is suppressed overall, while the mock quadrupole amplitude is higher at high $k$ than the data. The former effect can be compensated for by increasing $\log(M_0)$ (Figure~\ref{fig:plot_p0_M0_comparison}), but the latter cannot (Figure~\ref{fig:plot_p2_M0_comparison}). This indicates that populating satellite galaxies in subhalos is essential for accurately modeling the HETDEX [\ion{O}{2}] volume-limited samples.

\end{document}